\documentclass[12pt]{report}
\usepackage{urpasthesis}
\usepackage[square,sort&compress,comma,numbers]{natbib}
\usepackage{stmaryrd}
\usepackage[font=footnotesize,labelfont=bf]{caption}
\usepackage{enumerate}
\usepackage{graphicx}
\usepackage{amsmath}
\usepackage{amssymb}
\usepackage{amsthm}
\usepackage{amscd}
\usepackage{amsfonts}
\usepackage{makeidx}
\usepackage{bm}
\usepackage{tensor}
\usepackage{eucal}
\usepackage{epsfig}
\usepackage{latexsym}
\usepackage{mathrsfs}
\usepackage{verbatim,times,bbm}
\usepackage{dsfont}
\usepackage{algpseudocode}
\usepackage{algorithm}
\usepackage{mathtools}
\usepackage{hyperref}

\def\la{\langle}
\def\ra{\rangle}

\newcommand{\mean}[1]{\la#1\ra}                     
\newcommand{\ket}[1]{\vert#1\ra}                    
\newcommand{\bra}[1]{\la#1\vert}                    
\newcommand{\vect}[1]{\bm{\mathrm{#1}}}

\newcommand\iden{\ensuremath{\mathbbm{1}}}
\newcommand{\defeq}{\vcentcolon=}
\newcommand{\norm}[1]{\parallel #1 \parallel}

\theoremstyle{definition}
\newtheorem{definition}{Definition}[section]
\newtheorem{theorem}{Theorem}[section]
\theoremstyle{lemma}
\newtheorem{lemma}{Lemma}[section]

\hypersetup{
bookmarks=true,
pdfmenubar=true,
pdfstartview={FitH},
pdftitle={Characterizing High-Dimensional Optical Systems with Applications in Compressive Sensing and Quantum Data Locking},
pdfauthor={Daniel Jacob Lum},
pdfkeywords={Quantum Optics} {Compressive Sensing} {Quantum Data Locking} {FMCW LiDAR},
hidelinks,
pdfhighlight=/I}

\makeindex

\begin{document}

\title{Characterizing High-Dimensional Optical Systems with Applications in Compressive Sensing and Quantum Data Locking}
\author{Daniel Jacob Lum}
\thesissupervisor{Professor John C. Howell and Professor Robert W. Boyd}
\maketitle

\iftrue
\begin{dedication}
To my parents, Don and Genny Lum, for their endless encouragement and support.
\end{dedication}

\setcounter{tocdepth}{2}
\tableofcontents

\begin{biographicalsketch}
    The author was born and raised in the rural countryside just outside Clinton, Louisiana. He attended Louisiana State University (LSU) through the Louisiana Science Technology Engineering and Mathematics Scholarship where he earned a Bachelor of Sciences degree in Physics. While at LSU, he worked under the supervision of Professor Johnathan Dowling studying the results of photon loss within theoretical quantum-optics applications. During his summers, he gained experimental laboratory skills working as an intern in the labs of Professor Gregory Durgin at the Georgia Institute of Technology, Professor Robert Boyd at the University of Rochester, and both Dr. Kevin Silverman and Dr. Shellee Dyer while at the National Institute of Standards and Technology.

After graduating from Louisiana State University, he went to graduate school to study Physics at the University of Rochester under the supervision of Professor John Howell. There, he earned a Master's of Arts degree in Physics and actively published papers in the fields of quantum optics and computational imaging. The following publications resulted during his time at LSU and his doctoral study at the University of Rochester:

%
\noindent
\hangafter=1
\hangindent = 1.0cm
[1] \,\,\,  Petr~M. Anisimov, \underline{Daniel J. Lum}, S.~Blane McCracken, Hwang Lee, and
  Jonathan~P. Dowling.
\newblock An invisible quantum tripwire.
\newblock {\em New Journal of Physics}, 12(8):083012, 2010.

\noindent
\hangafter=1
\hangindent = 1.0cm
[2] \,\,\,  Gregory~A. Howland, \underline{Daniel J. Lum}, Matthew~R. Ware, and John~C.
  Howell.
\newblock Photon counting compressive depth mapping.
\newblock {\em Optics Express}, 21(20):23822--23837, Oct 2013.

\noindent
\hangafter=1
\hangindent = 1.0cm
[3] \,\,\,  Gregory~A. Howland, James Schneeloch, \underline{Daniel J. Lum}, and John~C.
  Howell.
\newblock Simultaneous measurement of complementary observables with
  compressive sensing.
\newblock {\em Physical Review Letters}, 112:253602, Jun 2014.

\noindent
\hangafter=1
\hangindent = 1.0cm
[4] \,\,\, Gregory~A. Howland, \underline{Daniel J. Lum}, and John~C. Howell.
\newblock Compressive wavefront sensing with weak values.
\newblock {\em Optics Express}, 22(16):18870--18880, Aug 2014.

\noindent
\hangafter=1
\hangindent = 1.0cm
[5] \,\,\,  \underline{Daniel J. Lum}, Samuel~H. Knarr, and John~C. Howell.
\newblock Fast {H}adamard transforms for compressive sensing of joint systems:
  measurement of a 3.2 million-dimensional bi-photon probability distribution.
\newblock {\em Optics Express}, 23(21):27636--27649, Oct 2015.

\noindent
\hangafter=1
\hangindent = 1.0cm
[6] \,\,\, James Schneeloch, Samuel~H. Knarr, \underline{Daniel J. Lum}, and John~C.
  Howell.
\newblock Position-momentum {B}ell nonlocality with entangled photon pairs.
\newblock {\em Physical Review A}, 93:012105, Jan 2016.

\noindent
\hangafter=1
\hangindent = 1.0cm
[7] \,\,\,  Gregory~A. Howland, Samuel~H. Knarr, James Schneeloch, \underline{Daniel J.
  Lum}, and John~C. Howell.
\newblock Compressively characterizing high-dimensional entangled states with
  complementary, random filtering.
\newblock {\em Physical Review X}, 6:021018, May 2016.

\noindent
\hangafter=1
\hangindent = 1.0cm
[8] \,\,\,  \underline{Daniel J. Lum}, John~C. Howell, M.~S. Allman, Thomas Gerrits,
  Varun~B. Verma, Sae~Woo Nam, Cosmo Lupo, and Seth Lloyd.
\newblock Quantum enigma machine: Experimentally demonstrating quantum data
  locking.
\newblock {\em Physical Review A}, 94:022315, Aug 2016.

\noindent
\hangafter=1
\hangindent = 1.0cm
[9$^\ast$] \,\, \underline{Daniel J. Lum}, Samuel~H. Knarr, and John~C. Howell.
\newblock Frequency-modulated continuous-wave {LiDAR} compressive
  depth-mapping.
\newblock {\em Pending} 

\noindent
\hangafter=1
\hangindent = 1.0cm
[10$^\ast$] Samuel~H. Knarr, \underline{Daniel J. Lum}, James Schneeloch, and John~C.
  Howell.
\newblock Compressive direct imaging of 268-million-dimensional optical
  phase-space.
\newblock {\em Pending}

\noindent
\hangafter=1
\hangindent = 1.0cm
[11$^\ast$] Thomas Gerrits, \underline{Daniel J. Lum}, Varun~B. Verma, John~C. Howell,
  Richard~P. Mirin, and Sae~Woo Nam.
\newblock Short-wave infrared compressive imaging of single photons.
\newblock {\em Pending}

\noindent
\hangafter=1
\hangindent = 1.0cm
[12$^\ast$] \underline{Daniel J. Lum}, Justin~M. Winkler, Samuel~H. Knarr, and John~C.
  Howell.
\newblock Slow-light interferometric frequency-modulated laser radar without a
  local oscillator.
\newblock {\em Pending}

\noindent
\hangafter=1
\hangindent = 1.0cm
[13$^\ast$] Justin~M. Winkler, \underline{Daniel J. Lum}, Samuel~H. Knarr, and John~C.
  Howell.
\newblock Measurement of kilohertz-level frequency shifts using a slow-light
  interferometer without a local oscillator.
\newblock {\em Pending}

\end{biographicalsketch}

\begin{acknowledgments}
    Both the work presented here and the opportunity to study Physics at the University of Rochester would not be possible without the help and guidance from numerous people.

I want to thank my parents, Don and Genny Lum, for their continual support in all my endeavors. I have my two sisters, Lauren and Beth, to thank for their constant encouragement and my two aunts, Sonia and Wilhelmina, for helping me to discover my passion for physics.

I thank my co-advisor, Professor John Howell, for continually encouraging me to seek new ideas, for helping to improving my lecturing and writing skills, and for the freedom he gave me to pursue my own research interests.

I also thank my other co-advisor Professor Robert Boyd and mentors -- Professor John Dowling, Professor Gregory Durgin, Dr. Kevin Silverman, and Dr. Shellee Dyer -- for granting me the opportunity to work in their labs or research groups, for teaching me invaluable skills, and for the strong letters of support that helped me get where I am today.

I am also indebted my lab colleagues. I would like to thank Gregory Howland for helping me to establish a productive graduate career by granting me the opportunity to work with him and for strengthening my programming skills. I thank James Schneeloch for helping me to understand information theory and for allowing me to work with him on his projects. I thank Samuel Knarr for several productive years in the lab, for the many thought-provoking discussions, and for the numerous invites to social gatherings outside the lab. I thank Justin Winkler for sharing his expertise with atomic systems for helping me to deepen my own understanding. Finally, I thank Curtis Broadbent, Bethany Little, Christopher Mullarkey, Juli{\'a}n Mart{\'\i}nez-Rinc{\'o}n, Gerardo Viza, and Joseph Choi for the useful discussions and for making the lab a pleasant experience.

I would like to thank the Department of Physics and Astronomy staff for their support and guidance. In particular, I would like to thank the graduate coordinator, Laura Blumkin, for her guidance and help with plotting a course designed to meet the school's requirements. I thank Mike Culver for his time and his uncanny ability to solve any lab or office problem and Connie Hendricks for her ability to make purchases and reimbursements a pleasant process.

Finally, I wish to thank Hannah for her ceaseless encouragement and compassion.

\end{acknowledgments}

\begin{abstract}
  High-dimensional systems are desired for their ability to transfer large amounts of information. This dissertation focuses on the characterization and usage of high-dimensional optical systems for computational imaging, high-dimensional entanglement, and efficient secure-information transfer. Within computational imaging systems, capturing the most spatial frequencies results in sharper images. Utilizing the correlations within high-dimensional entanglement offers signal-to-noise ratio enhancements over low-light imaging and spectroscopic systems. Finally, high-dimensional quantum channels offer a regime in which quantum data locking can encrypt information according to information-theoretic-secure standards more efficiently than classical systems.

While high-dimensionality offers certain performance gains, characterizing and then harnessing high-dimensional systems for computational imaging, entanglement-enhanced applications, and quantum-secure direct communication can be prohibitively difficult. This dissertation offers unique solutions to each of these problems. 

Compressive imaging is relied on heavily to improve measurement rates in limited-resource imaging systems. As such, compressive sensing is introduced in chapter 1 while entanglement and an experimental source of high-dimensional entangled photons is covered in chapter 2. Chapter 3 introduces Sylvester-Hadamard matrices for compressive measurement and efficient computational-recovery of high-dimensional correlations. Compressive imaging is then presented as an efficient means of converting a standard frequency-modulated continuous-wave LiDAR system into a high-resolution depth-imaging system in chapter 4. Chapter 5 introduces quantum data locking and presents one of the first experimental demonstrations made possible by the use of a large-area, high-efficiency, single-photon-counting detector array. For completeness, robust compressive sensing recovery algorithms using the alternating direction method of multipliers are presented in the appendix.

\end{abstract}

\begin{ContributorsAndFundingSources}
    Professors John C. Howell, Robert W. Boyd, Nicholas P. Bigelow, A. Nick Vamivakas, and Stephen L. Teitel served as members of the dissertation committee while Professor Zeljko Ignjatovic served as the dissertation committee's chair. 

To elaborate, Professor John Howell, from the Department of Physics and Astronomy, and Professor Robert Boyd, having a primary appointment at the Institute of Optics, were co-advisors. Professors Nicholas Bigelow and Stephen Teitel hold appointments in the Department of Physics and Astronomy while Professor Nick Vamivakas holds an appointment in the Institute of Optics. Professor Zeljko Ignjatovic holds an appointment in the Department of Electrical and Computer Engineering.

The results presented in chapters 3-5 were jointly produced and contributions are detailed in the following paragraphs:

Chapters 1-2 and the appendix serve as introductory material and summarize the works of others, and it should be noted that the SPDC derivation in chapter 2 was based on the derivation by James Schneeloch. This is explicitly emphasized within chapter 2 itself. Additionally, while the algorithms in the appendix were based on the works of others, I derived each equation in the iterative updates for each particular problem. Additionally, I modified the total-variation minimization algorithm to take advantage of fast-Hadamard transforms.

In chapter 3, Samuel Knarr helped with the experimental demonstration by acquiring the data. Both Samuel Knarr and Professor John Howell provided edits.

The compressive FMCW-LiDAR presented in chapter 4 was also aided by Samuel Knarr and Professor John Howell. They provided insightful comments to reduce the mathematical complexity and improved the overall clarity and of the article.

Chapter 5 is based on a theoretical description of quantum data locking published by Professor Seth Lloyd and Dr. Cosmo Lupo. Professor Howell and myself conceived of the idea for using spatial light modulators to phase modulate single photons as an experimental realization of quantum data locking. Dr. Sae-Woo Nam, Dr. Thomas Gerrits, and  Dr. M. S. Allman allowed me to use NIST facilities and an $8\times 8$ single-photon detecting nanowire array developed by Dr. Varun B. Verma at NIST. Dr. Lupo helped to derive the necessary key rates for secure communication. Dr. Nam, Dr. Gerrits, and Dr. Allman each helped build and automate the experiment while Dr. Nam managed the project. Additionally, Dr. Allman and Dr. Nam helped process the data.

The work presented here was funded by three grants:
The fast-Hadamard based compressive sensing paper was sponsored by the Air-Force Office of Scientific Research grant No. FA9550-13-1-0019. The compressive FMCW-LiDAR work was sponsored by the Air-Force Office of Scientific Research grant No. FA9550-16-1-0359. Finally, the quantum data locking collaboration was sponsored by DARPA Grant No. W31P4Q-12-1-0015.

\end{ContributorsAndFundingSources}

\listoffigures

\begin{acronyms}
\begin{tabular}{ll}
  ADMM & Alternating Direction Method of Multipliers \\
  AMCW & Amplitude Modulated Continuous Wave \\
  APD & Avalanche Photo-Diode \\
  BiBO & Bismuth Triborate \\  
  BS & Beam Splitter \\
  CS & Compressive Sensing \\
  DMD & Digital Micro-mirror Device \\
  EMCCD & Electron Multiplying Charge-Coupled Device \\
  EPR & Einstein-Podolsky-Rosen\\
  FMCW & Frequency-Modulated Continuous Wave \\
  HWP & Half-Wave Plate \\
  LiDAR & Light Detection and Ranging \\
  NIST & National Institute of Standards and Technology \\
  NSP & Null Space Property \\
  PBS & Polarizing Beam Splitter \\
  QDL & Quantum Data Locking \\
  QKD & Quantum Key Distribution\\
  QWP & Quarter-Wave Plate \\
  RIP & Restricted Isometry Property \\
\end{tabular}
\newpage
\begin{tabular}{ll}
  SLM & Spatial Light Modulator\\
  SPDC & Spontaneous Parametric Down-Conversion\\
  SNR & Signal-to-Noise Ratio \\
  TOF & Time of Flight \\
\end{tabular}

\end{acronyms}

\fi

\chapter{An Introduction to Compressive Sensing}\label{ch1}

\section{Introduction}

Compressive sensing (CS) is heavily relied on throughout this thesis. As such, it is beneficial to first answer the following questions that will undoubtedly arise. Typical questions include, What is compressive sensing? What is the minimum number of measurements needed to measure a signal? and How do I know I can trust the results from a compressive measurement, particularly in the presence of noise? These questions will be answered here.  The points presented in this chapter draw heavily from sources \cite{baraniuk2011introduction,vidyasagar2016introduction}. Additionally, this chapter is merely meant to introduce theorems used to gauge the reliability of CS. While the proofs for each theorem and lemma are not incomprehensible, adding them would add considerable length to the chapter. Because of this, the equivalent theorem number within \cite{baraniuk2011introduction} will be also be listed. Curious readers are encouraged to track down the online PDF and see the proof presented directly below each theorem and lemma. 

\section{Mathematical Prerequisites}

Before jumping into the fundamentals of CS, we first present commonly used notation. Notions of norms, sparsity, compressibility, the spark, and an indexing notation are presented here.

\subsection{Norms}
Normed vector spaces are vector spaces endowed with a norm.
\theoremstyle{definition}
\begin{definition}{}
A function $\parallel \cdot \parallel:\mathbb{R}^n\rightarrow\mathbb{R}_{+}$ is said to be a \textbf{norm} if it satisfies the following conditions:
\begin{enumerate}
    \item $\parallel \vect{x} \parallel\geq 0$ \,\,\,  $\forall\vect{x}\in\mathbb{R}^n$, \,\,\, and $\parallel \vect{x}\parallel = 0 \iff \vect{x} = \vect{0}$
    \item $\norm{c\vect{x}} = |c|\cdot \norm{\vect{x}} $, \,\,\,  $\forall c \in \mathbb{R}$, \,\,\,  $\forall\,\vect{x}\in\mathbb{R}^n$
    \item Triangle inequality: $\norm{\vect{x}+\vect{y}}\,\, \leq \,\, \norm{\vect{x}} + \norm{\vect{y}} \,\,\,\, \forall\,\vect{x},\vect{y}\in\mathbb{R}^n$
\end{enumerate}
\end{definition}
One such common norm is the $L^p$ norm defined as
\begin{equation}
    \norm{\vect{x}}_p = 
    \begin{cases}
        \left(\sum\limits_{i = 1}^n |x_i|^p\right)^{\frac{1}{p}} & p\in [1,\infty)\\
        \max\limits_{i = 1,2,...,n} |x_i| & p = \infty
    \end{cases}.
\label{eq:norm}
\end{equation}
For $p<1$, Eq. (\ref{eq:norm}) fails to satisfy the triangle inequality and is called a quasi-norm. 

\begin{definition}{}
The \emph{support} of a vector $\vect{x}$ is $\mathrm{supp}(\vect{x}) = \{i:x_i\neq0\}$ (i.e. the nonzero components of the vector $x$). Note that $\norm{\vect{x}}_0\,\coloneqq |\mathrm{supp}(\vect{x})|$, where $|\mathrm{supp}(\vect{x})|$ refers to the number of nonzero components. 
\end{definition}

Note that $\norm{\vect{x}}_0$ is not even a quasi-norm but arises in the limit $\lim\limits_{p\rightarrow 0}\norm{\vect{x}}_p^p$. Thus, the $L^0$-norm counts the number of nonzero components in a vector $\vect{x}$.

\begin{figure}
\centering\includegraphics[width=.5\textwidth]{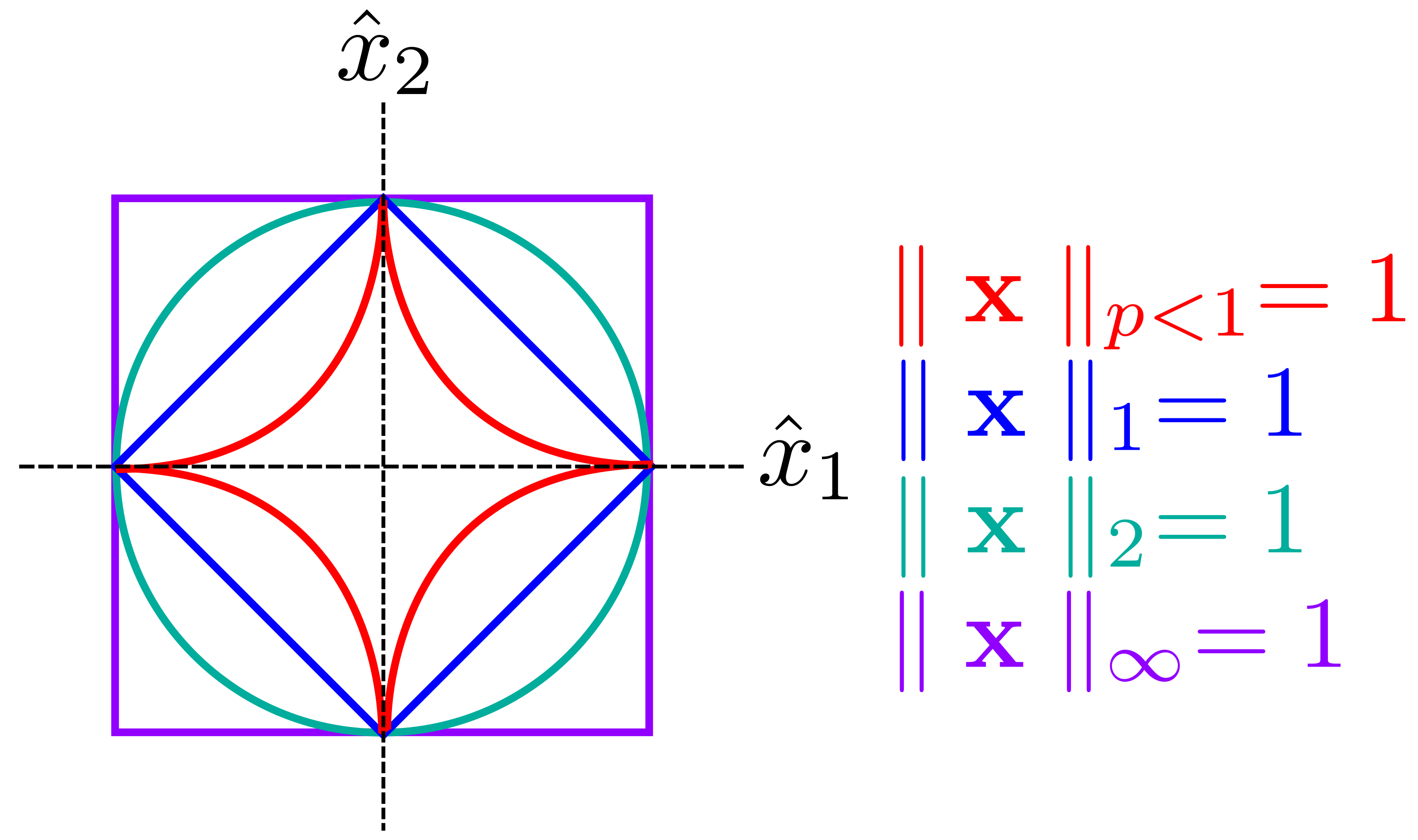}
\caption[Unit spheres for different norms]{Unit spheres for varying $L^p$-norms in a 2-dimensional space are presented here.}
\label{fig:unitsphere}
\end{figure}

The introduction of norms here will make more sense in the discussion of minimization. For a graphical intuition of norms, consider the unit spheres associated with each norm in $\mathbb{R}^2$. The unit spheres are plotted in Fig. \ref{fig:unitsphere} and are the result of fixing
\begin{equation}
\norm{\vect{x}}_p = ( |x_1|^p + |x_2|^p)^{1/p} = 1
\end{equation} 
for different values of $p$, $x_1$, and $x_2$. The $L^\infty$-norm forms a square, the $L^2$-norm (also known as the Euclidean norm) forms a circle with unit radius, the $L^1$-norm (also known as the taxi-cab norm) forms a diamond, and the $L^{p<1}$-norm approaches the shape of a $+$ as $p\rightarrow 0$.
As can be seen in Fig. \ref{fig:unitsphere}, different values of $p$ have unique properties that will be useful for quantifying error and justifying minimization techniques.

\subsection{Sparse and Compressible / Approximately-Sparse Signals}

\begin{definition}{}
A signal $\vect{x}$ is $K$-sparse when it has at most $K$ nonzeros: $\norm{\vect{x}}_0\,\leq K$
\end{definition}
Let
$\sum_K = \{\vect{x}:\norm{\vect{x}}_0\,\leq K\}$
denote the set of all $K$-sparse signals. As long as a signal is not just random noise, there exists a basis where the signal will have a sparse or, more commonly, approximately sparse representation. Signals that are approximately sparse are referred to as \emph{compressible} because they are well approximated by only a few signal components. For example, consider a signal $\vect{x}$ that has few nonzero components within a certain basis transformation $\Psi^{-1}$, i.e. $\vect{x}$ has a sparse representation $\vect{\alpha}$ such that $\vect{x} = \Psi \vect{\alpha}$. When using the basis vectors of $\Psi$, i.e. $\vect{\psi}_i$, which span the basis and are linearly independent, there exist unique coefficients $\{c_i\}_{i = 1}^n$ such that $\vect{x} = \sum_{i=1}^n c_i\vect{\psi}_i$. If $\vect{x}$ is compressible, the sorted coefficients (according to magnitude in descending order) $c_i$ exhibit a power-law decay:
\begin{equation}
|c_i| \leq C_1 i^{-q}, \,\,\, i = 1,2,...,
\end{equation}   
where $C_1$ and $q$ are constants. The larger $q$ is, the more compressible the signal is.
\begin{figure}
\centering\includegraphics[width=.8\textwidth]{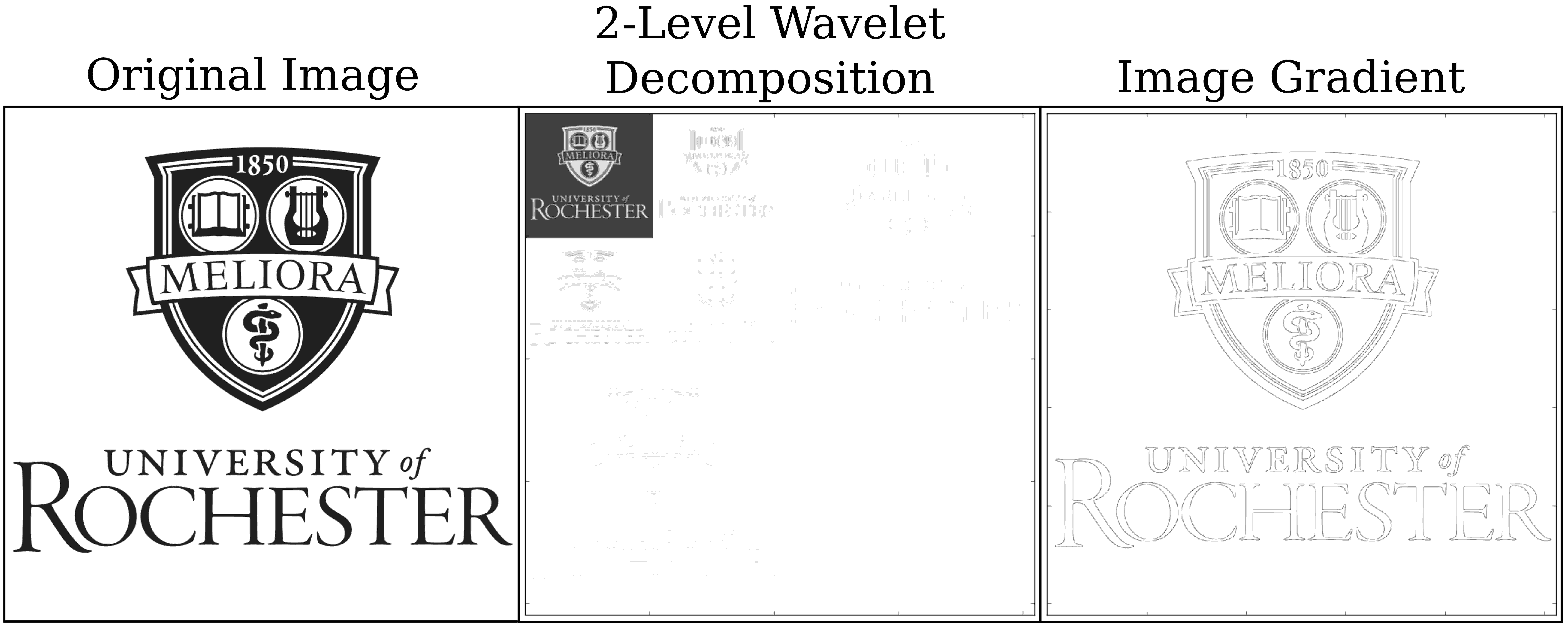}
\caption[Sparse basis-transform example]{From let to right: An image with its 2-level Haar-wavelet decomposition and image gradient.}
\label{fig:SparseExample}
\end{figure}
\begin{figure}
\centering\includegraphics[width=.9\textwidth]{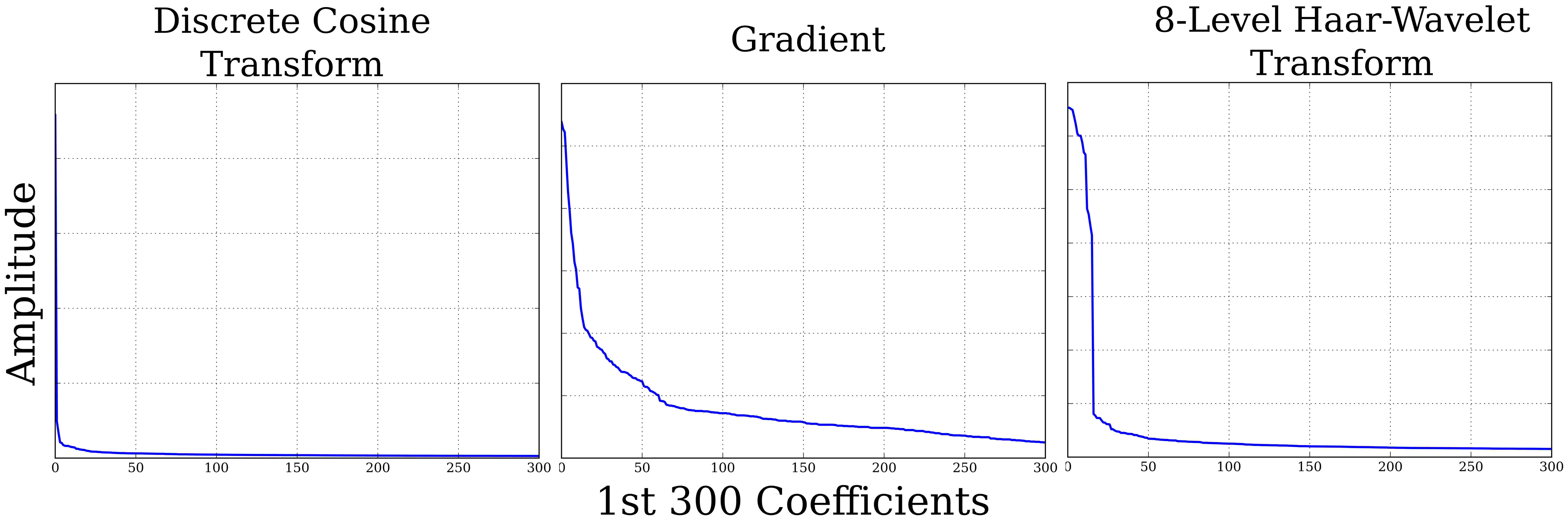}
\caption[Sorted sparse-transform coefficients example]{The coefficients of the original image within Fig. \ref{fig:SparseExample} associated with a discrete cosine transform, the gradient of the image, and an 8-level Haar-wavelet transform are shown.} 
\label{fig:Coefficients}
\end{figure}

As an example, consider Fig. \ref{fig:SparseExample}. The original image can be decomposed in a manner of ways. A two-level Haar wavelet transform \cite{chui1992wavelets,polikar1996wavelet} is presented along with the gradient of the image. The wavelet decomposition suggests that the image is well approximated by low-frequency components while the gradient suggests the image is well defined according to its edges. Thus, there exist several ways in which the image may be compressed.

Consider the sorted coefficients from a discrete cosine transform \cite{ahmed1974discrete}, the gradient, and an 8-level Haar wavelet transform of the original image in Fig. \ref{fig:SparseExample}. The sorted coefficients are presented in Fig. \ref{fig:Coefficients}. Each representation exhibits a power-law decay, with the discrete cosine transform having the largest $q$-value. Thus, this figure is highly compressible with respect to the cosine transform while the gradient appears to have the worst representation. In actuality, the gradient is not used as a sparse transform because it is not a unitary operation. However, as discussed in the appendix, an image can me minimized with respect to its gradient -- often with superior results compared to methods dependent on sparse unitary operations.

Figure \ref{fig:Coefficients} shows that a cosine transform exhibits better compression than a wavelet transform. JPEG compression is based on the discrete cosine transform, yet the new standard, JPEG2000, is based on wavelet transforms \cite{skodras2001jpeg}. One reason has to do with the differences between a cosine decomposition and a wavelet decomposition. A wavelet is confined spatially, having two tails that taper to 0 while a cosine is not confined spatially. Thus, any noise or adjustment to a cosine decomposition will affect a much larger area compared to a wavelet decomposition. For this reason, wavelet decompositions are more robust to noise and, oftentimes, better approximate the original signal when coefficients are neglected within lossy-compression protocols.  

Lossy compression is another way of thinking about approximate signal sparsity. Because the coefficients' magnitudes within the sparse basis, $|c_i|$, decay quickly in compressible signals, a signal can be accurately represented by $K\ll n$ coefficients. How we quantify the accuracy of the representation depends on our choice of the $L^p$ norm. The error between the true signal $\Psi\vect{\alpha}$ and the $K$-sparse approximation $\hat{\vect{x}}$ is
\begin{equation}
\sigma_K(\hat{\vect{x}})_p = \arg\min\limits_{\hat{\vect{x}}\in\sum_K}\norm{\Psi\vect{\alpha}-\hat{\vect{x}}}_p,
\label{eq:error}
\end{equation}
where $p$ is typically equal to either 1 or 2. When bounding reconstruction error in CS, the choice of $p$ in Eq. (\ref{eq:error}) can have serious implications for the number of required measurements when considering few outside assumptions.

\subsection{Spark of a matrix \texorpdfstring{$\vect{A}$}{A}}

\begin{definition}{}
The \textbf{spark} of a matrix $\vect{A}$, i.e. $\mathrm{spark}(\vect{A})$ , is the smallest number of columns of $\vect{A}$ that are linearly dependent. $\mathrm{spark}(\vect{A})=\min\limits_{\vect{h}\neq 0}\norm{\vect{h}}_0 \,\,\, \mathrm{s.t.}\,\,\,\vect{Ah}=0$ (meaning $\vect{h}$ resides within the null space of $\vect{A}$). 
\end{definition}
For example, let $\vect{A}\in\mathbb{R}^{2\times 4}$ be
\begin{equation}
\vect{A} = 
\left[
\begin{array}{cccc}
1 & 1 & -1 & 1 \\
1 & -1  & 0 & -2
\end{array}
\right].
\end{equation}
Letting $\vect{A}_i \,\, \textrm{for}\,\, i\in\{1,2,3,4\}$ be the columns of $\vect{A}$. Because $\vect{A}_1+\vect{A}_2+2\vect{A}_3 = \vect{0}$, $\mathrm{spark}(\vect{A})=3$.

\subsection{Indexing notation}
For $\vect{x}\in\mathbb{R}^n$, let $\Lambda \subseteq \{1,2,...,N\}$ denote the index set corresponding to the $K$ largest magnitude components of $\vect{x}$ (such that $|\Lambda| = K$). The remaining indices will form a set $\Lambda^C$ such that $\Lambda^C = \{1,2,...,N\} \setminus \Lambda$. To be clear, $|\Lambda|$ will denote the number of indices in $\Lambda$. 

As an example, let the vector $\vect{x} = [-10,9,8,0,-5,10]$ and $K=3$. The set $\Lambda$ will be
\begin{equation*}
\Lambda = \left\{ 1,2,6 \right\} \,\,\,\text{with}\,\,\, \Lambda^C=\{3,4,5\}.
\end{equation*}

\section{What is compressive sensing?}
Compressive sensing is a technique that trades a measurement problem for a computational reconstruction within limited-resource systems. An example of a limited-resource system would include imaging with a single-photon detector. Thus, we are mainly concerned with imaging applications. Perhaps the most well known example of a limited-resource system is the Rice single-pixel camera \cite{duarte2008single}. Instead of raster scanning a single-pixel to form an $n$-pixel resolution image $\vect{x}$, i.e. $\vect{x}\in\mathbb{R}^{n}$, an $n$-pixel digital micro-mirror device (DMD) takes $m\ll n$ random projections of the image. The set of all DMD patterns can be arranged into a sensing matrix $\vect{A}\in\mathbb{R}^{m\times n}$, and the measurement is modeled as a linear operation to form a measurement vector $\vect{y}\in\mathbb{R}^{m}$ such that $\vect{y}=\vect{Ax}$.

Once $\vect{y}$ has been obtained, we must reconstruct $\vect{x}$ within an undersampled system. As there are an infinite number of viable solutions for $\vect{x}$ within the problem $\vect{y} = \vect{Ax}$, CS requires additional information about the signal according to a previously-known function $g(\vect{x})$. The function $g(\vect{x})$ first transforms $\vect{x}$ into a sparse or approximately sparse representation within $\sum_K$, i.e. a representation with few, or approximately few, non-zero components. The function $g(\vect{x})$ then uses the $L^1$-norm to return a scalar. To simplify the notation slightly in the following sections, assume that $\vect{x}\in\sum_K$.

The next section will explain why $L^1$-minimization is used in the signal reconstruction and will present necessary and sufficient conditions for the sensing matrix $\vect{A}$ that will guarantee the uniqueness of a solution within the presence of noise.

\subsection{$L^1$-minimization}
Given a sampling matrix $\vect{A}\in\mathbb{R}^{m\times n}$, with $m<n$, and a measurement vector $\vect{y}\in\mathbb{R}^m$ such that $\vect{y} = \vect{Ax}$, we wish to find a solution $\hat{\vect{x}} = \vect{x}$, where $\vect{x}$ is the true solution and $\hat{\vect{x}}$ is our best estimate. To do so, we assume that the sparsest solution ($\arg\min_{\hat{\vect{x}}}\norm{\hat{\vect{x}}}_0$) that is consistent with the data ($\vect{y}=\vect{A}\hat{\vect{x}}$) is the correct solution. This is equivalent to minimizing the following equation:
\begin{equation}
\arg\min\limits_{\hat{\vect{x}}\in\sum_K}\norm{\hat{\vect{x}}}_0 \,\,\, \mathrm{subject \,\, to} \,\,\, \vect{y} = \vect{A}\hat{\vect{x}},
\label{eq:l0min}
\end{equation}
which is referred to as $L^0$-minimization.
Equation (\ref{eq:l0min}) requires searching for a $K$-sparse vector consistent with the data. For an $n$-dimensional space, this is equivalent to finding and then sorting $\binom{n}{K}$ possible solutions -- a prohibitively expensive operation. $L^0$-minimization does not have an efficient algorithm because the objective function is not convex.

\begin{figure}
\centering\includegraphics[width=.5\textwidth]{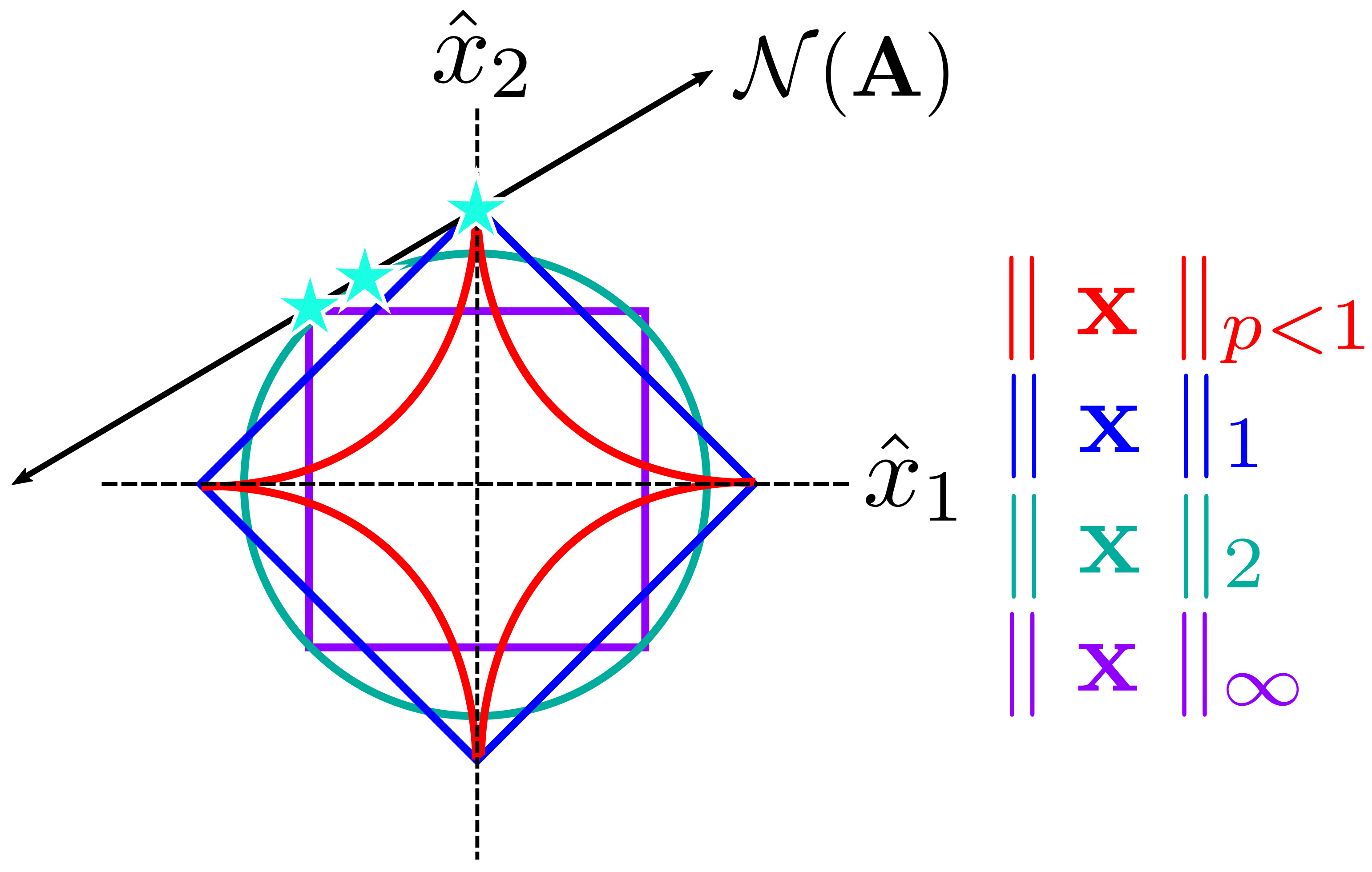}
\caption[Intersection of $L^p$-norm spheres with the null space of $\vect{A}$]{Spheres of the $L^p$-norm and their intersections (marked by stars) with the null space of a matrix $\vect{A}\in\mathbb{R}^{1\times 2}$, represented as $\mathcal{N}(\vect{A})$.}
\label{fig:L1Min}
\end{figure}

Alternatively, we can restate the $\vect{y}=\vect{A}\hat{\vect{x}}$ condition in terms of the null space of $\vect{A}$ -- represented as $\mathcal{N}(\vect{A})$. Given a solution $\hat{\vect{x}}$ consistent with the data, $\vect{y = \vect{A}\hat{\vect{x}}}$ means
\begin{align}
\vect{y} - \vect{A}\hat{\vect{x}}&=0 \\
\vect{Ax} - \vect{A}\hat{\vect{x}}&=0 \\
\vect{A}\left(\vect{x} - \hat{\vect{x}}\right)&=0.
\end{align}
Thus, all $\hat{\vect{x}}$ consistent with the data lie in $\mathcal{N}(\vect{A})$. Figure \ref{fig:L1Min} graphically depicts the result of finding the smallest spheres associated with $L^p$-norms in $\mathbb{R}^2$ that touch $\mathcal{N}(\vect{A})$. The minimum $L^p$-norm consistent with the data exists at the point where each unit sphere touches the solution set within $\mathcal{N}(\vect{A})$. The intersection points of each norm in Fig. \ref{fig:L1Min} are marked by a star. From the figure, we see that minimizing $\norm{\vect{x}}_{p<1}$ returns a sparse solution (having only 1 out of 2 nonzero vector components), but it is not a convex optimization problem. Alternatively, minimizing $\norm{\vect{x}}_{p\geq 2}$ returns a non-sparse solution. The only norm that is convex (allowing for efficient minimization) that returns a sparse solution is the $L^1$-norm. In fact, $L^1$-minimization is equivalent to the result obtained by $L^0$-minimization if the sensing matrix $\vect{A}$ meets certain criteria discussed below. Thus, CS is focused on solving the following problem:
\begin{equation}
\arg\min\limits_{\hat{\vect{x}}\in\sum_K}\norm{\hat{\vect{x}}}_1 \,\,\, \mathrm{subject \,\, to} \,\,\, \vect{y} = \vect{A}\hat{\vect{x}}.
\label{eq:L1Min}
\end{equation} 
Algorithms that are less sensitive to noise and measurement errors will solve the following equation:
\begin{equation}
\arg\min\limits_{\hat{\vect{x}}\in\sum_K}\norm{\hat{\vect{x}}}_1 \,\,\, \mathrm{subject \,\, to} \,\,\, \vect{y}-\vect{A}\hat{\vect{x}}\leq \vect{\epsilon},
\end{equation}
where $\vect{\epsilon}$ is a measurement error or noise.
A similar formulation, also relaxing the stringent $\vect{y} = \vect{A}\hat{\vect{x}}$ condition, requires solving the basis pursuit denoising problem \cite{gill2011crowd}, presented as
\begin{equation}
\arg\min\limits_{\hat{\vect{x}}\in\sum_K}\norm{\vect{y}-\vect{A}\hat{\vect{x}}}_2 + \lambda\norm{\hat{\vect{x}}}_1,
\label{eq:BPDN}
\end{equation}
where $\lambda\in\mathbb{R}_+$ is a constant that weights the $L^1$-regularization parameter. Note that Eq. (\ref{eq:BPDN}) is equivalent to solving the least absolute shrinkage and selection operator problem (LASSO) \cite{tibshirani1996regression}.

\subsection{Null Space Property}

\begin{figure}
\centering\includegraphics[width=.5\textwidth]{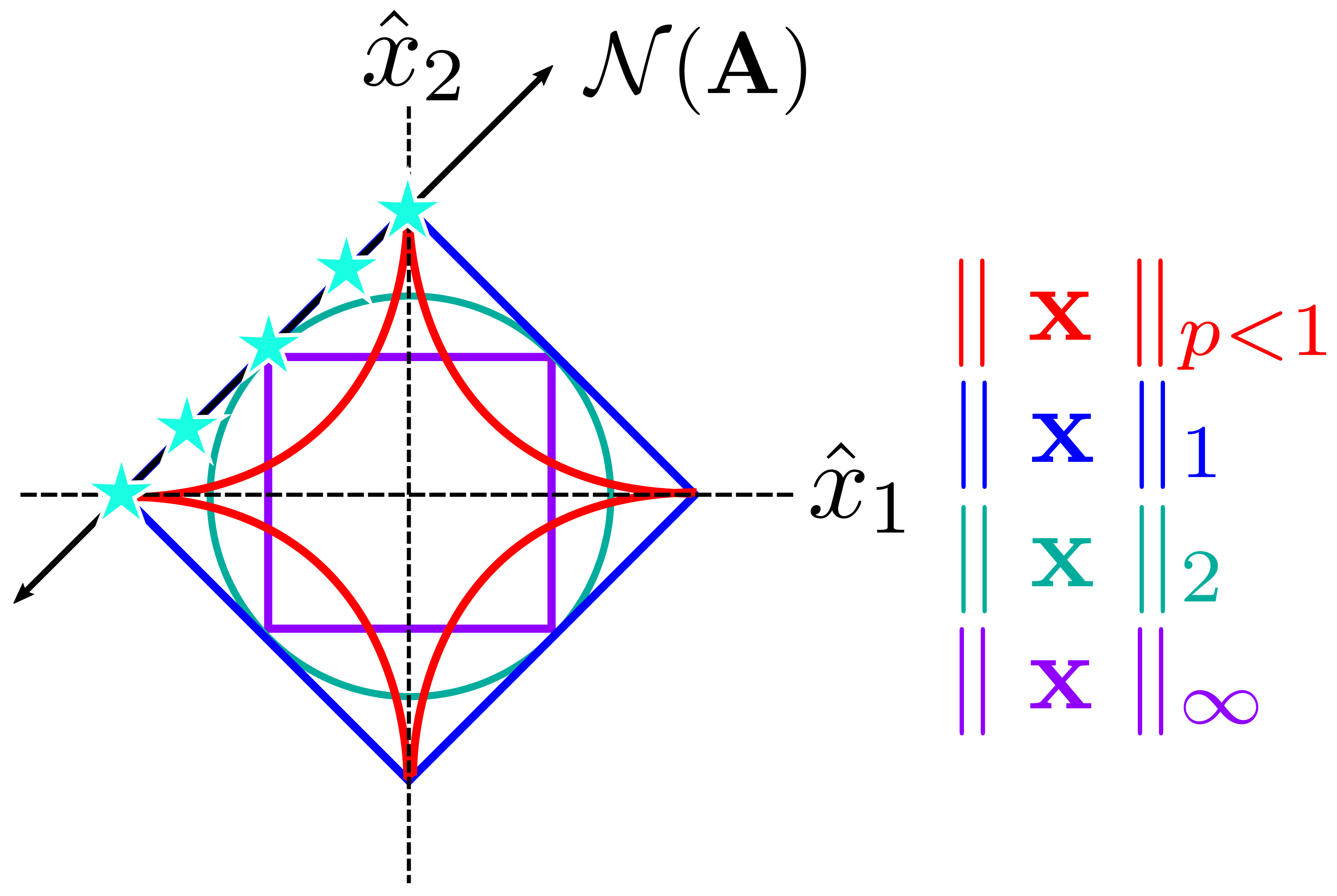}
\caption[Example of a matrix that does not satisfy the NSP]{Example of a matrix $\vect{A}$ that does not satisfy the null space property.}
\label{fig:NullSpace}
\end{figure}

The following sections will establish conditions that must be met by the sensing matrix $\vect{A}$ to guarantee a unique solution with and without measurement errors while also establishing the absolute minimum sample rate.

As previously stated, all possible solutions $\hat{\vect{x}}$ to the problem $\vect{y} = \vect{A}\hat{\vect{x}}$ exist in $\mathcal{N}(\vect{A})$, and we require a unique solution to the minimization problem presented in Eq. (\ref{eq:L1Min}). Before presenting the null space property, consider the null space presented in Fig. \ref{fig:NullSpace}. Again, stars denote the intersection points of a the $L^p$ spheres. Notice that $\mathcal{N}(\vect{A})$ does not intersect $\norm{\vect{x}}_1$ at a unique point. In addition to many non-sparse solutions, there exist two possible sparse solutions consistent with the data. We say that the matrix $\vect{A}$ in Fig. \ref{fig:NullSpace} does not satisfy the null space property.

\begin{theorem}
(Theorem 3.1 of \cite{baraniuk2011introduction}) For any vector $\vect{y}\in\mathbb{R}^m$, there exists at most one signal $\hat{\vect{x}}\in\sum_K$ (where $\sum_K$ is the set of all $K$-sparse vectors) such that $\vect{y}=\vect{A}\vect{\hat{x}}$  if  $\mathrm{spark}(\vect{A})\,>\,2K$.
\label{thm:nullspace} 
\end{theorem}
For any matrix $\vect{A}\in\mathbb{R}^{m\times n}$, $\mathrm{spark}(\vect{A})\in[2,m+1]$. Thus, theorem \ref{thm:nullspace} requires that $m\geq 2K$ -- which establishes a necessary minimum condition for any sensing matrix. 

Theorem \ref{thm:nullspace} is essential to guarantee uniqueness of sparse signals. If $\vect{x}$ and $\hat{\vect{x}}$ are each solutions consistent with the data, i.e. $\vect{Ax} = \vect{A}\hat{\vect{x}}$, then $\vect{A}(\vect{x}-\hat{\vect{x}})=0$ and $\vect{x}-\hat{\vect{x}}\in\sum_{2K}$. Thus, $\vect{A}$ uniquely represents all $\vect{x}\in\sum_{2K}$ iff $\mathcal{N}(\vect{A})$ contains no vectors in $\sum_{2K}$.

While theorem \ref{thm:nullspace} is necessary for sparse signals, most compressive sensing problems deal with compressible signals. This means we require more restrictions on $\mathcal{N}(\vect{A})$. In addition to sparse signals, we must also ensure that $\mathcal{N}(\vect{A})$ does not contain signals that are too compressible. Thus, we define the null-space property in the following definition.

\begin{definition}{}
(Definition 3.2 within \cite{baraniuk2011introduction}) A matrix $\vect{A}$ satisfies the null space property (NSP) of order $K$ if there exists a constant $C>0$ such that
\begin{equation*}
    \norm{\vect{h}_\Lambda}_2 \leq C \frac{\norm{\vect{h}_{\Lambda^C}}_1}{\sqrt{K}}
\end{equation*}
holds for all $\vect{h}\in\mathcal{N}(\vect{A})$ and for all $\Lambda$ such that $|\Lambda|\leq K$.
\label{def:nullspace}
\end{definition}
Definition \ref{def:nullspace} states that vectors in the null space of $\vect{A}$ should not be too concentrated on a small subset of indices. Another way to say this is to state that matrices having this property have very few sparse or compressible signals in $\mathcal{N}(\vect{A})$. For example, if $\vect{h}$ is exactly $K$-sparse, then there exists a $\Lambda^C$ such that $\norm{\vect{h}_{\Lambda^C}}_1 = 0$. Thus, if $\vect{A}$ satisfies the NSP, then the only $K$-sparse vector in $\mathcal{N}(\vect{A})$ is $\vect{h} = 0$. As previously stated, we actually require the NSP to be satisfied to order $2K$. How to design a matrix that satisfies the null space property of order $2K$ will be explained later.

For a direct implication of the null space property, consider the performance of a sparse recovery algorithm. Letting $\Delta:\mathbb{R}^m\rightarrow\mathbb{R}^n$ represent any recovery algorithm, we require
\begin{equation}
\norm{\Delta(\vect{Ax})-\hat{\vect{x}}}_2 \leq C\frac{\sigma_K(\vect{x})_1}{\sqrt{K}}
\label{eq:reconerror1}.
\end{equation}
Requirement (\ref{eq:reconerror1}) states that for a constant $C$ and a sparsity-approximation error given by Eq. (\ref{eq:error}), we can guarantee the recovery of exactly $K$-sparse signals as well as recover compressible signals that are well approximated by $K$-sparse signals up to a finite error. The following theorem relates Eq. \ref{eq:reconerror1} to the NSP.

\begin{theorem}
(Theorem 3.2 of \cite{baraniuk2011introduction}) Let $\vect{A}:\mathbb{R}^n\rightarrow\mathbb{R}^m$ denote a sensing matrix and $\Delta :\mathbb{R}^m\rightarrow\mathbb{R}^n$ denote any recovery algorithm. If the pair ($\vect{A}$,$\Delta$) satisfies Eq. (\ref{eq:reconerror1}), then $\vect{A}$ satisfies the NSP or order $2K$.
\label{thm:NSPorder2K} 
\end{theorem}

\subsection{Restricted Isometry Property}

The NSP provides necessary and sufficient conditions to satisfy Eq. (\ref{eq:reconerror1}) in \emph{noiseless} scenarios. To verify that CS will return a unique solution in the presence of noise, many theorems rely on a stronger condition provided by the restricted isometry property (RIP).

\begin{definition}{}
(Definition 3.3 of \cite{baraniuk2011introduction}) A matrix $\vect{A}$ satisfies the \textbf{restricted isometry property} (RIP) of order $K$ if there exists a $\delta_K\in (0,1)$ such that
\begin{equation*}
(1-\delta_K)\norm{\vect{x}}_2^2\,\,\leq\,\,\norm{\vect{Ax}}_2^2\,\,\leq \,\,(1+\delta_K)\norm{\vect{x}}_2^2
\end{equation*}
holds for all $\vect{x}\in\sum_K$.
\end{definition} 

The RIP of order $K$ is a measure for how well $\vect{A}$ approximately forms an orthonormal system for vectors in $\sum_K$ having a constant $\delta_K$ -- with a smaller $\delta_K$ being desired. In other words, the RIP is a measure for how well a matrix $\vect{A}$ maps $K$-sparse vectors from $\mathbb{R}^n$ to $\mathbb{R}^m$ (for $m<n$).

\begin{figure}
\centering\includegraphics[width=.7\textwidth]{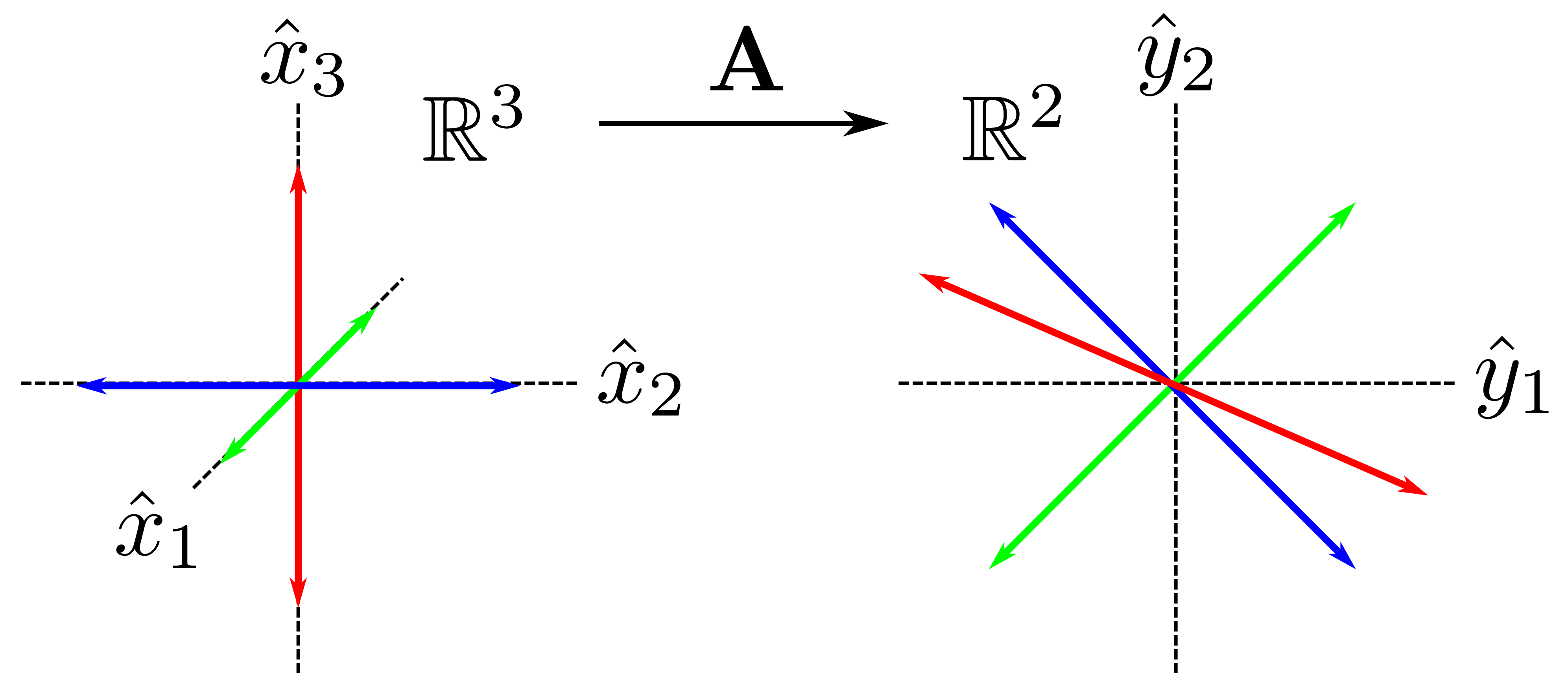}
\caption[Example of a matrix that satisfies the RIP]{A matrix $\vect{A}$ maps sparse vectors of $\sum_1$ in $\mathbb{R}^3$ to $\mathbb{R}^2$.}
\label{fig:JLLemma}
\end{figure}

To gain an intuitive understanding of the RIP, consider the result of mapping all possible sparse vectors of $\sum_1$ within $\mathbb{R}^3$ to $\mathbb{R}^2$, i.e $\vect{A}\in\mathbb{R}^{2\times 3}$.
\begin{equation}
\vect{A} = \left[
\begin{array}{ccc}
1 & 1  & 2 \\
1 & -1 & -1
\end{array}
\right]
\end{equation}
Thus, all sparse vectors in $\sum_1$ within $\mathbb{R}^3$ having vector components $[\hat{x}_1,\hat{x}_2,\hat{x}_3]$ are mapped to a measurement vector $\hat{\vect{y}}$ according to the equations
\begin{align}
\hat{y}_1 &= x_1 \hat{x} + x_2 \hat{y} + 2x_3 \hat{z} \\
\hat{y}_2 &= x_1 \hat{x} - x_2 \hat{y} - x_3 \hat{z}.
\end{align}
The mapping is depicted in Fig. \ref{fig:JLLemma}. Sparse vectors of $\sum_1$ in $\mathbb{R}^3$ are created by setting two elements within $[\hat{x}_1,\hat{x}_2,\hat{x}_3]$ to 0. Figure $\ref{fig:JLLemma}$ then plots the resulting components in $\mathbb{R}^2$. With this particular sensing matrix $\vect{A}$, all sparse vectors with $K=1$ in $\mathbb{R}^3$ are preserved when mapped to $\mathbb{R}^2$ and, consequently, can be transformed back to $\mathbb{R}^3$. Thus, this particular matrix $\vect{A}$ satisfies the RIP of order $2$ because it can uniquely preserve all $\vect{x}\in\sum_1$. In other words, $\mathcal{N}(\vect{A})$ contains no vectors in $\sum_2$.

In terms of uniqueness, if $\vect{A}$ satisfies the RIP of order $2K$, then $\vect{A}$ approximately preserves the distance between any pair of $K$-sparse vectors. This statement is akin to the NSP of order $2K$. In fact, the RIP is strictly stronger than the NSP.

\begin{theorem}
(Theorem 3.5 of \cite{baraniuk2011introduction}) Suppose that $\vect{A}$ satisfies the RIP of order $2K$ with $\delta_{2K}<\sqrt{2}-1$, then $\vect{A}$ satisfies the NSP or order $2K$ with constant
\begin{equation*}
C = \frac{\sqrt{2}\delta_{2K}}{1-\left(1+\sqrt{2}\right)\delta_{2K}}.
\end{equation*}
\end{theorem}

\subsection{Stability and Bounded Noise}

Within any compressive measurement, noise will be present and any compressive reconstruction algorithm must be \emph{stable} to noise. In other words, for any finite amount of noise, the impact on signal recovery should not be arbitrarily large. 

\begin{definition}{}
(Definition 3.4 of \cite{baraniuk2011introduction}) Let $\vect{A}:\mathbb{R}^n\rightarrow\mathbb{R}^m$ denote a sensing matrix and $\Delta :\mathbb{R}^m\rightarrow\mathbb{R}^n$ denote a recovery algorithm. The pair $(\vect{A},\Delta)$ is $C$-stable if for any $\hat{\vect{x}}\in\sum_K$ and any $\vect{e}\in\mathbb{R}^m$ when
\begin{equation*}
\norm{\Delta(\vect{Ax}+\vect{e})-\hat{\vect{x}}}_2 \leq C\norm{\vect{e}}_1.
\end{equation*}
\label{def:stability}
\end{definition}

Stability is an essential criteria within practical applications. While definition \ref{def:stability} works well for exactly $K$-sparse signals, it does not contain errors from sparsity approximations. The reconstruction error should also be bounded when handling compressible signals in the presence of measurement noise. While there are several methods of bounding this error, only one form will be presented here in the following theorem.

\begin{theorem}
(Theorem 4.2 of \cite{baraniuk2011introduction}) Let $\vect{A}$ satisfy the RIP of order $2K$ with $\delta_{2K}\leq\sqrt{2}-1$ and let $\vect{y}=\vect{Ax}+\vect{e}$ where $\norm{\vect{e}}_2\leq\epsilon$. Then the solution $\hat{\vect{x}}$ to the problem
\begin{equation*}
\arg\min\limits_{\hat{\vect{x}}}\norm{\hat{\vect{x}}}_1 \,\,\, \mathrm{subject \,\, to} \,\,\, \norm{\vect{A}\hat{\vect{x}}-\vect{y}}_2\leq\epsilon,
\end{equation*}
obeys 
\begin{equation*}
\norm{\hat{\vect{x}}-\vect{x}}_2\,\,\,\leq\,\,\,C_0\frac{\sigma_K(\vect{x})_1}{\sqrt{K}}+C_2\epsilon,
\end{equation*}
where
\begin{equation*}
C_0 = 2\frac{1-(1-\sqrt{2})\delta_{2K}}{1-(1+\sqrt{2})\delta_{2K}},\,\,\,
C_2 = 4\frac{\sqrt{1+\delta_{2K}}}{1-(1+\sqrt{2})\delta_{2K}}.
\end{equation*}
\label{thm:l1error}
\end{theorem}

Thus, it is crucial that $\vect{A}$ satisfies the RIP of order $2K$ with $\delta_{2K}\leq \sqrt{2}-1$. While it is difficult to deterministically construct a matrix row-by-row that satisfies the RIP, we will see it can be accomplished with high probability through randomness. This probabilistic nature extends to the reconstruction error -- being both stable and robust to measurement error and signal approximation with high probability.

\subsection{Satisfying the RIP}

We have already shown a unique solution for a $K$-sparse signal with $n$ components is only possible if $m \geq 2K$ for a measurement matrix $\vect{A}\in\mathbb{R}^{m\times n}$. We now show that by drawing the elements of $\vect{A}$ from a sub-Gaussian distribution, the RIP can be satisfied with high probability. 
\begin{definition}{}
A probability distribution of a random variable $X$ is \textbf{sub-Gaussian} if there exists a $\sigma >0$ such that
\begin{equation*}
\mathbb{E}\left[e^{tX}\right] \leq e^{\sigma^2 t^2/2}\,\,\, \forall \,\,\, t>0,
\end{equation*}
where $\mathbb{E}\left[\ast\right]$ is the moment generating function.
\label{def:subgauss}
\end{definition}
For discrete variables occurring with probability $p_i$, $\mathbb{E}\left[e^{tX}\right] = \sum_{i=1}^\infty e^{tx_i}p_i$. For continuous variables, $\mathbb{E}\left[e^{tX}\right] = \int_{-\infty}^\infty e^{tx}f(x)\,dx$.
Definition \ref{def:subgauss} is equivalent to saying that a distribution's tails decay at least as quickly as the tails of a Gaussian distribution. Thus, the Gaussian distribution, the Bernoulli distribution, and even uniform random variables chosen over a compact interval are sub-Gaussian. Alternatively, the Cauchy-Lorentz distribution, $(1/[\pi(1+x^2)])$, is not sub-Gaussian.

While we require at least $2K$ measurements to guarantee the uniqueness of a $K$-sparse signal, it is not sufficient to handle measurement noise or compressible signals. For example, consider a sensing matrix $\vect{A}\in\mathbb{R}^{m\times n}$ with $m = 2K$ and elements drawn from a Gaussian distribution. Any subset of $2K$ columns will be linearly independent and obey the RIP of order $2K$ for an unknown constant $\delta_{2K}$. To find the value of $\delta_{2K}$ within the RIP, we must consider all possible $K$-dimensional subspaces within $\mathbb{R}^n$. In other words, we must sift through $\binom{n}{K}$ possible combinations. Once we find $\delta_{2K}$, it is not guaranteed to be small. Alternatively, deterministically constructing matrices that satisfy the RIP of order $2K$ with a particular constant $\delta_{2K}$ is also difficult. 
The following theorem states that for a specified $\delta_{2K}$ of the RIP, we can still still randomly pick elements of $\vect{A}$ from a sub-Gaussian distribution if we include enough measurements $m$.
\begin{theorem}
(Theorem 4.5 of \cite{baraniuk2011introduction}) Fix $\delta\in(0,1)$ and let $\vect{A}\in\mathbb{R}^{m\times n}$ be a random matrix whose entries are i.i.d. and are drawn according to a sub-Gaussian distribution with $\sigma^2=1/m$ (using $\sigma$ from the sub-Gaussian definition \ref{def:subgauss}). If
\begin{equation*}
m \geq \kappa_1 K \log\left(\frac{n}{K}\right),
\end{equation*}
then \vect{A} satisfies the RIP of order $K$ with the prescribed $\delta$ with a probability exceeding $1-2e^{\kappa_2 m}$, where $\kappa_1$ is arbitrary and $\kappa_2 = \delta^2(1-\log(2))-\log(42e/\delta)/\kappa_1$. 
\label{theorem:subgauss}
\end{theorem}
Notice that theorem \ref{theorem:subgauss} states that for a smaller $\delta$ (as desired in the RIP), $m$ must be large enough to compensate for an adequate probability of satisfying the RIP.

\subsection{Required Measurements}
In the previous section, a restriction on the number of measurements arose when elements of the sensing matrix $\vect{A}$ were chosen randomly from a sub-Gaussian distribution. That particular measurement minimum is needed to ensure that $\vect{A}$ satisfies the RIP for a given $\delta_{2K}$. In this section, we consider the minimum number of measurements needed if we are given a sensing matrix $\vect{A}\in\mathbb{R}^{m\times n}$ and are informed that $\vect{A}$ satisfies the RIP of order $2K$ with constant $\delta_{2K}\in (0,1/2]$. 
\begin{theorem}
(Theorem 3.4 of \cite{baraniuk2011introduction}) Suppose $\vect{A}\in\mathbb{R}^{m\times n}$ satisfies the RIP of order $2K$ with constant $\delta_{2K}\in (0,1/2]$. Then
\begin{equation*}
m \geq cK\log\left(\frac{n}{K}\right),
\end{equation*}
where
\begin{equation*}
c = \frac{1}{2\log\left(\sqrt{24}+1\right)}\approx 0.28.
\end{equation*}
\end{theorem}
Note that the form presented here is similar to the form presented in theorem \ref{theorem:subgauss}.
Thus, to account for sources of noise and error, we design our systems to recover $K$-sparse vectors of length $N$ by using $\mathcal{O}(K\log(n/K))$ measurements -- meaning we multiply $K\log(n/K)$ by a constant $1<c<10$ dependent on the system and noise level.

Notice the efficiency of compressive sampling with random matrices. The compressive sampling rate is on the order of the information rate (or entropy) of the system; i.e. it is nearly optimal. For example, if trying to find the locations of $K$ items within $n\gg K$ bins, we require, on average, $K\log_2(n/K)$ yes/no questions. Finding the items is akin to finding the $K$ important elements of an $n$-dimensional vector while CS must \emph{also} find the $K$ amplitudes.

\subsection{Coherence}

The previous sections demonstrate how the spark, NSP, and RIP can be used to guarantee the recovery of a unique sparse signal that is stable under noise and measurement errors. However, because the presented theorems are dependent on the RIP and because deterministically constructing a matrix that satisfies the RIP of order $2K$ with constant $\delta_{2K}\in (0,1/2]$ is practically infeasible, we can only make these 
``guarantees" with high probability. For this reason, a significant amount of time has been dedicated to alternate, more tangible, results using the coherence of a matrix.

\begin{definition}{}
The \textbf{coherence} of a matrix $\vect{A}$, $\mu(\vect{A})$, is the largest absolute inner product between any two columns of $A_i$, $A_j$ of $\vect{A}$ such that
\begin{equation*}
\mu(\vect{A}) = \max\limits_{1\leq i\leq j \leq n}\frac{|\mean{A_i|A_j}|}{\norm{A_i}_2\norm{A_j}_2}
\end{equation*}
\end{definition}

In contrast to the RIP, the coherence of a sensing matrix is easily calculated and the coherence resides in the range $\mu(\vect{A})\in\left[\sqrt{\frac{n-m}{m(n-1)}},1\right]$. Similarly, the objective is to provide conditions for uniqueness of sparse recovery while also bounding the recovery error of noisy and approximately sparse (compressible) signals. Without going into detail, we present two results relating the coherence to the spark of a matrix and present conditions for uniqueness in ideal scenarios.

\begin{lemma}
(Lemma 3.5 of \cite{baraniuk2011introduction}) For any matrix $\vect{A}$,
\begin{equation*}
\mathrm{spark}(\vect{A})\geq 1 + \frac{1}{\mu (\vect{A})}.
\end{equation*}
\end{lemma}

Thus, any condition using the spark of a matrix can be represented in terms of the coherence -- namely, guarantees of uniqueness in the following theorem.

\begin{theorem}
(Theorem 3.8 of \cite{baraniuk2011introduction}) If 
\begin{equation*}
K < \frac{1}{2}\left(1+\frac{1}{\mu(\vect{A})}\right),
\end{equation*}
then for each measurement vector $\vect{y}\in\mathbb{R}^m$ there exists a unique signal $\hat{\vect{x}}$ such that $\vect{y}=\vect{A}\hat{\vect{x}}$.
\end{theorem}

The two results above state that matrices with little coherence, i.e. incoherent, are best for CS. Thus, we design our matrices to be as random as possible. Intuitively, this makes sense from a measurement point of view. When subsampling a signal, the sensing matrix must be able to partially sample all of the basis vectors to avoid leaving out information. Matrices that are incoherent have column entries that are dissimilar and can effectively sample all the basis vectors. Sensing matrices generated from randomness are naturally incoherent. The two results above can further be extended to include noise and approximately-sparse signals with the hope of arriving at smaller measurement bounds. This particular problem is an active area of research.  

\section{Conclusion}

To summarize, CS is a measurement technique that samples a signal $\vect{x}\in\mathbb{R}^n$ incoherently using $m < n$ projections from an incoherent sensing matrix $\vect{A}\in\mathbb{R}^{m\times n}$. The projections yield a measurement vector $\vect{y}\in\mathbb{R}^{m}$ from a linear operation $\vect{y}=\vect{Ax}$. Finally, $L^1$-minimization is used to recover the $K$ largest coefficients of a sparse signal representation of $\vect{x}$. To guarantee the uniqueness of a reconstructed solution $\hat{\vect{x}}$ in the event of infinite SNR, the NSP states that the sensing matrix must contain a minimum of $M = 2K$ measurements. Less than ideal applications will include measurement noise. Fortunately, the RIP states that error induced by both measurement noise and signal compression will be bounded, assuming we have $M = \mathcal{O}(K\log(n/K))$ measurements. Incidentally, this is also the number of measurements needed to construct a random sensing matrix that obeys the RIP with high probability. Additionally, we have restricted the discussion to real signals and sensing matrices for simplicity. Similar promises, such as sample ratios and error bounds, can also be obtained for complex valued signals and sensing matrices.

Up this point, both matrix construction and signal reconstruction algorithms have not been addressed. Matrix construction is the motivation of chapter \ref{ch3} and compressive imaging is the subject of chapter \ref{ch4}. Matrix construction is an important issue because the dimension of the sensing matrix scales quadratically with the dimension of the image. A measurement can be obtained compressively, but image reconstruction may be unfeasible if the sensing matrix is poorly designed.
Numerous reconstruction algorithms are presented in the literature. However, we focus on a particularly robust minimization technique using the alternating direction method of multipliers (ADMM) \cite{boyd2011distributed}. Due to the specificity of the reconstruction algorithm, this material may be found in the appendix. ADMM solvers are considered state-of-the-art because of their robustness and ease of implementation -- particularly with respect to minimizing complicated objective functions. Several minimization algorithms are presented in the appendix and include cases for both unitary and non-unitary signal transforms. The last algorithm presented is of a fast total-variation minimization reconstruction algorithm. Readers are encouraged to visit the appendix to gain an understanding of the method as well as to learn how to apply it to a specific problem.

\chapter{Entanglement from Spontaneous Parametric Down Conversion}\label{ch2}  

\section{Introduction}

Over the past few decades, the field of quantum information has been on the forefront of research and technological advancement. It began as a fundamental study in how quantum mechanics might impact the fields of computer science, information theory, and cryptography \cite{bennett1998quantum}. With the apparent realization of quantum information's potential, subsequent technologies and research areas including quantum secure communication \cite{wiesner1983conjugate,bennett1984quantum}, superdense coding \cite{PhysRevLett.69.2881}, quantum teleportation \cite{bennett1993teleporting,knill2001scheme,barrett2004deterministic}, quantum imaging \cite{lugiato2002quantum,boyd2012quantum}, and quantum computation \cite{gershenfeld1998quantum,benioff1980computer,feynman1982simulating,deutsch1985quantum,nielsen2002quantum} quickly emerged. Quantum information is the study of how information is held in a quantum system. Some of the unique characteristics that set quantum information apart from classical information result from the effects of quantization and quantum correlations. For example, refer to chapter \ref{ch5} to see how classical versus quantum communication channels can lead to drastic differences in security requirements.

This chapter will focus on quantum correlations within multipartite states, specifically those arising from entangled photons generated through spontaneous parametric down-conversion (SPDC). A multipartite state is describe as \emph{entangled} if the state cannot be factored as a product of individual single particle states \cite{PhysRevLett.80.2245}. 
Quantum entanglement has a rich history in the development of quantum mechanics, particularly with regard to the EPR paradox \cite{PhysRev.47.777}. As such, the historical context of the EPR paradox will be briefly covered before introducing entanglement. Because the work in chapter \ref{ch3} pertains to the characterization of high-dimensional correlations exhibited by position-momentum entanglement, we present a theoretical framework for photons generated through SPDC and show how they approximate the original EPR state \cite{howell2004realization}. 

\section{EPR Paradox}

In 1935, Einstein, Podolsky, and Rosen (EPR) proposed a \emph{gedanken} experiment in which they argued that quantum mechanics cannot be a complete theory \cite{PhysRev.47.777}. They were dissatisfied with an interpretation that arose from the Heisenberg uncertainty principle. A prominent interpretation of quantum mechanics is that the results of non-commuting observables need not be simultaneous elements of reality. If observables are \emph{elements of reality}, we mean they are well defined measurable properties that are independent of the observer. For non-commuting observables, a measurement of one observable will affect the other observable. This concept arose from the Heisenberg uncertainty principle which says that non-commuting observables, such as position and momentum, are not simultaneously well defined to infinite precision. More formally, if $\sigma_x = \sqrt{\mean{x^2}-\mean{x}^2}$ is the standard deviation in the measured position of a particle and $\sigma_p = \sqrt{\mean{p^2}-\mean{p}^2}$ is the standard deviation of the momentum of a particle, then
\begin{equation}
\sigma_x \sigma_p \geq \frac{\hbar}{2},
\end{equation}
where $\hbar$ is Plank's constant divided by $2\pi$. The EPR paper sought to prove that quantum mechanics does not present a complete description of reality by considering the results of measurement on the following two-particle EPR state:
\begin{equation}
\ket{\mathrm{EPR}} \equiv \int\limits_{-\infty}^{\infty}\ket{x_1,x_1}\,dx_1 \propto \int\limits_{-\infty}^{\infty}\ket{p_1,-p_1}\,dp_1,
\label{eq:EPR}
\end{equation}
where position and momentum are perfectly correlated (s.t. $x_2 = x_1$ and $p_2 = -p_1$). Critical to their argument, they assumed locality -- which states that interactions at one point do not immediately affect distant locations. Mathematically, the EPR paper presents the equivalent argument. Let the eigenvalue equations be
\begin{align}
\hat{x}\ket{x} &= x\ket{x} \\
\hat{p}\ket{p} &= p\ket{p}, 
\end{align}
where $\hat{x}$ and $\hat{p}$ are the position and momentum measurement operators, respectively, that return an observable position $x$ or an observable momentum $p$. Let the position of the first particle be measured by projecting it onto the state $\ket{x'}$.
\begin{align}
\mean{x'|\mathrm{EPR}} &= \int\limits_{-\infty}^{\infty}\mean{x'|x_1}\ket{x_1}dx_1 \\
&= \int\limits_{-\infty}^{\infty}\delta({x'-x_1})\ket{x_1}dx_1 \\
&= \ket{x'},
\end{align}
where $\delta$ a the Dirac-delta function. The unmeasured particle is immediately projected into a well-defined position state.
Alternatively, let the momentum of the first particle be measured by projecting it into the state $\ket{p'}$.
\begin{align}
\mean{p'|\mathrm{EPR}} &= \int\limits_{-\infty}^{\infty}\mean{p'|p_1}\ket{-p_1}dp_1 \\
&= \int\limits_{-\infty}^{\infty}\delta({p'-p_1})\ket{-p_1}dp_1 \\
&= \ket{-p'}
\end{align}
The unmeasured state is now in a well-defined momentum state. The EPR argument states that, because of locality, a measurement on the first particle has no impact on the second particle. Because of the state's correlated properties, we can infer the position or momentum of the second particle without measuring it. Thus, the second particle must have a well-defined position and momentum that exists as an element of reality. The EPR paper concludes by stating quantum mechanical theory must be incomplete and alluded to the existence of a complete theory.

The EPR paradox resulted in decades of theoretical and experimental research into quantum measurement. It showed that a measurement of a particle could be obtained by only interacting with its entangled partner. It also led to the discovery that a measurement of one particle would lead to randomness of the entangled particle's measured conjugate variable; quantum mechanics appeared to behave non-locally.
The EPR paradox was later recast into a discrete variable formalism using spin-1/2 particles by Bohm \cite{bohm1951quantum}. This led to a significant amount of work attempting to explain the correlations between entangled systems and resulted in the development of local hidden variable theories \cite{bohm1952suggested} that attempted to preserve the concept of locality \cite{blaylock2010epr}. Finally, Bell's theorem \cite{bell1995einstein}, which bounds the degree of correlations that can exist between classical systems, stated that no theory of local hidden variables could reproduce the predictions of quantum mechanics \cite{bell2004speakable} and numerous experiments have testified to this \cite{PhysRevLett.47.460,PhysRevLett.49.91,PhysRevLett.49.1804}. Today, most physicists accept that quantum mechanics is a complete nonlocal theory, as long as it does not violate causality through faster-than-light communication \cite{dieks1982communication}. Additionally, entangled states are the only systems that can violate a Bell inequality \cite{RevModPhys.81.865,werner2001bell}. Thus, entanglement is presented as a resources for nonclassical correlations and is essential to quantum information. As such, entanglement is more formally introduced in the next section.

\section{Entanglement}

Previously, we said that a multipartite state is describe as \emph{entangled} if it cannot be factored as a product of individual single-particle states. Because states can be described as pure or mixed, the mathematical definition of entanglement must take the state into consideration. Please note that a more complete guide to the formalism presented in the following section can be found in the following citations: \cite{wootters2001entanglement,PhysRevLett.80.2245,barnett2009quantumbook}.

\subsection{Pure and Mixed States}

A \textbf{pure state} $\ket{\psi}$ is defined as a vector of unit length in a complex Hilbert space $\mathcal{H}$ such that $\mean{\psi|\psi}=1$.
We often say that a pure state can be represented by a normalized wavefunction $\ket{\psi}$.
As an example, consider the superposition of pure states $\ket{0}$ and $\ket{1}$:
\begin{equation}
\ket{\Psi_{\pm}} = \frac{1}{\sqrt{2}}\left(\ket{0}\pm\ket{1}\right),
\label{eq:superposition}
\end{equation}
such that the result $\Psi$ is also a pure state.

However, not all quantum systems can be described by a pure state. Instead, we must often consider a classical ensemble of pure states, i.e. \emph{mixed states}. A mixed state is represented by a Hermitian, positive-semidefinite density matrix $\rho$ such that
\begin{equation*}
\rho = \sum\limits_i \rho_i \ket{\psi_i}\bra{\psi_i},
\end{equation*}
where $\rho_i$ is the probability for the system to be in the pure state $\ket{\psi_i}$. As such, $\sum_i\rho_i = 1$ infers the trace of the density matrix is $\mathrm{Tr}(\rho) = 1$. Mixed states can be differentiated from pure states easily due to the following property:
\begin{align*}
\mathrm{Tr}(\rho^2) &= 1 \,\,\, \text{for pure states} \\
\mathrm{Tr}(\rho^2) &< 1 \,\,\, \text{for mixed states}.
\end{align*}

As an example, consider the classical mixture of Eq. (\ref{eq:superposition}), with probability $1/2$ of being either $+$ or $-$:
\begin{equation*}
\rho = \frac{1}{2}\ket{\Psi_+}\bra{\Psi_+} + \frac{1}{2}\ket{\Psi_-}\bra{\Psi_-} 
= \frac{1}{2}\left(
\begin{array}{cc}
1 & 0 \\
0 & 1
\end{array}
\right),
\end{equation*}
when using the basis vectors $\ket{0}$ and $\ket{1}$ for the matrix.
For clarity, we can also represent the density matrix of Eq. (\ref{eq:superposition}) as
\begin{equation*}
\rho_{\pm} = \frac{1}{2}\left(
\begin{array}{cc}
1 & \pm 1 \\
\pm 1 & 1
\end{array}
\right).
\end{equation*}

\subsection{Entangled Pure States}

Here, we discuss entanglement of pure states. First consider two noninteracting systems, labeled $A$ and $B$, that have their own respective Hilbert spaces, labeled $\mathcal{H}_A$ and $\mathcal{H}_B$. The composite system is the tensor product of the Hilbert spaces such that $\mathcal{H}_{AB} = \mathcal{H}_A\otimes\mathcal{H}_B$. Given two states $\ket{\psi}_A$ and $\ket{\phi}_B$, each belonging to systems $A$ and $B$ respective, a pure state $\ket{\Psi}_{AB}$ is said to be \emph{entangled} if there exists a function $G(\psi,\phi)$ that cannot be factored into two functions of different variables such that $G(\psi,\phi) \neq g_A(\psi)g_B(\phi)$ within the state
\begin{equation}
\ket{\Psi}_{AB} = \iint\limits_{-\infty}^{\infty} G(\psi,\phi) \ket{\psi}\ket{\phi} \,\, d\psi \, d\phi,
\label{eq:entangledef}
\end{equation}
where it is understood that $\ket{\psi}\ket{\phi} = \ket{\psi}\otimes\ket{\phi}$.

As an example of an entangled pure state, consider let $\ket{\psi} = \ket{x_1}$ and $\ket{\phi} = \ket{x_2}$ and $G(x_1,x_2) = \delta(x_2-x_1)$. This results in the state
\begin{align*}
\ket{\Psi}_{AB} &= \iint\limits_{-\infty}^{\infty} \delta(x_2 - x_1) \ket{x_1}\ket{x_2} \,\, dx_1 \, dx_2 \\
&= \int\limits_{-\infty}^{\infty} \ket{x_1}\ket{x_1} \,\, dx_1 \\
&= \ket{\mathrm{EPR}}
\end{align*} 
Using the closure relation, $\int_{-\infty}^\infty \ket{k}\bra{k} dk = 1$, the fact that $p = \hbar k$, and $\mean{x|k}=\exp(ikx )/\sqrt{2\pi}$, we can show the equivalent momentum entangled state through the following:
\begin{align*}
\ket{\mathrm{EPR}} &= \int\limits_{-\infty}^{\infty} \ket{x_1}\ket{x_1}\,\,dx_1 \\
&= \iiint\limits_{-\infty}^{\infty} \ket{k_1}\mean{k_1|x_1}\ket{k_2}\mean{k_2|x_1}\,\,dx_1 \, dk_1 \, dk_2 \\
&= \frac{1}{2\pi} \iiint\limits_{-\infty}^{\infty} e^{ix_1(k_1+k_2)} \ket{k_1}\ket{k_2}\,\,dx_1 \, dk_1 \, dk_2 \\
&= \frac{1}{2\pi} \iint\limits_{-\infty}^{\infty} \delta(k_1+k_2) \ket{k_1}\ket{k_2}\,\, \, dk_1 \, dk_2 \\
&= \frac{1}{2\pi} \int\limits_{-\infty}^{\infty} \ket{k_1}\ket{-k_1}\,\, \, dk_1 \\
&= \frac{1}{2\pi\hbar^3}\int\limits_{-\infty}^{\infty} \ket{p_1}\ket{-p_1}\,\, \, dp_1
\end{align*}
Thus, the correlations are still present, despite the basis change. This trait is the hallmark of entanglement. The EPR state is an example of a pure state that exhibits continuous variable entanglement -- exhibiting correlations in continuous-variable degrees of freedom such as position, momentum, energy, and time. For a discrete form, convert the integral to a summation in Eq. (\ref{eq:entangledef}). A discrete variable form for $G(\psi,\phi)$ is now a vector $\vect{G}_{\psi,\phi}$. As an example, consider the combination of spin up / down pure states associated with two particles with $\mathcal{H}_A,\mathcal{H}_B\in\mathbb{C}^2$. Letting
\begin{equation*}
\vect{G}_{\psi,\phi} = \left[0_{\uparrow\uparrow};\, 1_{\uparrow\downarrow};\, -1_{\downarrow\uparrow};\, 0_{\downarrow\downarrow} \right],
\end{equation*}
the normalized discrete variable state becomes
\begin{equation*}
\ket{\Psi}_{AB} = \sum\limits_{\psi,\phi}\vect{G}_{\psi,\phi}\ket{\psi}_A\ket{\phi}_B = \frac{1}{\sqrt{2}}\Big(\ket{\uparrow}\ket{\downarrow}-\ket{\downarrow}\ket{\uparrow}\Big).
\end{equation*}
Because $\vect{G}_{\psi,\phi}$ cannot be factored into a tensor product of two vectors, i.e. $\vect{G}_{\psi,\phi}\neq \vect{g}_{\psi}\otimes\vect{g}_{\phi}$ for $\vect{g}_{\psi},\, \vect{g}_{\phi}\in\mathbb{C}^{2}$, the state is entangled. This particular example actually gives the largest quantum mechanical violation of Bell's theorem.

For a definitive way of establishing whether or not a state can be factored, it is helpful to consider the Schmidt decomposition and the Schmidt number.
\begin{theorem}
($\mathrm{Schmidt \,decomposition}$) Let $\mathcal{H}_A$ and $\mathcal{H}_B$ be two Hilbert spaces and $\ket{\Psi}_{AB}$ a normalized state in $\mathcal{H}_A\otimes\mathcal{H}_B$. Then there exists orthonormal basis sets $\{\ket{\psi_A}\}\in\mathcal{H}_A$, $\{\ket{\psi_A}\}\in\mathcal{H}_B$ such that
\begin{equation*}
\ket{\Psi}_{AB}= \sum\limits_\psi \sqrt{p_\psi}\ket{\psi_A}\otimes\ket{\psi_B},
\end{equation*}
where the $p_\psi$ are the non-zero eigenvalues of $\rho_A = \mathrm{Tr}_B(\ket{\Psi}\bra{\Psi})$. 
\end{theorem}
Note that $\mathrm{Tr}_B(\ast)$ is the partial trace and is defined in the following manner. Letting $\ket{a_i}$ be a basis of $\mathcal{H}_A$ and $\ket{b_i}$ be a basis of $\mathcal{H}_B$, any density matrix $\rho_{AB}$ on $\mathcal{H}_A\otimes\mathcal{H}_B$ can be decomposed as
\begin{equation*}
\rho_{AB} = \sum\limits_{ijkl} c_{ijkl}\ket{a_i}\bra{a_j}\otimes\ket{b_k}\bra{b_l}.
\end{equation*} 
The partial trace over system $B$ is then
\begin{equation*}
\mathrm{Tr}_B\left(\rho_{AB}\right) = \sum\limits_{ijkl}c_{ijkl}\ket{a_i}\bra{a_j}\mean{b_l|b_k}.
\end{equation*}
As an example of the partial trace, consider the state
$\ket{\Psi}_{AB}=(\ket{\uparrow}\ket{\downarrow}-\ket{\downarrow}\ket{\uparrow})/\sqrt{2}$. The corresponding density matrix is
\begin{align*}
\rho &= \frac{1}{2}\Big(\ket{\uparrow}\ket{\downarrow}-\ket{\downarrow}\ket{\uparrow}\Big)\Big(\bra{\uparrow}\bra{\downarrow}-\bra{\downarrow}\bra{\uparrow}\Big) \\
&= \frac{1}{2}\Big(\ket{\uparrow\downarrow}\bra{\uparrow\downarrow}+\ket{\downarrow\uparrow}\bra{\downarrow\uparrow}-\ket{\uparrow\downarrow}\bra{\downarrow\uparrow}-\ket{\downarrow\uparrow}\bra{\uparrow\downarrow}\Big),
\end{align*}
where we have assumed $\ket{a}\ket{b}=\ket{ab}$ for simplicity.
Tracing over system $B$, we obtain
\begin{align*}
\rho_A = \mathrm{Tr}_B(\rho) &= \frac{1}{2}\Big(\ket{\uparrow}\bra{\uparrow}\mean{\downarrow | \downarrow}+\ket{\downarrow}\bra{\downarrow}\mean{\uparrow | \uparrow}-\ket{\uparrow}\bra{\downarrow}\mean{\downarrow | \uparrow}-\ket{\downarrow}\bra{\uparrow}\mean{\uparrow | \downarrow}\Big) \\ 
&= \frac{1}{2}\Big(\ket{\uparrow}\bra{\uparrow}+\ket{\downarrow}\bra{\downarrow}\Big).
\end{align*} 
Referring back to the Schmidt decomposition, notice that the $p_\psi$ are the nonzero eigenvalues of $\rho_A$. We could also have used the partial trace $\rho_B$ in the definition since $\rho_A$ and $\rho_B$ have the same nonzero eigenvalues. The Schmidt decomposition is useful for obtaining the eigenvalues of subsystems $A$ and $B$. From the spectral decomposition, we can obtain the Schmidt number for a pure state -- defined as the number of non-zero eigenvalues for $\rho_A$ and $\rho_B$. The state $\ket{\Psi}_{AB}$ is entangled / non-separable if its Schmidt number is larger than 1 -- meaning it can be represented as a separable state $\ket{\psi_A}\otimes\ket{\psi_B}$ using the Schmidt decomposition.

We have introduced a separability criterion, but we should also discuss the degree of entanglement. Consider the following definition:
\begin{definition}{}
Let $\ket{\Psi}_{AB}\in\mathcal{H}_A\otimes\mathcal{H}_B$ be a pure state with $\mathrm{dim}(\mathcal{H}_A) \leq \mathrm{dim}(\mathcal{H}_B)$. Then $\ket{\Psi}_{AB}$ is \textbf{maximally entangled} if
\begin{equation*}
\rho_A = \frac{1}{\mathrm{dim}(\mathcal{H}_A)}\iden,
\end{equation*}
where $\iden$ is the identity matrix.
\end{definition}
Thus, a pure, separable state will have a reduced density matrix $\rho_A$ with only one nonzero eigenvalue (equal to 1) while a maximally entangled reduced density matrix will have all eigenvalues equal to $1/\mathrm{dim}(\mathcal{H}_A)$. To characterize the entanglement of pure states with values less than maximally mixed, we use an entropy metric. The most widely used metric for entanglement $E(\Psi)$ -- being equal to 0 for pure, separable states and having a dimension-dependent maximum for maximally entangled states -- is the von Neumann entropy $S$ of the reduced density matrix. Using the density matrix $\rho_{AB}=\ket{\Psi}\bra{\Psi}$, entanglement entropy is
\begin{equation}
E(\Psi) = S\left(\mathrm{Tr}_B\left(\rho_{AB}\right)\right) = S\left(\mathrm{Tr}_A\left(\rho_{AB}\right)\right) = -\sum\limits_{\psi=1}^n p_\psi \log_b (p_\psi),
\end{equation}
where the $p_\psi$ are the nonzero eigenvalues of the reduced density matrices.
As such, a non-separable state, having a reduced density matrix with one eigenvalue ($p_\psi =1$), has an entropy of 0 ebits (entangled bits when $b=2$). Alternatively, maximally entangled states $\Psi_\mathrm{max}$ will have $n = \mathrm{dim}(\mathcal{H}_A) < \mathrm{dim}(\mathcal{H}_B)$ nonzero eigenvalues, each with value $1/\mathrm{dim}(\mathcal{H}_A)$. Thus, $E(\Psi_\mathrm{max}) = \log_b(\mathrm{dim}(\mathcal{H}_A))$. Examples of maximally entangled pure states are the Bell states, presented as
\begin{align*}
\ket{\Phi^+} &= \frac{1}{\sqrt{2}}\Big(\ket{\downarrow}_A\ket{\downarrow}_B+\ket{\uparrow}_A\ket{\uparrow}_B\Big) \\
\ket{\Phi^-} &= \frac{1}{\sqrt{2}}\Big(\ket{\downarrow}_A\ket{\downarrow}_B-\ket{\uparrow}_A\ket{\uparrow}_B\Big) \\
\ket{\Psi^+} &= \frac{1}{\sqrt{2}}\Big(\ket{\downarrow}_A\ket{\uparrow}_B+\ket{\uparrow}_A\ket{\downarrow}_B\Big) \\
\ket{\Psi^-} &= \frac{1}{\sqrt{2}}\Big(\ket{\downarrow}_A\ket{\uparrow}_B-\ket{\uparrow}_A\ket{\downarrow}_B\Big).
\end{align*}

An intuitive way of thinking about the entropy of the state is to consider the number of ebits in the maximally entangled singlet state $\ket{\Psi^-}$. The singled state contains one ebit. In terms of resources, if a state $\rho_\Phi$ distributed between multiple parties has $E(\Phi) = n$ ebits, then the party has an entangled state with correlations that can be duplicated with the distribution of $n$ singlet states. Alternatively, they require $n$ maximally mixed Bell states to make a copy of $\ket{\Phi}$.

 
\subsection{Entangled Mixed States}

There has been much research into characterizing the entanglement that may exist within mixed states. This is because the quantum correlations must be distinguished from the classical correlations. The von Neumann entropy of a subsystem is a poor metric for mixed states because each subsystem has a non-zero entropy. Here, we introduce the \emph{entropy of formation} as introduced by Wooters \cite{wootters2001entanglement}.

Imagine trying to create $n$ copies of the mixed state $\rho$. The mixed state can be expressed as a decomposition of pure states according to
\begin{equation}
\rho = \sum\limits_{j=1}^{N} p_j \ket{\Phi_j}\bra{\Phi_j}, \,\,\, \text{such that} \,\,\, \sum\limits_{j=1}^{N}p_j = 1,
\end{equation}
such that the $\ket{\Phi_j}$'s are pure states that are not necessarily orthogonal with $p_j>0$. To make $n$ copies of $\rho$ requires that we make $np_j$ copies of the state $\ket{\Phi_j}$. If constructing each $\ket{\Phi_j}$ from the singlet states (as discussed in the last section), then we need $n p_jE(\Phi_j)$ singlet states. Finally, we collect the states into an ensemble and discard any information that could associate the index $j$ with each individual pure state. Thus, each pure state $\ket{\Phi_j}$ can exist with probability $p_j$ -- the definition of the mixed state $\rho$. From this argument, the number of singlet states $n_{\text{singlets}}$ needed to generate $n$ copies of the mixed state $\rho$ is
\begin{equation}
n_{\text{singlets}} = n\sum\limits_{j=1}^{N}p_jE(\Phi_j).
\label{eq:numsinglets}
\end{equation}
While Eq. (\ref{eq:numsinglets}) appears relatively straightforward, we have neglected to include the numerous possible decompositions of $\rho$. 
For example, we wish to make $n$ copies of the state $\rho_\mathrm{ex} = (1/2)(\ket{00}\bra{00}+\ket{11}\bra{11})$ using two spin $1/2$ particles. This mixed state $\rho$ could be decomposed into an equal mixture of $\ket{00}$ and $\ket{11}$ states. These pure states are separable and do not require the use a singlet state. Alternatively, $\rho$ could be decomposed into an equal mixture of $(1/\sqrt{2})(\ket{00}+\ket{11})$ and $(1/\sqrt{2})(\ket{00}-\ket{11})$ and would therefore require 2 singlet states. Taking the possible decompositions into account, the entropy of formation asks for the minimum number of singlet states needed to generate a copy of $\rho$. Finding the minimum number of singlet states requires that we find the greatest lower bound, i.e. the \emph{infimum}, of all possible pure state decompositions. For an example of the infimum, the positive real number line has no minimum -- since every number can be further subdivided -- yet, it has an infimum of 0. The entanglement of formation is then
\begin{equation}
E_f(\rho) = \mathrm{inf}\sum_j\rho_jE(\Phi_j).
\end{equation}
Note that $E_f(\rho)$ is zero if and only if $\rho$ can be written as a mixture of product states, meaning $\rho$ is separable. In the example where $\rho_\mathrm{ex} = (1/2)(\ket{00}\bra{00}+\ket{11}\bra{11})$, the state can be decomposed into an equal mixture of nonentangled states ($\ket{00}$ and $\ket{11}$), meaning $E_f(\rho_{\mathrm{ex}})=0$.

Overall, finding the entanglement of formation for a general state is a difficult task. However, a general formula exists for states consisting of a pair of quibits which we will discuss briefly. The entanglement of formation can be bounded from below by $\mathcal{E}(\mathcal{C}(\rho))\leq E_f(\rho)$, which is defined as
\begin{align}
\mathcal{E}(\mathcal{C}(\rho)) &= h\left(\frac{1+\sqrt{1-\mathcal{C}^2}}{2}\right) \,\,\, \text{with} \\
h(x) &= -x\log_2(x)-(1-x)\log_2(1-x),
\end{align}
where $\mathcal{C}(\rho)$ is the \emph{concurrence}. The concurrence is a metric for entanglement in itself and is defined for a pure state as $\mathcal{C}(\Phi) = |\mean{\Phi|\tilde{\Phi}}|$. The state $\ket{\tilde{\Phi}}$ is obtained by applying the ``spin-flip" operation, perhaps more commonly known as the Pauli operator $\hat{\sigma}_y$, to each particle. The operator $\hat{\sigma}_y$ is
\begin{equation*}
\hat{\sigma}_y = \left(
\begin{array}{cc}
0 & -i \\
i &  0
\end{array}
\right)
\end{equation*}
and is applied to each particle within the complex conjugate of $\ket{\Phi}$ (i.e. $\ket{\Phi^\ast}$) such that $\ket{\tilde{\Phi}} = (\sigma_y\otimes\sigma_y)\ket{\Phi^\ast}$. To see why the concurrence is a metric for entanglement (non-separability), observe the pure state below:
\begin{equation}
\ket{\Phi} = a\ket{00} + b\ket{01} + c\ket{10} + d\ket{11} = \left[
\begin{array}{c}
a \\
b \\
c \\
d
\end{array}
\right].
\label{eq:PureState}
\end{equation}
$\ket{\Phi}$ can only be factored if
\begin{equation*}
\Phi = \left[
\begin{array}{c}
a \\
b \\
c \\
d
\end{array}
\right] = 
\left[
\begin{array}{c}
\alpha \\
\beta 
\end{array}
\right] \otimes
\left[
\begin{array}{c}
\gamma \\
\theta
\end{array}
\right] =
\left[
\begin{array}{c}
\alpha \gamma \\
\alpha \theta \\
\beta \gamma \\
\beta \theta
\end{array}
\right],
\end{equation*}
which means the product of the first and last term $(\alpha\gamma)(\beta\theta)$ must be equal to the product of the second and third term $(\alpha\theta)(\beta\gamma)$, or $ad=bc$. Applying the concurrence to the state $\ket{\Phi}$ in Eq. (\ref{eq:PureState}) results in the expression $\mathcal{C}(\Phi)=2|ad-bc|$ and it becomes apparent that the concurrence, having values $0 \leq \mathcal{C} \leq 1$ , is a monotonically increasing function for how far from separable the pure state is. 

The concurrence is also defined for mixed states $\rho = \mathrm{inf}\sum_j p_j\ket{\Phi}\bra{\Phi}$ such that
\begin{equation}
\mathcal{C}(\rho) = \text{inf}\sum\limits_j p_j \mathcal{C}(
\Phi_j).
\end{equation}
An explicit formula for $\mathcal{C}(\rho)$ is
\begin{equation}
\mathcal{C}(\rho) = \max\{0,\lambda_1-\lambda_2-\lambda_3-\lambda_4\},
\end{equation}
where $\lambda_i$'s are the square roots of the eigenvalues of the density matrix $\rho\tilde{\rho}$ indexed in descending order ($\lambda_1 > \lambda_2 >...$). The matrix $\tilde{\rho}$ is defined as
\begin{equation}
\tilde{\rho} = (\sigma_y\otimes\sigma_y)\rho^\ast(\sigma_y\otimes\sigma_y).
\end{equation}
It should be noted that while $\mathcal{E}(\mathcal{C}(\rho))\leq E_f(\rho)$ for mixed states, the metric reaches equality for pure states: $\mathcal{E}(\mathcal{C}(\Phi))= E_f(\Phi)$.

To conclude this section, various characterization metrics for entanglement have been introduced. It should be mentioned that these metrics are useful for discerning the degree of entanglement, assuming one has access to the density matrix. Unfortunately, finding the density matrix can be a difficult task experimentally. Easier experimental metrics exist if one merely wishes to determine if a system exhibits non-separability. This can be accomplished through the violation of Bell and steering inequalities. Bell tests are not covered in this thesis, but steering inequalities are briefly introduced in chapter 3 -- mainly because mapping high-dimensional position / momentum entangled systems to a lower-dimensional form for a Bell test is not a trivial endeavor \cite{PhysRevA.93.012105}.   

\section{Spontaneous Parametric Down Conversion}

Now that entanglement and various characterization methods have been introduced (assuming one has access to the density matrix), we present a theoretical treatment for one of the most well known methods for generating sources of entanglement experimentally. While, entanglement has been experimentally demonstrated in many forms, e.g. electron spin \cite{hensen2015loophole}, neutrinos \cite{PhysRevLett.117.050402}, and superconducting circuits \cite{PhysRevLett.119.180511}, perhaps the most well known form of entanglement arises from the generation of photon pairs through spontaneous parametric down-conversion (SPDC). SPDC is a particularly simple mechanism for generating entangled photon pairs -- exhibiting entanglement in both discrete and continuous degrees of freedom. Discrete variable entanglement examples include the generation of the Bell states using polarization correlations present in type-II SPDC \cite{brida2007two,rangarajan2009optimizing,jeong2016bright}. Continuous variable entanglement, such as in energy-time and position-momentum, can be created through type-I and type-II SPDC \cite{howell2004,PhysRevLett.92.127903,PhysRevLett.75.4337}.

Much of our time in the lab has been spent attempting to characterize the correlations in high-dimensional position-momentum entanglement -- similar to the correlations first proposed within the EPR paradox \cite{PhysRevLett.112.253602,PhysRevX.3.011013,PhysRevX.6.021018,lum2015fast}. As such, we present a derivation of the position-momentum correlated photons that arise from the SPDC mechanism. This introduction is based on the work done by Schneeloch et al. \cite{schneeloch2016introduction} and Walborn et al. \cite{walborn2010spatial}, both of which are arguably the best and thorough investigations of SPDC as derived from first principles. It is highly recommended that the reader read those sources for a more detailed treatment of SPDC.

SPDC is a $\vect{\chi}^{(2)}$ nonlinear optical process occurring in a nonlinear, non-centrosymmetric medium \cite{boyd2003nonlinear} whereby a high-energy ``pump" photon is spontaneously converted into two lower-energy daughter photons, often referred to as ``signal" and ``idler". The process is parametric, i.e. energy conserving, which means no energy is absorbed by the crystal. Consequently, the daughter photons are highly correlated in their energy and momentum according to energy and momentum conservation. 

The derivation of the position and momentum correlations will follow this basic outline. We first introduce the electromagnetic field Hamiltonian and only consider contributions from the first two terms in the electric field's polarization Taylor expansion. We then replace the electric fields by the quantum mechanical field observables to arrive at the Hamiltonian needed to generate the state associated with SPDC after operating on the vacuum state. With a few assumptions with respect to the crystal's dimensions, the pump beam's diameter and strength, and the conservation of energy and momentum, we will arrive at an approximation to the EPR state.   





To begin, we present the electric field Hamiltonian as
\begin{equation}
H_{EM}\!(t) = \frac{1}{2}\int\limits_{V} d\vect{r}\,\left[\vect{D}\!(\vect{r},t)\cdot\vect{E}\!(\vect{r},t)+\vect{B}\!(\vect{r},t)\cdot\vect{H}\!(\vect{r},t)\right],
\end{equation}
where $\vect{D}$ is the displacement vector, $\vect{E}$ is the electric field, $\vect{B}$ is the magnetic induction, and $\vect{H}$ is the magnetic field. We can assume the material is nonmagnetic and ignore the $\vect{B}\cdot\vect{H}$ term. We also assume the displacement vector has the form $\vect{D} = \epsilon_0\vect{E}+\vect{P}$, where $\vect{P}$ is the electric field polarization. Our experiments pass a coherent optical field (the pump) through a nonlinear, non-centrosymmetric crystal. Because the electric field amplitude is significantly smaller than the electric field binding the atoms within the crystal, the electric field effectively ``tickles" or minutely perturbs the crystal's atoms according to the material's susceptibility $\vect{\chi}$. Thus, we can Taylor expand the field polarization about the linear component \cite{boyd2003nonlinear} so that
\begin{equation}
\vect{P} = \epsilon_0\left[\vect{\vect{\chi}}^{(1)}\vect{E}+\vect{\chi}^{(2)}\vect{E}^2+\vect{\chi}^{(3)}\vect{E}^3+\dotsc\right].
\end{equation}
Because the field interaction strength drops off as a power law, we keep only the first and second term within the expansion such that
\begin{align}
\vect{P}(\vect{r},t) &\approx \epsilon_0\vect{\vect{\chi}}_{ij}^{(1)}\!\vect{E}_j\!(\vect{r},t)
+ \epsilon_0\vect{\chi}_{ijl}^{(2)}\vect{E}_j(\vect{r},t)\vect{E}_l(\vect{r},t) \\
&= \vect{P}_{\mathrm{L}}(\vect{r},t) + \vect{P}_{\mathrm{NL}}(\vect{r},t), 
\end{align}
where we have specified a linear $\vect{P}_{\mathrm{L}}$ and a nonlinear $\vect{P}_{\mathrm{NL}}$ contribution. Here,
$E_j\!(\vect{r},t)$ is the $j$ component of the electric field vector at position $\vect{r}$ and time $t$ while $\vect{\chi}^{(1)}$ and $\vect{\chi}^{(2)}$ are the first and second order susceptibility tensors, respectively. Also note the use of the Einstein summation notation and the fact that $\vect{\chi}^{(n)}$ is an $n+1$ rank tensor. This approximation results in the Hamiltonian having linear and nonlinear components;
\begin{equation}
H_{EM}\!(t) = H_{L}\!(t) + H_{NL}\!(t),
\end{equation}
where
\begin{equation}
H_{NL} = \frac{\epsilon_0}{2}\int\limits_V d\vect{r} \, \vect{\chi}_{ijl}^{(2)} \vect{E}_i\!(\vect{r},t)\vect{E}_j\!(\vect{r},t)\vect{E}_l\!(\vect{r},t).
\end{equation}


The electric field can be quantized by assuming each field will result from operators that generate plane waves with positive and negative wave-vector contributions, confined within a box of volume $V$, with the relation
\begin{equation}
\vect{E}\!(\vect{r},t) = \hat{E}^+\!(\vect{r},t) + \hat{E}^-\!(\vect{r},t),
\end{equation}
where
\begin{align}
\hat{E}^{+}(\vect{r},t) &= \frac{1}{\sqrt{V}}\sum\limits_{\vect{k},s}i\sqrt{\frac{\hbar\omega\!(\vect{k},\vect{s})}{2\epsilon_0n^2\!(\vect{k},s)}}\vect{\epsilon}_{\vect{k},\vect{s}}e^{\left(i\vect{k}\cdot\vect{r}-\omega t\right)} \hat{a}_{\vect{k},s}\!(t) \label{eq:fieldop} \\
&= \left[\hat{E}^-\!(\vect{r},t)\right]^\dagger. \nonumber 
\end{align}
Within Eq. (\ref{eq:fieldop}), $V$ is the quantization volume (normally a cavity that contains the field modes), $n^2\!(\vect{k},s)$ is momentum and polarization dependent refractive index, $\hat{a}_{\vect{k},s}\!(t)$ is a photon annihilation operator at time $t$ that annihilates a photon with wave-vector $\vect{k}$ and frequency $\omega\!(\vect{k})$, and $\vect{\epsilon}_{\vect{k},s}$ is a unit-valued polarization vector having $s$ as an index for the polarization component. Of course $\hbar$ is Plank's constant divided by $2\pi$. The quantization volume $V$ will eventually be taken to infinity to describe free-space propagation.

Replacing the electric field variables results in the Hamiltonian
\begin{equation}
\hat{H} = \sum\limits_{\vect{k},s}\hbar\omega\!(\vect{k})\hat{a}_{\vect{k},s}^{\dagger}\!(t)\hat{a}_{\vect{k},s}\!(t) 
+ \frac{\epsilon_0}{2}\int\limits_V d\vect{r}\vect{\chi}_{ijl}^{(2)}\hat{E}_i\!(\vect{r},t)\hat{E}_j\!(\vect{r},t)\hat{E}_l\!(\vect{r},t).
\label{eq:OppHamiltonian}
\end{equation}
With each electric field operator having 2 terms, the nonlinear component of the Hamiltonian in Eq. (\ref{eq:OppHamiltonian}) actually contains 8 terms, yet not all terms are energy conserving. Because we are only interested parametric processes, we only include terms that preserve the relation $\omega_i(\vect{k_p}) = \omega_j(\vect{k}_1)+\omega_l(\vect{k}_2)$. Here, we have assigned $\omega_i$ to the pump, $\omega_j$ to the signal, and $\omega_l$ to the idler photons. Energy conservation means we can only keep terms having $a^\dagger_i a_j a_l$ and $a_i a_j^\dagger a_l^\dagger$ operators. The resulting Hamiltonian is
\begin{align}
\hat{H} &= \sum\limits_{\vect{k}_p,s}\hbar\omega\!(\vect{k}_p)\hat{a}_{\vect{k}_p,s}^{\dagger}\!(t)\hat{a}_{\vect{k}_p,s}\!(t) 
+ \frac{\epsilon_0}{2}\int\limits_V d\vect{r} \, \vect{\chi}_{ijl}^{(2)}\hat{E}_i^+\!(\vect{r},t)\hat{E}_j^-\!(\vect{r},t)\hat{E}_l^-\!(\vect{r},t) + \mathrm{H.C.} \nonumber\\
&= \hat{H}_{\mathrm{L}} + \hat{H}_{\mathrm{NL}},
\label{eq:EnergyConservHamil}
\end{align}
where $\mathrm{H.C.}$ is the Hermitian conjugate.

At this point, we can make several approximations to simplify Eq. (\ref{eq:EnergyConservHamil}). We can assume that $n^2(\vect{k},s) \approx 1$. The rate of pair generation is minuscule in comparison to the rate at which pump photons are passing through the crystal. In a typical experiment, roughly 1 pair will be generated for every $10^8-10^{10}$ pump photons \cite{PhysRevA.77.043834}, depending on the crystal used. Hence, we can replace the pump field operator by the classical field in the undepleted-pump approximation such that $\hat{E}^{\pm}_i(\omega(\vect{k_p}))\rightarrow \vect{E}_i(\omega(\vect{k_p}))$. Following in the direction of \cite{schneeloch2016introduction} to make the expressions less complicated later on, we define the transverse momenta for the pump, signal, and idler as $\vect{q}_p$, $\vect{q}_j$, and $\vect{q}_l$, respectively. Each $\vect{q}$ is a projection of each $\vect{k}$ into the $\hat{x}$ and $\hat{y}$ coordinate plane. We then assume the pump's frequency is sufficiently stable / narrow-band to factor out the time dependence -- a practical assumption, particularly for continuous-wave sources. These assumptions allow us to express the classical pump field as an integral over plane waves:
\begin{equation}
\vect{E}_i(\vect{r},t) = \frac{1}{2\pi}\int\limits_A d\vect{q}\, \vect{E}_i(\vect{q}_p,t)e^{i(\vect{q}_p\cdot\vect{r})}e^{i(k_{zp}z-\omega_p t)},
\end{equation}
where the integral is taken over the 2D plane within the $\hat{x}$ and $\hat{y}$ coordinates.
Next, we factor the vector polarization component out of the pump's electric field such that $\vect{E}_i(\vect{q}_p,t)=E_i(\vect{q}_p,t)\vect{\epsilon}_{\vect{k}_p,s_p}$. Putting it all together, we have the Hamiltonian
\begin{align}
\hat{H} = \hat{H}_{\mathrm{L}}
&+ \frac{\epsilon_0}{4\pi}\int\limits_V d\vect{r} \,\int\limits_A d\vect{q}
\frac{-1}{V}\sum\limits_{\vect{k}_1,s_1}\sum\limits_{\vect{k}_2,s_2}
\vect{\chi}_{ijl}^{(2)}(\vect{\epsilon}_{\vect{k}_1,s_1})_j(\vect{\epsilon}_{\vect{k}_2,s_2})_l(\vect{\epsilon}_{\vect{k}_p,s_p})_i  \nonumber \\
&\times \sqrt{\frac{\hbar^2\omega\!(\vect{k}_1)\omega\!(\vect{k}_2)}{4\epsilon_0^2}}
e^{-i(\Delta\vect{q})\cdot\vect{r}}e^{-i(\Delta k_z z)}e^{-i\omega_p t}
E_i\! (\vect{q}_p,t)a_{\vect{k}_1,s_1}^\dagger\!(t)a_{\vect{k}_2,s_2}^\dagger\! (t)+ \mathrm{H.C.}, \label{eq:AlmostThere}
\end{align}
where $\Delta\vect{q} \equiv \vect{q}_1+\vect{q}_2-\vect{q}_p$ and $\Delta k_z \equiv k_{1z}+k_{2z}-k_{pz}$. Note that, in general
\begin{equation}
\vect{\chi}_{ijl}^{(2)} = \vect{\chi}_{ijl}^{(2)}\big(\vect{r};\omega\!(\vect{k}_p),\omega\!(\vect{k}_1),\omega\!(\vect{k}_2)\big).
\end{equation}
We introduce additional assumptions to simplify Eq. (\ref{eq:AlmostThere}). First, we assume the nonlinear crystal is isotropic so that $\vect{\chi}$ has no $\vect{r}$ dependence. This allows us to carry out the integral over $\vect{r}$. We assume that the crystal has a anti-reflective coating to prevent multiple reflections. The volume integral assumes the crystal is centered at $\vect{r}=\vect{0}$ within a Cartesian coordinate system, i.e. $d\vect{r}= dx\,dy\,dz$. Remembering that
\begin{equation}
\int\limits_{-L_x/2}^{L_x/2}e^{ikx}dx = L_x\mathrm{sinc}\left(\frac{kL_x}{2}\right),
\end{equation}
where $\mathrm{sinc(x)\equiv sin(x)/x}$, evaluating the volume integral results in the Hamiltonian
\begin{align}
\hat{H} = \hat{H}_{\mathrm{L}}
&+ \frac{\epsilon_0}{4\pi}\int\limits_A d\vect{q}
\frac{-L_x L_y L_Z}{V}\sum\limits_{\vect{k}_1,s_1}\sum\limits_{\vect{k}_2,s_2}
\vect{\chi}_{ijl}^{(2)}(\vect{\epsilon}_{\vect{k}_1,s_1})_j(\vect{\epsilon}_{\vect{k}_2,s_2})_l(\vect{\epsilon}_{\vect{k}_p,s_p})_i  \nonumber \\
&\times \sqrt{\frac{\hbar^2\omega\!(\vect{k}_1)\omega\!(\vect{k}_2)}{4\epsilon_0^2}}
\mathrm{sinc}\!\left(\frac{\Delta q_x L_x}{2}\right)\mathrm{sinc}\!\left(\frac{\Delta q_y L_y}{2}\right)\mathrm{sinc}\!\left(\frac{\Delta k_z L_z}{2}\right)e^{-i\omega_p t} \nonumber \\
&\times E_i\! (\vect{q}_p,t)a_{\vect{k}_1,s_1}^\dagger\!(t)a_{\vect{k}_2,s_2}^\dagger\! (t)+ \mathrm{H.C.},
\end{align}

The Hamiltonian can be simplified further through a few more approximations. First, we will assume that evaluating the sums of the tensor components with the basis vectors will result in an experimentally defined constant $\chi$. Second, we can assume that the polarizations of the signal and idler photons are fixed, meaning the sums over $s$ are no longer needed. We will also logically assume that the crystal's transverse dimensions are significantly larger than the wavelengths of our electric fields. This allows us to replace the summation over $\vect{k}$ with 
\begin{equation}
\lim\limits_{V\rightarrow\infty} \frac{1}{V}\sum\limits_{\vect{k}_i} = \frac{1}{(2\pi)^3}\int d^3k_i.
\end{equation}

Because the pair generation rate associated with $\hat{H}_{\mathrm{NL}}$ is so small, we can use first-order time-dependent perturbation theory within the interaction picture to model the state's evolution at time $t$. If the state at $t=0$ is $\ket{\psi(0)}$, then the it will evolve at time $t$ to
\begin{equation}
\ket{\psi(t)} = \exp\left(\frac{1}{i\hbar}\int\limits_0^t d\tau H_{NL}\!(\tau)\right) \ket{\psi(0)}.
\end{equation}
We can expand the time evolution operator to first order to obtain
\begin{equation}
\ket{\psi(t)} \approx \left(1 + \frac{1}{i\hbar}\int\limits_0^t d\tau H_{NL}\!(\tau)\right)\ket{\psi(0)}.
\label{eq:TimeEvolution}
\end{equation}
Letting the initial state be the vacuum state, $\ket{\psi(0)} = \ket{0}_1\ket{0}_2$, we can neglect the Hermitian conjugate component of $\hat{H}_{\mathrm{NL}}$ because $\hat{a}\ket{0}=0\ket{0}$.
Additionally, the raising operators will evolve as $a^\dagger(t)\rightarrow e^{i\omega t}a^\dagger(0).$
As a result, the simplified nonlinear Hamiltonian is
\begin{align}
\hat{H}_{\mathrm{NL}} 
&= \chi\frac{- L_x L_y L_z \hbar}{64\pi^4}\int\limits_A d\vect{q}
\int d\vect{k}_1 \int d\vect{k}_2 \\
&\times \sqrt{\omega\!(\vect{k}_1)\omega\!(\vect{k}_2)}
\left[\prod\limits_{m = 1}^{3}\mathrm{sinc}\!\left(\frac{\Delta q_m L_m}{2}\right)\right]e^{i\Delta\omega t} \nonumber \\
&\times E\! (\vect{q}_p,t)a^\dagger\!(\vect{k}_1)a^\dagger\! (\vect{k}_2)
\end{align}
where the product is taken over $m = \{1,2,3\} = \{x,y,z\}$ coordinates and $\Delta\omega \equiv \omega_1+\omega_2-\omega_p$. To simplify the notation here within the product $\prod$, we let $\Delta k_z = \Delta q_z$.
Finally, if we assume that the pump intensity is relatively constant (as with continuous-wave pump sources or with respect to the time it takes a nanosecond pulse to propagate through a centimeter length crystal), then $E(\vect{q},t)$ essentially has, on average, no time dependence and is proportional to the square root of the pump intensity $I_p$ times the transverse mode structure of the pump beam $\mathrm{A}\!(\vect{q})$ such that $E(\vect{q},t)\rightarrow \mathrm{A}\!(\vect{q})\sqrt{I_p}$.
To find the time evolved state at time $T$, i.e the time it takes light to traverse the crystal, we must evaluate the integral in \ref{eq:TimeEvolution} from time $t = 0$ to $t = T$. Because the only time-dependent term is $e^{i\Delta\omega}$, the integral evaluates to
\begin{equation}
\int\limits_0^Te^{i\Delta\omega t} = \frac{e^{i\Delta\omega T}-1}{\Delta\omega} = \frac{T}{2}e^{i\frac{\Delta\omega T}{2}}\mathrm{sinc}\left(\frac{\Delta\omega T}{2}\right).
\end{equation}

Thus we arrive at the down-conversion state
\begin{align}
\ket{\Phi}_{\mathrm{SPDC}} &\approx C_0\ket{0_1,0_2} +C_1 \sqrt{I_p}T\iint d\vect{k}_1\,d\vect{k}_2\,\Phi(\vect{k}_1,\vect{k}_2) \\
&\times \sqrt{\omega\!(\vect{k}_1)\omega\!(\vect{k}_2)}e^{i\frac{\Delta\omega T}{2}}\mathrm{sinc}\left(\frac{\Delta\omega T}{2}\right) \hat{a}^\dagger\!(\vect{k}_1)\hat{a}^\dagger\!(\vect{k}_2)\ket{0_1,0_2},
\end{align}
where
\begin{equation}
\Phi(\vect{k}_1,\vect{k}_2)\equiv \int\limits_A d\vect{q}_p 
\left[\prod\limits_{m = 1}^{3}\mathrm{sinc}\!\left(\frac{\Delta q_m L_m}{2}\right)\right]
\mathrm{A}\!(\vect{q}_p).
\label{eq:phik}
\end{equation}
Note that $\Phi(\vect{k}_1,\vect{k_2})$ relies on the transverse momentum components of the signal and idler photons. Because it cannot be factored, $\Phi(\vect{k}_1,\vect{k_2})\neq\phi(\vect{k}_1)\phi(\vect{k_2})$, the signal and idler photons are entangled in their transverse momentum components. For down-converted photons with a large transverse momentum, i.e. a large $\Delta q_x$ and $\Delta q_y$, the $\mathrm{sinc}(\Delta q L/2)$ approximates a Dirac-delta function and becomes a suitable source for generating an approximate EPR state in the lab.

\chapter{Fast Hadamard Transforms for Compressive Sensing of Joint Systems}\label{ch3}

In this chapter, the reconstruction side of compressively characterizing position correlations between down-converted photon pairs is presented. As will be shown, the computational overhead of storing a sensing matrix for even modestly-sized correlation characterizations can quickly become nonviable. This chapter presents work published in \cite{lum2015fast} that eliminates the computer-memory scaling problem and also decreases the reconstruction time by carefully designing sensing matrices based on fast-transform operations. Specifically, Sylvester-Hadamard matrices are used to sample each subsystem in a way that enables the use of fast transforms in high-dimensional reconstructions. This work is also the key to the high resolution characterizations obtained in \cite{PhysRevLett.112.253602}.   

\section{Introduction}
Characterizing high-dimensional joint systems, such as continuous variable entanglement in time / energy or position / momentum, is a difficult problem due to experimental impracticalities such as long measurement times, low flux, or insufficient computing resources. Yet, continuous-variable entanglement is becoming a valuable resource in quantum technologies \cite{masada2015continuous,o2009photonic,braunstein2005quantum,steane1998quantum,huver2008entangled,giovannetti2011advances}. As discussed in chapter \ref{ch2}, SPDC is a readily available sources of continuous-variable entangled photons. To determine if the system is entangled, both the bi-photon joint position and joint momentum probability distributions must be measured through correlation measurements.
 
Much experimental work has been done to characterize high-dimensional time-energy and position-momentum entanglement \cite{kwiat1993high,maclean2018direct,howell2004,schneeloch2013violation,schneeloch2014demonstrating,lvovsky2009continuous,giovannini2013,bolduc2015direct,shabani2011efficient}. 
This chapter focuses on characterizing position-momentum entangled photons.
Characterizations are done by measuring signal and idler pixel correlations in either an image plane of the crystal (constituting a position measurement) or a Fourier-transform plane of the crystal (constituting a momentum measurement) through coincidence counting.
 
Correlated measurements are typically done by raster scanning through all possible projections of the two-particle state. This means that, within an image or Fourier plane of the entanglement generating source, correlations from every pixel of the signal plane must be measured against every pixel in the idler plane. The time required to complete a raster scan with single-photon detectors quickly becomes impractical for certain scans -- with the number of measurements scaling as $N^2$ for $N$ pixes in each signal and idler planes. Imaging these distributions with an electron multiplying charge-coupled device (EMCCD) has been shown in \cite{edgar2012imaging,fickler2013real}, yet EMCCDs often introduce a significant amount of noise that can mask the correlations while also being quite expensive.

Recently, compressive sensing (CS) \cite{donoho2006compressed,duarte2008single} techniques were introduced as an alternative to raster scanning for characterizing a high-dimensional entangled system \cite{howland2013efficient,tonolini2014}. While the data-acquisition time is drastically reduced, it comes at the cost of computational complexity, requiring a computational reconstruction of the signal. Performing CS on high-dimensional signals is not a new problem, and several clever solutions exist for utilizing separable compressive sensing matrices combined by a Kronecker product \cite{rivenson2009practical,duarte2012kronecker}. However, these methods are ill suited for sampling the correlations in a joint space. Because the sensing matrix dimensions grow as $N^4$, eliminating the need to store the sensing matrix in computer memory is of paramount importance. 
 
These problems can be overcome with Hadamard-based sensing matrices. Specifically, we show how the Kronecker relation in Sylvester-type Hadamard matrices enables the use of fast-Hadamard transforms for joint-space reconstructions while eliminating the need to store the sensing matrix in computer memory. Compressive sensing based on fast-Hadamard transforms has already been shown to drastically reduce CS reconstruction times \cite{shishkin2012fast}. Using the randomization techniques outlined in \cite{li2011compressive,li2009user}, pseudo-random Hadamard matrices offer tremendous speed enhancements in many reconstruction algorithms. This chapter explains how to take advantage of the speed enhancement by carefully designing the single-particle sensing matrices. 
 
To demonstrate the effectiveness of our method, we reconstruct a $16.8\times10^6$ dimensional joint-space distribution from only $20,000$ measurments in only a few minutes. Within the joint-space exists a 3.2 million-dimensional probability distribution from which we measure the degree of transverse correlations using the mutual information. Additionally we show that the information in the signal and idler marginal distributions can aid in the joint-space reconstructions.

The experimental realization here is closely related to the work performed in \cite{howland2013efficient}. The experiment in this article is merely meant to demonstrate how structured randomness enables efficient reconstructions of the joint space distribution at even higher dimensions. In theory, this increase in resolution allows for an increase in the amount of measurable mutual information.

\subsection{Motivation}

Characterizing the degree of non-separability in continuous-variable entanglement is a difficult task both theoretically and experimentally. In chapter \ref{ch2}, we showed that the Schmidt number and the concurrence are two such ways the degree of separability can be measured. However, those metrics are hinged on the assumption that the quantum state is well characterized -- having access to the density matrix. In many instances, acquiring enough measurements needed to obtain a density matrix is impractical. However, at times is still beneficial to merely \emph{witness} entanglement.

Two entanglement witnesses that are most commonly employed include Bell inequalities \cite{bell2001einstein,PhysRevLett.49.91,clauser1969proposed} and steering inequalities \cite{PhysRevLett.106.130402,PhysRevLett.98.140402,PhysRevA.40.913,PhysRevA.87.062103}. A Bell test is a means of testing elements of locality and local realism when the detectors for constituent particles are spatially separate, especially when the detection events occur outside of each particle's light cone \cite{hensen2015loophole,PhysRevLett.115.250401}. Alternatively, an entanglement criterion weaker than Bell-nonlocality, yet stronger than non-separability \cite{werner1989quantum}, is available through the violation of a steering inequality. 

Steering inequalities test for correlations strong enough to demonstrate the EPR paradox, but are not strong enough to rule out all local hidden variable models. By exploiting quantum correlations, two parties that share constituent particles within an entangled state can \emph{steer} the measurement results of the other party in a way that is impossible in classical mechanics. In particular, we focus our attention on the EPR steering inequality presented in \cite{PhysRevLett.110.130407}. There, an inequality for continuous-variable systems using only on discrete probability distributions (as measured in the lab) is presented. Letting the signal vectors $\vect{X}$ represent the discretized position distribution and $\vect{K}$ represent the discretized momentum distribution, then the steering inequality reads
\begin{equation}
I(\vect{X}_A,\vect{X}_B) + I(\vect{K}_A,\vect{K}_B) \leq 2\log_2\left(\frac{n \Delta \!x \Delta \!k}{\pi e}\right),
\label{eq:steering}
\end{equation} 
where $\Delta k$ is the detector resolution of each particle's momentum, $\Delta x$ is the detector resolution of each particle's position, $n$ is the number of pixels in each distribution, and $I(\vect{X}_A,\vect{X}_B)$ [$I(\vect{K}_A,\vect{K}_B)$] represents the mutual information that exits between particle A and B in their position [momentum] degrees of freedom.

The mutual information $I(A,B)$ that exits between parties $A$ and $B$ is defined as
\begin{align}
I(A,B) &\equiv H(A)-H(A|B)\\
&\equiv H(B)-H(B|A) \\
&\equiv H(A,B) - H(A|B) - H(B|A),
\end{align}
where $H$ is the classical Shannon entropy and is defined as
\begin{equation}
H(X) =  -\sum\limits_{i=1}^n p(x_i)\log_b\big(p(x_i)\big).
\label{eq:classentropy}
\end{equation}
Within Eq. (\ref{eq:classentropy}), the entropy is calculated by summing over the random variables $x_i\in X$ for $i = 1,2,...,n$ that occur with probability $p(x_i)$. Note that the logarithm's base was left as a variable $b$. The entropy is in units of bits if $b=2$. The mutual information is a metric for how much information can be obtained about $A$ when only having access to $B$, and vice-versa. As such, is a common metric used to quantify the information capacity of a communication channel \cite{shannon1949communication}. The mutual information is symmetric and positive. Note that $H(A|B)$ is is the entropy of $A$ conditioned on $B$ and $H(A,B)$ is the joint entropy. Each entropy term, in turn, is based on the corresponding conditional or joint probabilities of variables $A$ and $B$.

The point in introducing Eq. (\ref{eq:steering}) is to show the dependence on the position and momentum distributions $\vect{X}$ and $\vect{K}$ associated with particles $A$ and $B$. The methods introduced below enable the efficient compressive measurment of these distributions. 

\section{Measuring a non-separable joint system}

We apply CS to measure the joint position probability distribution of the down-converted signal and idler photons from SPDC. Quantum mechanics tells us that the bi-photon state exists in a Hilbert space composed of the tensor product of the individual signal and idler photon Hilbert spaces. In order make a measurement, we approximate the state as living in a finite dimensional Hilbert space. We can therefore represent a bi-photon operator matrix in terms of Kronecker products of individual signal and idler photon operator matrices. For CS, we can let these operators be projection operators and manipulate them such that they form the rows of a sensing matrix $\mathbf{A}$. We designate the set of projections for each subspace as $\mathbf{P}_S$ and $\mathbf{P}_I \in\mathbb{R}^{M\times N}$ for signal and idler, respectively. In this manner, the sensing matrix is written as
\begin{equation} \label{eq:sensing_matrix} 
\mathbf{A} =  \left[ \begin{array}{c} \mathbf{P}_S[1] \otimes \mathbf{P}_I[1] \\ 
\mathbf{P}_S[2] \otimes \mathbf{P}_I[2] \\ \vdots \\ \mathbf{P}_S[M] \otimes 
\mathbf{P}_I[M] \end{array} \right] 
\end{equation}
for $i \in 1...M$ where 
$\mathbf{P}[i]$ represents the $i^{th}$ row of $\mathbf{P}$.
 
The Kronecker product $\otimes$ in this article operates on matrices and vectors such that if $\mathbf{a}$ is of dimension $m\times n$ and $\mathbf{b}$ is of dimension $p\times q$, then their Kronecker is of dimension $mp\times nq$ represented as
\begin{equation}
\mathbf{a}\otimes\mathbf{b} = 
\left[ 
\begin{array}{ccc} 
a_{11}\mathbf{b} & \cdots & a_{1n}\mathbf{b} \\ 
\vdots & \ddots & \vdots \\ 
a_{m1}\mathbf{b} & \cdots & a_{mn}\mathbf{b}
\end{array}
\right].
\end{equation}

\begin{figure}
\centering
\includegraphics[width=.8\textwidth]{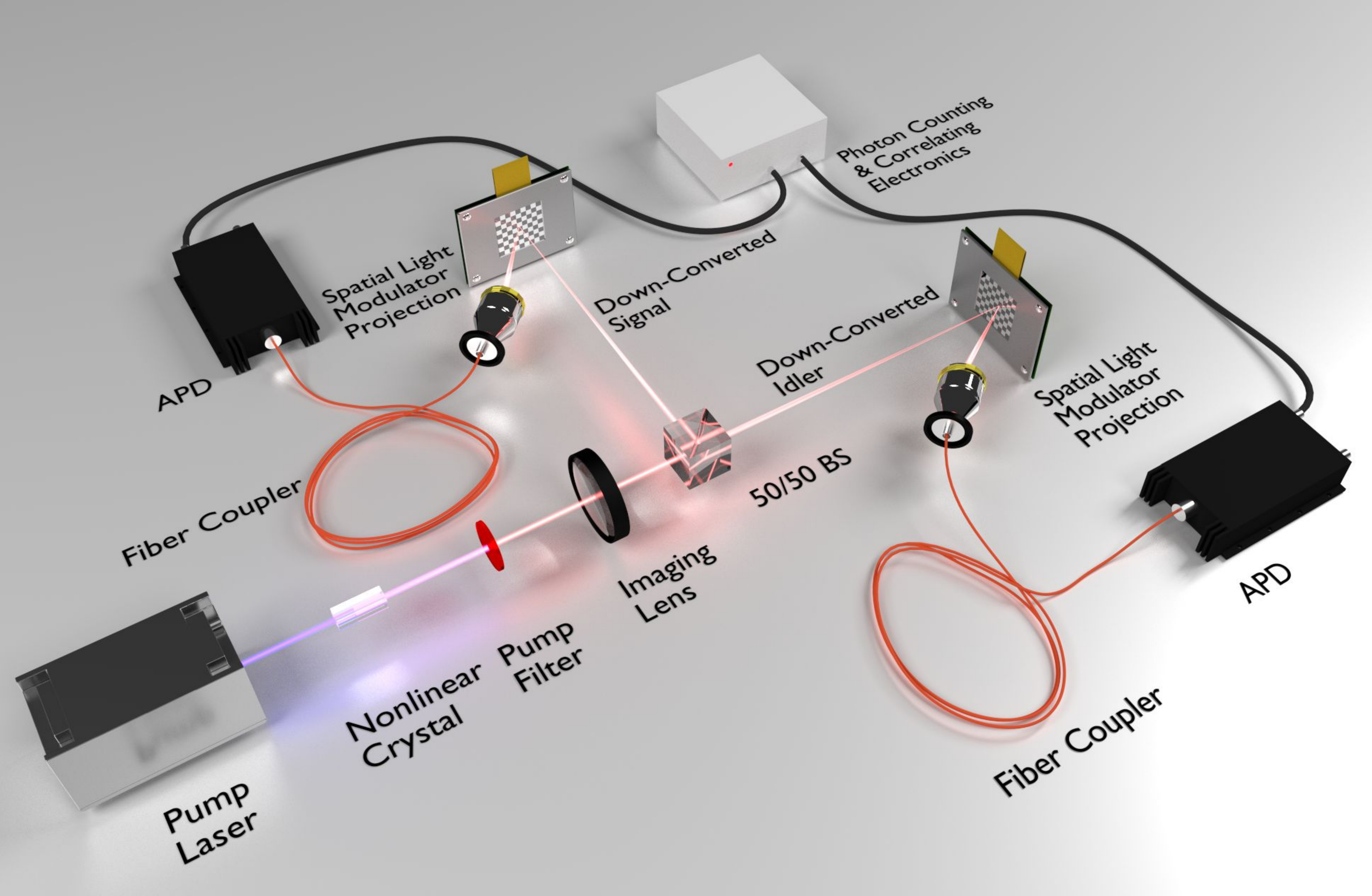} 
\caption[Experimental setup to compressively measure correlations]{The above 
experimental diagram measures the correlations in a joint two-particle system. Random projections of signal and idler spatial distributions are taken with a spatial light modulator
within an image plane of the crystal. An avalanche photodiode detector (APD) and correlation electronics detect photon arrivals in coincidence.} \label{fig:Experiment} 
\end{figure}

An experimental 
diagram is presented in 
Fig. \ref{fig:Experiment}. To simplify 
the CS formalism, we represent the signal and idler spaces as one-dimensional living in $\mathbb{R}^N$ and the 
joint-space distribution as a vector $\mathbf{x}\in\mathbb{R}^{N^2}$. As outlined in \cite{howland2013efficient}, compressive sensing is experimentally 
accomplished by taking random projections with patterns composed of 
$\left[\sqrt{N}\times\sqrt{N}\right]$ pixels within each subspace and then measuring the resulting coincidences in photon counts.

During reconstruction, the bi-photon probability 
distribution is already sparse in the joint-space due to the tight pixel 
correlations in position and momentum. This eliminates the need to define a sparse-basis.

Random binary matrices were used in \cite{howland2013efficient}, and consequently limited the subspace measurements to $32\times 32$ pixel resolutions. Additionally, the experiment required a significant amount of time to reconstruct the data. We wish to avoid these limitations by using \emph{structured} matrices for reconstruction purposes. In \cite{howland2013efficient}, using properties of Kronecker products enabled relatively efficient computations of the reconstruction operations $\mathbf{A}\mathbf{x}$ and $\mathbf{A}^{T}\mathbf{y}$, where $\mathbf{A}^{T}$ is the transpose of $\mathbf{A}$, because $\mathbf{A}$ never needed to be computed explicitly \cite{horn1991matrix}. However, structured randomness can enable the use of fast transforms which are even more efficient.

\section{Fast-Hadamard-transform based sensing matrices}

\subsection{Randomly sampled \& permuted Hadamard sensing matrices}

Sylvester-Hadamard matrices have a structure that is particularly advantageous to the CS framework. 
These matrices are generated from a simple recursion relation defined by a Kronecker product.
\begin{eqnarray*} \mathbf{H}_1 & = & [1] \\ \mathbf{H}_2 & = & \left[ \begin{array}{cc} 1 
& 1 \\ 1 & -1 \end{array} \right].\\ \end{eqnarray*} From these, any Sylvester-Hadamard
matrix can be decomposed as follows:
\begin{equation} 
\label{eq:kronecker}
\mathbf{H}_{2^k} = 
\mathbf{H}_2 \otimes 
\mathbf{H}_{2^{k-1}} =
\left[ 
\begin{array}{cc} 
\mathbf{H}_{2^{k-1}} & \mathbf{H}_{2^{k-1}} \\ 
\mathbf{H}_{2^{k-1}} & - \mathbf{H}_{2^{k-1}} 
\end{array} \right],
\end{equation}
for $k > 1$. Because of this structure, Sylvester-Hadamard matrices are restricted to powers of two but can be used to build patterns and a sensing matrix that utilizes the speed and 
efficiency of a fast-Hadamard transform $\mathscr{H}[\ast]$. We use a normally-ordered 
fast transform in this work. Its algorithm is similar to that of a fast-Fourier 
transform, but it consists of only additions and subtractions. Hence, it performs the reconstruction operations $\mathbf{A} 
\mathbf{x}$ and $\mathbf{A}^T \mathbf{y}$ in $\mathcal{O}\left(N^2\log N 
\right)$ time -- significantly faster than an explicit matrix-vector multiplication of $\mathcal{O}\left(N^4\right)$ 
time. A thorough overview of Hadamard matrices, fast-Hadamard 
transforms, and their applications to signal and image processing may be found in 
\cite{yarlagadda1997hadamard}.

To construct $\mathbf{P}_S$, $\mathbf{P}_I$, and $\mathbf{A}$ (within Eq. (\ref{eq:sensing_matrix})) from Hadamard matrices, the 
subspace Hadamard matrices must be randomized in both their rows and columns. 
The sensing matrices 
must be both incoherent with the image yet span the space in which the signal resides. Random sensing matrices perform this task well, yet Hadamard matrices 
naturally contain much structure.
To begin, each sensing matrix must be formed by taking specific rows from a 
Hadamard matrix with the correct dimensions. The joint space sensing matrix $\mathbf{A}$ is constructed from 
$\mathbf{H}_{N^2}$ while the subspace sensing matrices $\mathbf{P}_S$ and $\mathbf{P}_I$ are constructed from 
$\mathbf{H}_{N}$. Because of the relation in Eq. (\ref{eq:sensing_matrix}), the rows 
of $\mathbf{A}$ will be determined by the rows of $\mathbf{P}_S$ and $\mathbf{P}_I$.

The randomization of the Hadamard matrix rows is accomplished by
defining two vectors $\mathbf{r}_S$ and $\mathbf{r}_I \in\mathbb{R}^{M}$ for each signal 
and idler system composed of $M$ randomly chosen integers on the interval [2,N]. 
The values in $\mathbf{r}$ state which rows should be extracted from $\mathbf{H}_N$ when constructing $\mathbf{P}_S$ and $\mathbf{P}_I$. Note 
that the interval begins at 2 because the first row of a Hadamard matrix is composed 
entirely of ones. The interval may begin at 1 if the total photon flux on a detector is desired.
Also, note that 
$\mathbf{A}\in\mathbb{R}^{M\times N^2}$ where $M << N^2$. This condition allows for 
scenarios where $\mathbf{P}_S, \mathbf{P}_I\in\mathbb{R}^{M\times N}$ such that $M > N$ 
in the individual subspaces, meaning rows of $\mathbf{H}_N$ may be repeated when 
constructing $\mathbf{P}_S$ and $\mathbf{P}_I$.

The randomization of the Hadamard columns is accomplished by defining permutation vectors 
$\mathbf{p}_S$ and $\mathbf{p}_I \in\mathbb{R}^N$ that randomly permute the $N$ columns 
of $\mathbf{H}_N$. Once $\mathbf{r}$ and $\mathbf{p}$ have been defined for both the 
signal and idler subspaces, patterns are constructed by the following equations: 
\begin{eqnarray} \label{eq:Patterns} \mathbf{P}_S & = & 
\mathbf{H}_{N}[\mathbf{r}_S,\mathbf{p}_S ] \nonumber \\ \mathbf{P}_I & = & 
\mathbf{H}_{N}[\mathbf{r}_I,\mathbf{p}_I] \end{eqnarray} where the $y$ and $x$ components 
of $\mathbf{H}[y,x]$ refer to the rows and columns of $\mathbf{H}$ respectively.

The manner in which $\vect{P}_S$ and $\vect{P}_N$ combine to 
manipulate a Hadamard matrix $\mathbf{H}_{N^2}$ that spans the joint space is detailed in 
the next section.

\subsection{Joint-space Sylvester-Hadamard sensing matrices}

Once $\mathbf{r}$ and $\mathbf{p}$ have been defined for the individual signal and idler subspaces, they may be used to construct 
the corresponding joint space row-selection and permutation 
vectors, $\mathbf{r}_{SI}$ and $\mathbf{p}_{SI}$ such that $\vect{A} = \vect{H}_{N^2}[\vect{r}_{SI},\vect{p}_{SI}]$. Consider the construction of $\mathbf{r}_{SI}$ first.  By Eq. (\ref{eq:kronecker}), $\mathbf{A}$ is formed by the row-wise 
Kronecker product of the subspace sensing matrices $\mathbf{P}_S$ and $\mathbf{P}_I$. As 
$\mathbf{r}_S$ and $\mathbf{r}_I$ determine the ordering of the Hadamard rows within 
these patterns, $\mathbf{r}_{SI}$ must also be a subset of a Kronecker product of 
$\mathbf{r}_S$ and $\mathbf{r}_I$.  Knowing that the Kronecker product of $\mathbf{P}_S$ 
and $\mathbf{P}_I$ will form ``blocks" of size $[M\times N]$, it is straightforward to 
show that 
\begin{equation}\label{eq:PicksN^2} \mathbf{r}_{SI}[i] = 
N\left(\mathbf{r}_S[i]-1\right)+\mathbf{r}_I[i] 
\end{equation} 
for $i \in \{1,2,...,M\}$ where 
$\mathbf{r}[i]$ represents the $i^{th}$ component of $\mathbf{r}$. Note that element-wise 
counting in this article starts at 1.

Because $\mathbf{r}_S$ and $\mathbf{r}_I$ are chosen at random and will often be 
over-complete, $M > N$, and $\mathbf{r}_{SI}$ will probably have repeating units, and a row 
within $\mathbf{A}$ will appear more than once. This is equivalent to taking the same 
projection more than once and offers no additional information. To prevent this, compare each value within $\mathbf{r}_{SI}$ and eliminate 
repeating values. If $\mathbf{r}_{SI}[i]$ is a repeated value, we eliminate 
$\mathbf{r}_{SI}[i]$ along with the components $\mathbf{r}_S[i]$ and $\mathbf{r}_I[i]$. 
In this way, the number of samples $M$ will decrease yet contain the same amount of 
information.

The formation of $\mathbf{p}_{SI}$ follows a similar form as $\mathbf{r}_{SI}$, yet it 
will be of length $N^2$. Although it is not a simple Kronecker product, it does follow 
from the structure in Eq. (\ref{eq:sensing_matrix}). The structure of 
$\mathbf{p}_{SI}$ takes the form 
\begin{equation}\label{eq:PermN^2} 
\mathbf{p}_{SI}[N(i-1)+j] = N\left(\mathbf{p}_S[i]-1\right)+\mathbf{p}_I[j] 
\end{equation}
for $i \in \{1,2,...,N\}$ and $j \in \{1,2,...,N\}$. Generating randomized Hadamard matrices using 
$\mathbf{r}$ and $\mathbf{p}$ for each signal, idler, and joint space are summarized 
below: 
\begin{eqnarray} \label{eq:PatternsA} \mathbf{P}_S & = & 
\mathbf{H}_{N}[\mathbf{r}_S,\mathbf{p}_S ] \nonumber\\ \label{eq:Patterns} \mathbf{P}_I 
& = & \mathbf{H}_{N}[\mathbf{r}_I,\mathbf{p}_I]\\ \label{eq:A} \mathbf{A} & = & 
\mathbf{H}_{N^2}[\mathbf{r}_{SI},\mathbf{p}_{SI}] \nonumber \end{eqnarray} 
where 
the $y$ and $x$ components of $\mathbf{H}[y,x]$ refer to the rows and columns of 
$\mathbf{H}$ respectively. The construction of $\mathbf{A}$ presented in Eq. (\ref{eq:PatternsA}) allows us to use fast transforms as explained in the next section.

\subsection{Joint-space fast-Hadamard transform operations}

Keeping track of the randomization operations allows the use of fast-Hadamard transforms when computing $\mathbf{A}\mathbf{x}$ and $\mathbf{A}^T\mathbf{y}$.
This is accomplished by reordering either $\mathbf{x}$ or $\mathbf{y}$ according to $\mathbf{p}$, taking the fast Hadamard 
transform, and then picking specific elements from the final 
result according to $\mathbf{r}$. The manner in which they are rearranged and picked 
depends upon the operation $\mathbf{A} \mathbf{x}$ or $\mathbf{A}^T \mathbf{y}$ 
in either the data acquisition or reconstruction processes.

Starting with the data-taking procedure $\mathbf{y} = \mathbf{A} \mathbf{x}$, 
projections are taken in each signal and idler system by first constructing individual patterns. Pattern construction is done by fast Hadamard transforming basis vectors and then 
permuting them. Because of the symmetric nature of a Hadamard matrix ($\mathbf{H} = \mathbf{H}^T$), a fast Hadamard 
transform of a basis vector $\boldsymbol{\alpha}(i)$ (in which the $i^{th}$ component is 
equal to one and the rest zeros) is equal to the $i^{th}$ row of the Hadamard matrix 
$\mathbf{H}[i,:]$. In short, $\mathscr{H}\left[\boldsymbol{\alpha}[i]\right] = 
\mathbf{H}[i]$. Hence, every $i^{th}$ pattern $\mathbf{P}[i,:]$ can be built according 
to a fast transform by 
\begin{eqnarray} \label{eq:fastPattern}
\mathbf{H}[\mathbf{r}\!\left[i\right],:] &=& 
\mathscr{H}\left[\boldsymbol{\alpha}\!\left(\mathbf{r}[i]\right)\right] \nonumber \\ 
\mathbf{P}[i] &=& \mathbf{H}[\mathbf{r}\!\left[i\right],\mathbf{p}] \end{eqnarray} for $i \in 1...M$ where 
$\boldsymbol{\alpha}(\mathbf{r}[i])$, a basis vector whose $\mathbf{r}[i]^{th}$ component 
is equal to 1, is fast-Hadamard transformed and then permuted according to $\mathbf{p}$.

To take experimental data, many spatial light modulators (SLMs), such as digital micro-mirror devices (DMDs), are operated 
in a binary on / off fashion -- transmitting light either to or away from a detector. 
If only using one detector per subspace, at any given moment a pattern may only be 
composed of $(0,1)$ while Hadamard matrices are composed of $(1,-1)$. To display 
the full Hadamard pattern with one detector per subspace, the data-taking operations must 
be split into positive and negative operations. $\mathbf{H}_{N^2}$ may be decomposed into 
a sum of positive $\mathbf{H}^+_N$ (composed of $(0,1)$) and negative $\mathbf{H}^-_N$ (composed of $(-1,0)$) matrices such that
\begin{equation} 
\label{eq:CorrelationMeas} \begin{split} \mathbf{H}_{N^2} = \left(\mathbf{H}_N^+\otimes 
\mathbf{H}_N^+\right) \, + \, \left(\mathbf{H}_N^-\otimes \mathbf{H}_N^-\right) \, + \, \left(\mathbf{H}_N^+\otimes 
\mathbf{H}_N^-\right) \, + \, \left(\mathbf{H}_N^-\otimes \mathbf{H}_N^+\right). \end{split} 
\end{equation} 
Consequently, every element in $\mathbf{y}$ will 
require four coincidence measurements. Even though $4M$ coincidence measurements are required when using one 
detector per subspace, the drastic sampling performance gained through CS methods is such 
that $4M\ll N^2$. Alternatively, if two detectors are used per subspace, the 
detection process requires only $M$ projections.

In reconstruction, fast-Hadamard transforms may be utilized by CS reconstruction 
algorithms to perform the operations $\mathbf{A}\mathbf{x}$ and $\mathbf{A}^T\mathbf{y}$. The operation 
$\mathbf{A}\mathbf{x}$ first requires that $\mathbf{x}$ be 
inverse-permuted, fast Hadamard transformed, and then have the correct $M$ elements extracted from the final result. The inverse-permutation is done by 
defining an inverse permutation vector $\mathbf{q}$ as \begin{equation}\label{eq:InvPerm} 
\mathbf{q}\left[\mathbf{p}[i]\right] = i \end{equation} for all $i$ elements in 
$\mathbf{p}$. Hence, $\mathbf{A}\mathbf{x}$ is realized with the 
following operations 
\begin{eqnarray}
\mathbf{y}' & = & 
\mathscr{H}\left[\mathbf{x}[\mathbf{q}_{SI}]\right] \nonumber \\ \mathbf{y} & = & 
\mathbf{y}'[\mathbf{r}_{SI}]. \end{eqnarray}

The operation $\mathbf{A}^T \mathbf{y}$ requires that a vector 
$\boldsymbol{\beta}$ composed of $N^2$ zeros be filled with the elements of $\mathbf{y}$ 
according to $\mathbf{r}_{SI}$, fast-Hadamard transformed, and then permuted according to 
$\mathbf{q}$ as follows: 
\begin{eqnarray}\label{eq:fastAty}
\boldsymbol{\beta}[\mathbf{r}_{SI}] & = & 
\mathbf{y} \nonumber \\ \mathbf{x}[\mathbf{q}_{SI}] & = & 
\mathscr{H}\left[\boldsymbol{\beta}\right]. 
\end{eqnarray}

Again, these operations work because Hadamard matrices are symmetric. However, it should be noted that the true inverse operation is $\mathbf{H}^{-1}_{N}=\mathbf{H}^{T}_{N}/N$. When taking the fast transform operation in Eq. (\ref{eq:fastAty}), we are explicitly taking the forward fast transform and neglecting the normalization term. Because of this structure, the operations $\mathbf{A}\mathbf{x}$ and 
$\mathbf{A}^T\mathbf{y}$ can be utilized by most reconstruction algorithms to 
operate more efficiently.

\section{Compressive measurement of a $16.8\times 10^6$-dimensional correlated space}

To demonstrate the practicality of the previous results, we compressively measure and quickly reconstruct a $16.8\times 10^6$ dimensional joint-space probability distribution.

\subsection{Theoretical expectations}

Before reporting our experimental results, it is useful to first estimate the theoretical maximum amount of possible mutual information as derived from first principles based on the crystal and the pump-laser specifications. That result should then be compared with the maximum possible information we could measure given our SLM resolution. A thorough calculation characterizing degenerate SPDC is done in \cite{schneeloch2015introduction} in one transverse spatial dimension. Assuming a double Gaussian bi-photon state, the mutual information in the position domain between down-converted photons ($\vect{X}_S$ and $\vect{X}_I$) is
\begin{equation}
\label{eq:maxmutualinfo}
I(\vect{X}_S , \vect{X}_I)=\text{log}_2\left(\frac{9\pi\sigma_p^2+L_z\lambda_p}{2\sigma_p\sqrt{9\pi L_z\lambda_p}}\right)
\end{equation}
where $L_z$, $\lambda_p$, and $\sigma_p$ represent the length of the nonlinear crystal, the pump-laser wavelength, and the standard deviation of the Gaussian intensity pump-laser profile, respectively.
We use a 325 nm pump laser and a 1 mm length nonlinear crystal. The maximum $1/e^2$ pump diameter is listed as 1.2 mm resulting in a $\sigma_p$ that is four times smaller, i.e. $\sigma_p = 3\times10^{-4}$ m. The experiment uses approximately degenerate down-converted light because of the paraxial nature of beam propagation and the use of narrow-band filters. Our pump laser is approximately Gaussian in two dimensions resulting in a mutual information twice as large as reported in Eq. (\ref{eq:maxmutualinfo}). Hence, the theoretical maximum mutual information between signal and idler photons is 10.9 bits. However, when moving from an infinite-dimensional Hilbert space to a finite dimensional space, dictated by the resolution of the SLMs, the resulting measurable mutual information must be less than or equal to to the continuous variable case of 10.9 bits.

\subsection{Iterative soft-thresholding reconstruction algorithm}

Not yet knowing of of the newer and more efficient ADMM methods presented in appendix, we used an iterative thresholding algorithm \cite{besk2009afastiterative} that computes the mutual information at each iteration. The program exits when the mutual information no longer increases with thresholding. 

The algorithm we use is summarized as follows:
\begin{eqnarray}
\label{Eq:Algorithm}
\mathbf{x}_0 &=& \mathbf{c} \nonumber \\
\mathbf{x}_{t+1} &=& \hat{\eta}_{2}[ \mathbf{x}_t\cdot\lbrace\hat{\eta}_1\left[\mathbf{A}^T\cdot\left(\mathbf{y}-\mathbf{A}\cdot
\mathbf{x}_t\right)\right]\rbrace+\mathbf{x}_t - \text{min}\left(\mathbf{x}_t\right)]
\end{eqnarray}
where $\mathbf{c}$ is a constant vector (with all elements equal to $c$) and min($\mathbf{x}_t$) is a vector composed entirely of ones times the smallest element in $\mathbf{x}_t$.
At each iteration, we take a projection of the current result with the term $\hat{\eta}_1\left[\mathbf{A}^T\left(\mathbf{y}-\mathbf{A}
\mathbf{x}_t\right)\right]$. The function $\hat{\eta}_1[\ast]$ is an operator that performs soft thresholding on everything within its brackets using a biorthogonal 4.4 wavelet transform with a two-level decomposition \cite{selesnick2005dual,daubechies1990wavelet,antonini1992image}. 

Soft thresholding in the wavelet basis, often referred to as wavelet shrinkage, is performed using the universal threshold of Donoho and Johnstone \cite{donojohnstone1994ideal,donoho1995denoising}. Because image noise is often concentrated in the high-frequency components of the wavelet transform, wavelet shrinkage estimates the noise according the variance in the highest frequency wavelet coefficients and then performs soft-thresholding -- effectively shrinking the amplitude of the wavelet coefficients. 

The filtered signal is then inverse transformed back to the pixel basis, and $\hat{\eta}_2[\ast]$ performs a hard thresholding in the pixel basis before normalizing the final result. Hard thresholding, as opposed to soft thresholding, sets elements below the threshold to zero. The threshold of $\hat{\eta_2}$ gradually increases with each iteration. 
While $\mathbf{y}-\mathbf{A}\mathbf{x}_t$ approaches zero, it never truly vanishes due to the effects of measurement noise. To prevent injecting random noise into each filtered iteration, we take the projection of the noisy term with the current clean solution. We then add this projection back to the current solution to prevent discarding current signal components. After hard thresholding the first iteration, $\text{min}\left({\mathbf{x}_{t>0}}\right) = \mathbf{0}$ where $\mathbf{0}$ is a null vector.

\subsection{Experimental results}

As an experimental demonstration, we compare how the measured mutual information from compressive measurements compares to the theoretical value of 10.9 bits.
With binary patterns on each SLM, having a resolution of $N = 4096$, we measured coincidence counts at a rate of $4\times 10^3$ counts per second. Since each SLM reduces the incoming flux by approximately $50\%$, the total number of coincidences $\Phi$ was approximately four times larger, or $\Phi = 1.6\times 10^4$ coincidences per second. Using only two detectors, all four projections per $\mathbf{y}$-element required a total of 8 seconds due to power constraints. The resulting ratio of the standard deviation of $\vect{y}$ relative to the shot noise was 2.4.

To discern how many measurements $M$ are needed for reconstructions in the joint space, we estimated that the number of signal components $k \approx N = 4096$ due to tight correlations. Thus, we assigned $M = \mathcal{O}\left(k\log_2(N^2/k)\right) = \mathcal{O}(10^4)$ measurements. We chose the number of measurements to be $2\times 10^4$ ($M/N^2\approx .001$) as a reasonable compromise between the total integration time and reconstruction quality. 

\begin{figure} \includegraphics[width=1\textwidth]{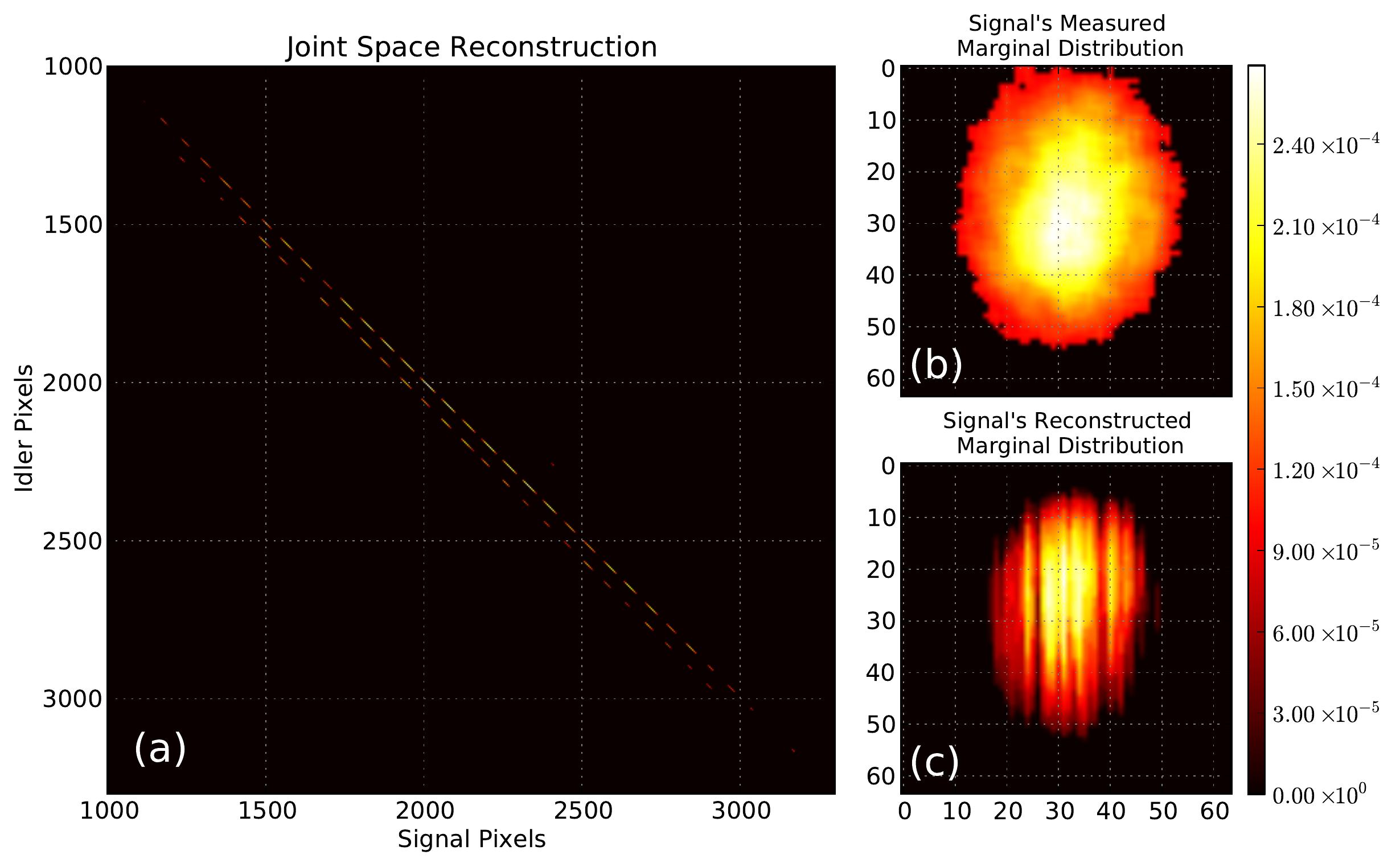} 
\caption[Joint-space reconstructed correlations]{Image (a) depicts a zoomed-in view that captures the largest components of the reconstructed $16.8 \times 10^6$ joint space distribution. The largest components are contained within the resulting 4-megapixel image shown. The $1/e^2$ intensity profile, as seen by the signal's SLM, is presented as the signal's measured marginal distribution in (b). The reconstructed marginal distribution was obtained through the reconstructed joint space distribution and is displayed in (c) for comparison. This data was obtained compressively with $M = 
20,000$ ($.00119\times 64^4$) samples in approximately 44 hours and was reconstructed 
in under ten minutes.} 
\label{fig:Data} \end{figure}

The resulting scan took just over 44 hours. To compare these values to a raster scan, the signal-to-noise ratio (SNR) goes as $\sqrt{\Phi \, t/N}$, assuming perfect pixel correlations and uniform illumination. Here, $t$ is the integration time per pixel. When raster scanning in an $N^2$ dimensional joint space, the total integration time goes as $N^2 t = N^3\, \mathrm{SNR}^2/\Phi$ for shot-noise limited signals. Therefore, a raster scan operating under perfect conditions would require a  minimum of 50 days to achieve a $\text{SNR} = 1$ for $N = 4096$. The reconstructed joint space probability distribution is presented in in Fig. \ref{fig:Data}. 

From the same data set, we analyzed the recovered mutual information as a function of the number of measurements $M$. Unfortunately, the mutual information is \emph{highly} dependent on the noise. As the iterative thresholding algorithm seeks to find the optimal threshold, measurement noise results in sparse speckle patterns for low SNR signals -- resulting in false correlations. To stress the severity of this flaw, we consistently recover about 8 bits of mutual information with $M = 10$ projections. 

\begin{figure} \includegraphics[width=.8\textwidth]{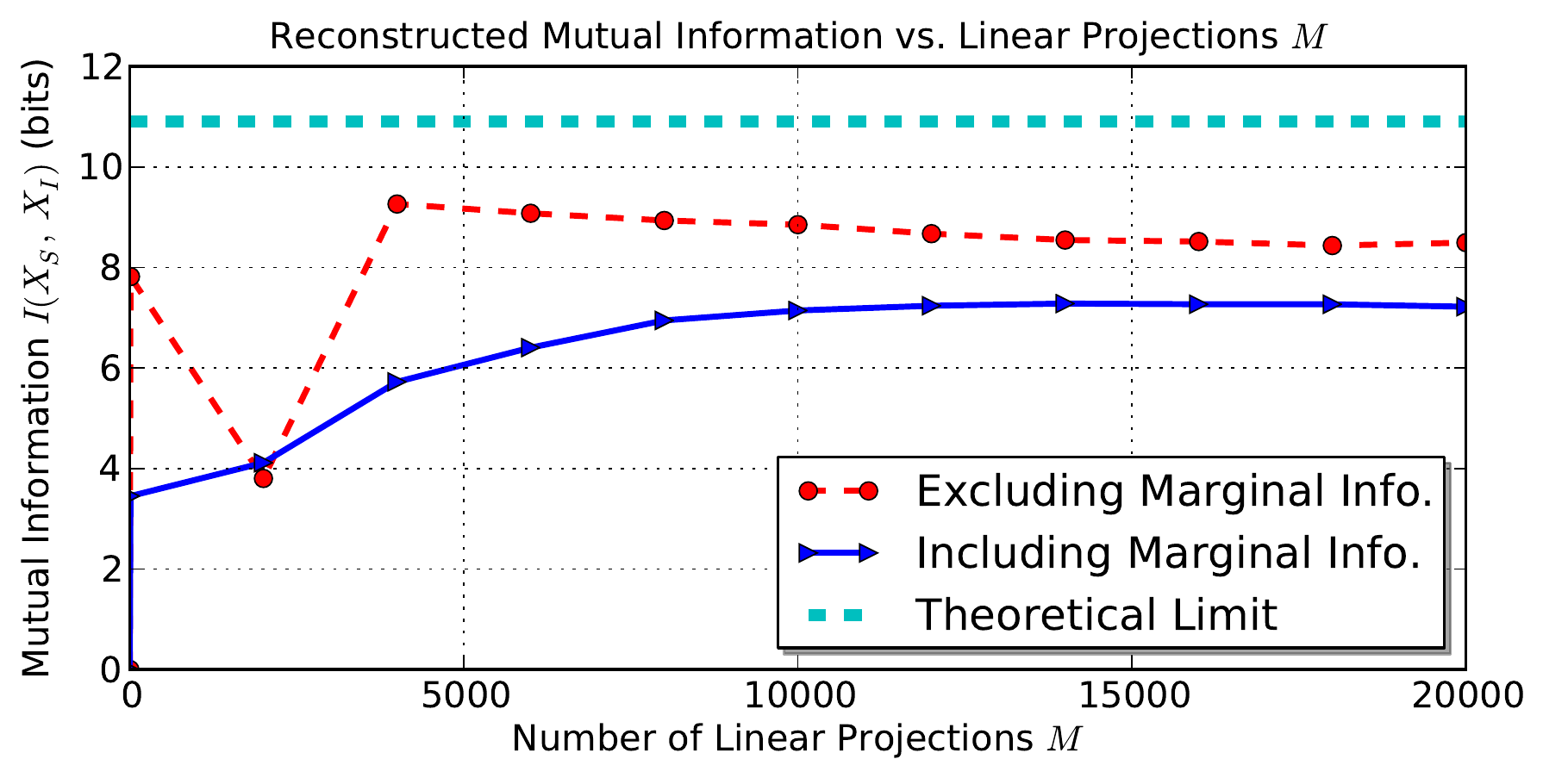}
\centering
 \caption[Reconstructed mutual information as a function of $M$]{The mutual information, obtained through reconstructions of the joint-space distribution, as a function of the number of projections $M$ is depicted above. The figure lists the theoretical maximum mutual information and the corresponding measurement results either utilizing or neglecting available information from the marginal distributions.} 
\label{fig:MvsI} \end{figure}

However, there exists information in the marginal probability distributions that can reduce this error. One marginal distribution is shown in Fig. \ref{fig:Data} and was recovered from $2\times 10^4$ projections. Figure \ref{fig:Data} plots everything within the $1/e^2$ intensity beam diameter while thresholding the background intensity to zero. The marginal distributions indicate that there exist regions where correlations should not exist. This information is used used to reduce the reconstruction error, and the results are summarized in Fig. \ref{fig:MvsI}. 

It is evident from Fig. \ref{fig:MvsI} that background noise significantly alters the results, fabricating a larger mutual information. 
These results stress the importance of adequate sampling when characterizing these distributions.

\section{Conclusion} 
Using a laptop with 8 GB of memory, we perform reconstructions of a $16.8\times 10^6$ dimensional joint space in under ten minutes with satisfactory results using fast-Hadamard transforms with the algorithm presented in Eq. (\ref{Eq:Algorithm}).
As an example, we compressively measured 7.21 bits, in comparison to the theoretical value of 10.9 bits, of mutual information while also demonstrating how to improve the accuracy of these results using information contained within the marginal distributions. We believe these 
methods will be invaluable for other correlation-based CS implementations.

\chapter{Frequency-Modulated Continuous-Wave Compressive LiDAR}\label{ch4}

In this chapter, compressive sensing is used within a classical LiDAR system to decrease the acquisition time needed to form high-resolution depth-maps. This time-reduction coupled with high resolution mapping will be of use in both ranging, object identification, and topography applications. 
A theoretical investigation is presented and suggests the proposed method should work in practical applications with enviornmental noise. Extracting 3D information from a 2D projection is normally a complicated task, yet the method outline here simplifies the 3D reconstruction problem by decomposing it into two easier problems. The total-variation minimization algorithm presented in the appendix is used once and then a fast iterative least-squares algorithm then takes over. The result is a high-resolution depth map with minimial computational overhead. Finally, an analysis of the signal reconstruction quality versus measurment noise is included for completeness. 

\section{Introduction}

Depth-mapping, also known as range-imaging or 3D imaging, is defined as a non-contact system used to produce a 3D representation of an object or scene \cite{cheok20103d}. 
Depth-maps are typically constructed through either triangulation and projection methods, which rely on the angle between an illumination source and a camera offset to infer distance, or on time-of-flight (TOF) LiDAR (light detection and ranging) techniques. LiDAR systems are combined with imaging systems to produce 2D maps or images where the pixel values correspond to depths.
 
Triangulation and projection methods, such as with Microsoft's Kinect \cite{smisek20133d}, are inexpensive and can easily operate at video frame rates but are often confined to ranges of a few meters with sub-millimeter uncertainty. Unfortunately, the uncertainty increases to centimeters after roughtly ten meters \cite{cheok20103d,khoshelham2012accuracy}. Phase-based triangulation systems also suffer from phase ambiguities \cite{geng2011structured}. Alternatively, LiDAR-based methods can obtain better depth-precision with less uncertainty, to within micrometers \cite{cheok20103d,behroozpour2017electronic}, or operate up to kilometer ranges. Thus, LiDAR based systems are chosen for active remote sensing applications when ranging accuracy or long-range sensing is desired. Additionally, technological advances in detectors, lasers, and micro-electro-mechanical devices are making it easier to incorporate depth-mapping into multiple technologies such as automated guided vehicles \cite{frank1993laser}, planetary exploration and docking \cite{amzajerdian2011lidar,stettner2010compact,4526302}, and unmanned aerial vehicle surveillance and mapping \cite{remondino2011uav}. Unfortunately, the need for higher resolution accurate depth-maps can be prohibitively expensive when designed with detector arrays or prohibitively slow when relying on raster scanning. 

Here we show how a relatively inexpensive, yet robust, architecture can convert one of the most accurate LiDAR systems available \cite{amann2001laser} into an efficient high-resolution depth-mapping system. Specifically, we present an architecture that transforms a frequency-modulated continuous-wave LiDAR into a depth-mapping device that uses compressive sensing (CS) to obtain high-resolution images. CS is a measurement technique that compressively acquires information significantly faster than raster-based scanning systems, trading time limitations for computational resources \cite{donoho2006compressed,duarte2008single,eldar2012compressed}. With the speed of modern-day computers, the compressive measurement trade-off results in a system that is easily scalable to higher resolutions, requires fewer measurements than raster-based techniques, obtains higher signal-to-noise ratios than systems using detector arrays, and is potentially less expensive than other systems having the same depth accuracy, precision, and image resolution. 

CS is already used in both pulsed and amplitude-modulated continuous-wave LiDAR systems (discussed in the next section). CS was combined with a pulsed time-of-flight (TOF) LiDAR using one single-photon-counting avalanche diode (SPAD), also known as a Geiger-mode APD, to yield $64\times 64$ pixel depth-images of objects concealed behind a burlap camouflage \cite{howland2011photon,colacco2012compressive}. CS was again combined with an SPAD-based TOF-LiDAR system to obtain $256\times 256$  pixel depth-maps, while also demonstrating background subtraction and $32\times 32$ pixel resolution compressive video \cite{howland2013photon}. Replacing the SPAD with a photodiode within a single-pixel pulsed LiDAR CS camera, as in \cite{sun2016single}, allowed for $128\times 128$ pixel resolution depth-maps and real-time video at $64\times 64$ pixel resolution. 
In addition to pulsed-based 3D compressive LiDAR imaging, continuous-wave amplitude-modulated LiDAR CS imaging has been studied with similar results \cite{6289148,conde2017compressive}. The architecture we presented here, to our knowledge, is the first time CS has been applied to a frequency-modulated continuous-wave LiDAR system. 

Furthermore, we show how to efficiently store data from our compressive measurements and use a single total-variation minimization followed by two least-squares minimizations to efficiently reconstruct high-resolution depth-maps. We test our claims through computer simulations using debilitating levels of injected noise while still recovering depth maps. The accuracies of our reconstructions are then evaluated at varying levels of noise. Results show that the proposed method is robust against noise and laser-linewidth uncertainty.

\section{Current LiDAR and Depth-Mapping Technologies}

There are many types of LiDAR systems used in depth-mapping, yet there are two general categories for obtaining TOF information: pulsed LiDAR and continuous wave (CW) LiDAR \cite{horaud2016overview,remondino2013tof} (also known as discrete and full-waveform LiDAR, respectively \cite{lim2003lidar,mallet2009full}). 

Pulsed LiDAR systems use fast electronics and waveform generators to emit and detect pulses of light reflected from targets. Timing electronics measure the pulse's TOF directly to obtain distance. Depth resolution is dependent on the pulse width and timing resolution -- with more expensive systems obtaining sub-millimeter precision. Additionally, the pulse irradiance power is significantly higher than the background which allows the system to operate outside and over long range. Unfortunately, the high-pulse power can be dangerous and necessitates operating at an eye-insensitive frequency, such as the far-infrared, which adds additional expense. Pulsed LiDAR range-imaging cameras include flash-LiDAR cameras and TOF-range scanners \cite{horaud2016overview}. Flash-LiDAR cameras illuminate an entire scene with a bright pulse of light and use an array of photo-detectors to obtain a depth-image. 
TOF-range scanners scan one or more pulsed lasers across a scene, often relying on fast micro-electro-mechanical mirrors. 
Generally, APDs or SPADs detect returning low-flux radiation and require raster scanning to form a depth-map. More expensive SPAD-arrays can quickly acquire depth-maps at the expense of SNR, but are currently limited to resolutions of $512\times 128$ \cite{Burri:14}.

Alternatively, CW-LiDAR systems continuously emit a signal to a target while keeping a reference signal, also known as a local oscillator. Because targets are continuously illuminated, they can operate with less power, compared to the high peak-power of pulsed systems. CW-LiDAR systems either modulate the amplitude while keeping the frequency constant, as in amplitude-modulated continuous-wave (AMCW) LiDAR, or they modulate the frequency while keeping the amplitude constant, as in frequency-modulated continuous wave (FMCW) LiDAR. 

AMCW-LiDAR systems typically require high-speed radio frequency (RF) electronics to modulate the laser's intensity. However, low-speed electronics can be used to measure the return-signal after it has been demodulated. Examples of AMCW-LiDAR systems include phase-shift LiDAR, where the laser intensity is sinusoidally \cite{amann2001laser} or randomly modulated \cite{Takeuchi:86}. The TOF is obtained by convolving the demodulated local oscillator with the demodulated time-delayed return signal. Sinusoidally modulated systems, in particular, are hindered by a phase ambiguity \cite{amann2001laser,conde2017phase}. Phase-shift LiDAR cameras can often utilize high-resolution CMOS/CCD arrays \cite{lange1999time}, again, at the expense of SNR. 
Another popular AMCW systems is one that uses a linear-chirp of the laser's intensity \cite{stann1996intensity,batet2010intensity}. Linearly chirped AMCW-LiDAR systems require demodulation via electronic heterodyne detection to generate a beat note that is directly proportional to the TOF. While more complicated than phase-shift methods, linearly-chirped AMCW-LiDAR systems do not suffer from phase ambiguities. Thus, higher bandwidth chirps can be used to obtain better depth resolution. These systems can operate at long-range with a precision that increases with increasing chirp bandwidth or decreasing chirp duration \cite{amann2001laser}. However, linearly chirped systems are more expensive, requiring fast RF electronics, but show promise in long-range object detection and tracking \cite{redman2006chirped}.

FMCW-LiDAR is the most precise LiDAR system available. The laser frequency is linearly chirped with a small fraction of the beam serving as a local oscillator. The modulation bandwidth is often larger than in linearly-chirped AMCW LiDAR -- yielding superior depth resolution. Yet, high-speed electronics are not required for signal detection. Instead, detection requires an optical heterodyne measurement, using a slow square-law detector, to demodulate the return signal and generate a beat-note frequency that can be recorded by slower electronics. Unfortunately, FMCW-LiDAR is limited by the coherence length of the laser \cite{amann2001laser}. To obtain precision at longer ranges, the local oscillator can be time-delayed such that the system can operate up to 100 km -- but with relatively poor resolution \cite{Herschel:12}. For marginally shorter distances, 5 cm precision in the field at 39.2 km and sub-centimeter precision at 40 km in the lab was demonstrated in \cite{Ito:12}. 

Both AMCW and FMCW linearly-chirped LiDAR systems can obtain similar or better depth accuracy compared to pulsed systems while operating with cheaper electronics. This is because the heterodyne beat note can be engineered to reside within the bandwidth of slow electronics and more precisely measured. For this reason, high resolution AMCW based systems are more popular when used in conjunction with relatively slow CCD cameras because the pixels can be modulated directly. Because CCD cameras are relatively slow, the dynamic range over which ranging is possible is shortened. The working range is often lengthened at the expense of depth resolution. Additionally, higher resolution is obtained at the expense of SNR. Thus, a linearly-chirped FMCW-LiDAR that uses a high-bandwidth detector array for superior accuracy and larger dynamic range is desirable. Yet, a high-bandwidth detector array is prohibitively expensive and raster scanning with a readily available high-bandwidth detector is too slow. Our solution to this problem is to build a compressive camera \cite{duarte2008single} linearly-chirped FMCW-LiDAR system.

\section{FMCW-LiDAR Compressive Depth-Mapping}
\subsection{Single-Object FMCW-LiDAR}

Within an FMCW-LiDAR system, a laser is linearly swept from frequency $\nu_0$ to $\nu_f$ over a period $T$. The electric field $E$ takes the form
\begin{equation}
E(t) = A \exp\left(2\pi i\left[\nu_0 + \frac{\nu_f-\nu_0}{2T}t\right]t\right),
\label{eq:sweep}
\end{equation}
where $A$ is the field amplitude. It can be easily verified, by differentiation of Eq. (\ref{eq:sweep}), that the instantaneous laser frequency is $\nu_0+[(\nu_f-\nu_0)/T]t$. 

A small fraction of the laser's power is split off to form a local oscillator ($E_{\text{LO}}$) while the remainder is projected towards a target at an unknown distance $d$. The returning signal reflected off the target, with electric field $E_{\text{Sig}}(t-\tau)$ and time delay $\tau$, is combined with $E_{\text{LO}}$ and superimposed on a square-law detector. The resulting signal, as seen by an oscilloscope ($P_{\text{Scope}}$) and neglecting detector efficiency, is
\begin{equation}
P_{\text{Scope}}(t) = \frac{\epsilon_0 c}{2}\left[A_{\text{Sig}}^2 + A_{\text{LO}}^2 +
 2A_{\text{LO}}A_{\text{Sig}}\sin\left(2\pi\frac{\Delta\nu\tau}{T}t+\phi\right)\right],
\label{eq:singleLIDAR} 
\end{equation}
where $\phi = \nu_0\tau - \frac{\Delta\nu}{2T}\tau^2$ is a constant phase, $\epsilon_0$ is the permittivity of free space, $c$ is the speed of light in vacuum, and $\Delta\nu = \nu_f-\nu_0$. The alternating-current (AC) component within Eq. (\ref{eq:singleLIDAR}) oscillates at the beat-note frequency $\nu = \Delta\nu\tau/T$ from which a distance-to-target can be calculated from
\begin{equation}
d = \frac{\nu T c}{2\Delta\nu}.
\label{eq:dist}
\end{equation}

\subsection{Multiple-Object FMCW-LiDAR}

The detection scheme should be altered when using a broad illumination profile to detect multiple targets simultaneously. Just as beat notes exist between the local oscillator and signal, there now exist beat notes between reflected signals at varying depths that will only inject noise into the final readout. Balanced heterodyne detection is used to overcome this additional noise source. A 50/50 beam-splitter is used to first mix $E_{\text{LO}}$ and $E_{\text{Sig}}$. Both beamsplitter outputs, containing equal powers, are then individually detected and differenced with a difference-detector. 

Mathematically, $E_{\mathrm{LO}}$ is the same as before, but the return signal from $j$ objects at different depths takes the form
\begin{equation}
\sum\limits_{j} E_{\text{Sig}}(t-\tau_j) = \sum\limits_{j} A_{j}\exp\left(2\pi i\left[\nu_0+\frac{\Delta\nu}{2T}\left(t-\tau_j\right)\right]\left(t-\tau_j\right)\right),
\end{equation} 
where we have labeled each object with an identifier index $j$ that introduces a time delay $\tau_j$ and amplitude $A_j$, depending on the reflectivity of the object. After mixing at the beamsplitter, the optical power from the difference detector, as seen by an oscilloscope ($P_{\text{Scope}}$), again neglecting detector efficiency, is  
\begin{align}
P_{\text{Scope}}(t)&=\frac{\epsilon_0 c}{2}\left(\left|\frac{E_{\text{LO}}\left(t\right) + i\sum_{j} E_{\text{Sig}}\left(t-\tau_j\right)}{\sqrt{2}}\right|^2 - \left|\frac{iE_{\text{LO}}\left(t\right) + 
\sum_{j} E_{\text{Sig}}\left(t-\tau_j\right)}{\sqrt{2}}\right|^2\right) \\
&= \epsilon_0 c\sum\limits_{j} A_{\text{LO}}A_{j}\sin\left(2\pi\frac{\Delta\nu\tau_j}{T}t + \phi_j\right),
\label{eq:heterodyneSig}
\end{align}
where $\phi_j = \nu_0\tau_j - \frac{\Delta\nu}{2T}\tau_j^2$ is a constant phase. The sum-frequency signal components are filtered by the slow-bandwidth detectors. The resulting balanced heterodyne signal in Eq. (\ref{eq:heterodyneSig}) is still amplified by the local oscillator and contains AC signals at only the frequencies $\nu_j = \Delta\nu\tau_j/T$. Again, these frequencies can be converted to distance using Eq. (\ref{eq:dist}).

\subsection{Compressive Imaging Background}

Compressive imaging is a technique that trades a measurement problem for a computational reconstruction within limited-resource systems. Perhaps, the most well known example of a limited-resource system is the Rice single-pixel camera \cite{duarte2008single}. Instead of raster scanning a single-pixel to form an $n$-pixel resolution image $\vect{x}$, i.e. $\vect{x}\in\mathbb{R}^{n}$, an $n$-pixel digital micro-mirror device (DMD) takes $m\ll n$ random projections of the image. The set of all DMD patterns can be arranged into a sensing matrix $\vect{A}\in\mathbb{R}^{m\times n}$, and the measurement is modeled as a linear operation to form a measurement vector $\vect{y}\in\mathbb{R}^{m}$ such that $\vect{y}=\vect{Ax}$.

Once $\vect{y}$ has been obtained, we must reconstruct $\vect{x}$ within an undersampled system. As there are an infinite number of viable solutions for $\vect{x}$ within the problem $\vect{y} = \vect{Ax}$, CS requires additional information about the signal according to a previously-known function $g(\vect{x})$. The function $g(\vect{x})$ first transforms $\vect{x}$ into a sparse or approximately sparse representation, i.e. a representation with few, or approximately few, non-zero components. The function $g(\vect{x})$ then uses the $L^1$-norm to return a scalar.

Within image-reconstruction problems, $g(\vect{x})$ is often the total-variation (TV) operator because it provides a sparse representation by finding an image's edges. Anisotropic TV is defined as $\text{TV}(\vect{x}) = \parallel [\nabla_x^T, \nabla_y^T]^T \vect{x}\parallel_1$, where $\nabla$ is a finite difference operator that acts on an image's Cartesian elements ($x$ and $y$) and $\ast^T$ is the transpose of $\ast$. Note that the $L^p$-norm of $\vect{x}$ is defined as $\parallel \vect{x} \parallel_p = (\sum_i |\vect{x}_i|^p )^{1/p}$. Thus, we solve the following TV-minimization problem:
\begin{equation}
\text{arg}\,\min\limits_{\hat{\vect{x}}\in\mathbb{R}^n}\parallel \vect{A}\hat{\vect{x}} - \vect{y} \parallel_2^2 + \alpha \mathrm{TV}\left(\hat{\vect{x}}\right),
\label{eq:TV}
\end{equation}
where the first term is a least-squares fitting parameter consistent with the measurements $\vect{y}$, the second term is a sparsity promoting parameter, and $\alpha$ is a weighting constant.

\subsection{Compressive FMCW-LiDAR depth-mapping theory} \label{ssec:ReconTheory}

\begin{figure}
\centering\includegraphics[width=.8\textwidth]{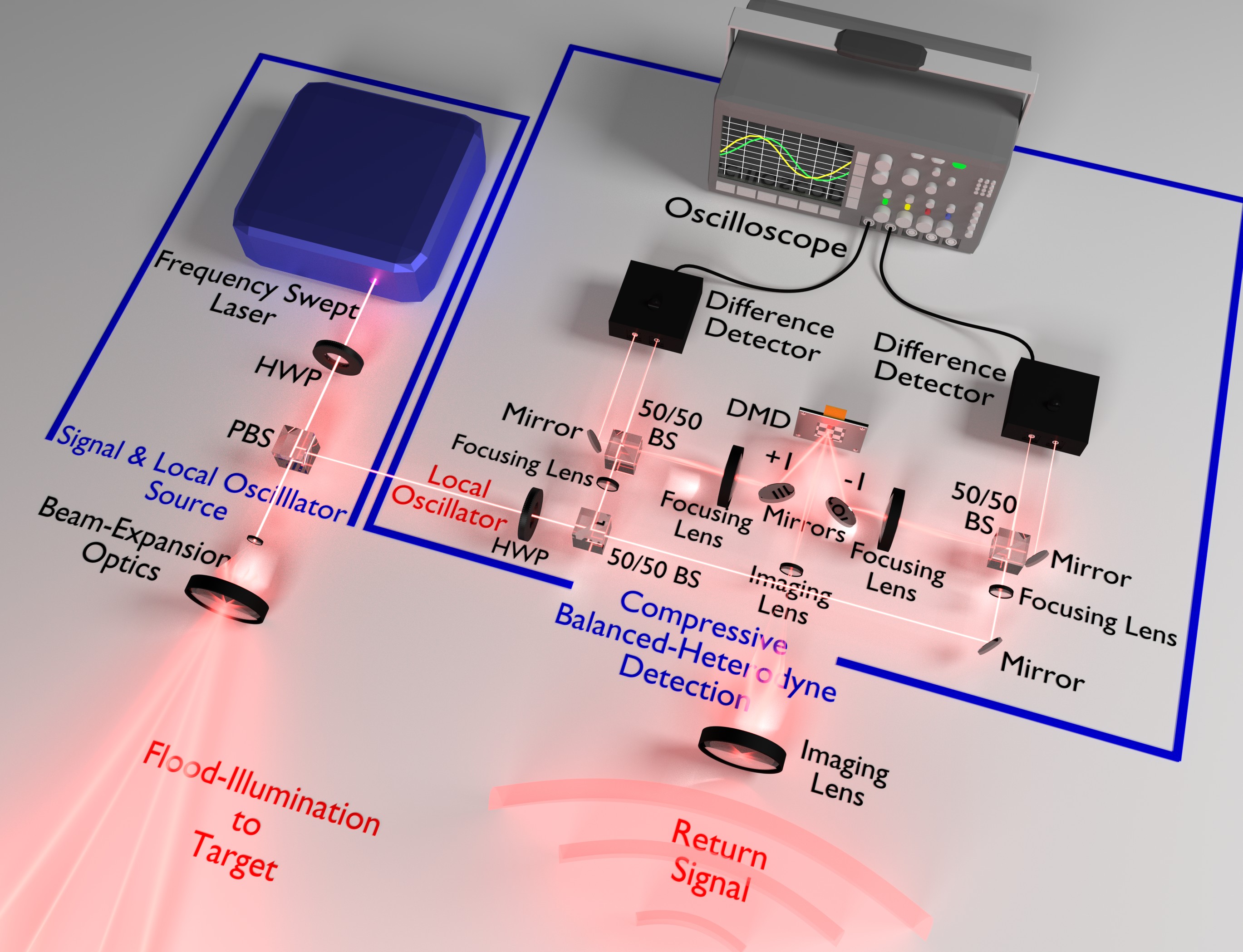}
\caption[Proposed compressive FMCW-LiDAR experiment]{\textbf{Proposed Experiment: } HWP: half-wave plate, PBS: polarizing beam-splitter, DMD: digital micro-mirror device, BS: beam-splitter. A linearly chirped laser is split into two beams, designated as a local-oscillator and a signal, via a HWP and a PBS. The signal illuminates a target scene and the reflected radiation is used to image the scene onto a DMD. The DMD takes pseudo-random spatial projections, consisting of $\pm 1$ pixel values, and directs the projections to balanced-heterodyne detectors using the local-oscillator.}
\label{fig:setup}
\end{figure}

The proposed experimental diagram is shown in Fig. \ref{fig:setup}. A balanced-heterodyne-detection based FMCW-LiDAR system for multiple-object ranging is combined with a DMD-based compressive-imaging system to obtain transverse spatial information more efficiently than raster scanning. A linear-frequency chirped laser broadly illuminates a scene and the reflected radiation is imaged onto a DMD. 
The DMD can take projections using $\pm 1$ pixel values, meaning the sensing matrix $\vect{A}$ is limited to $\pm 1$ values. Because the 1 and -1 DMD pixel values reflect light in different directions, we split the acquisition process into separate $(1,0)$ and $(0,1)$ balanced heterodyne detections. A readout instrument, such as an oscilloscope, records the two signals. The time-dependent readout signals are then Fourier transformed -- keeping only the positive frequency components and the amplitudes are differenced to arrive at a $(1,-1)$ balanced heterodyne projection frequency representation. 
The measurement vector construction and depth-map reconstruction processes are detailed in this section.

Let a three-dimensional scene $\vect{x}$ of depth $L$ be discretized into $N$ depths and $n$ transverse spatial pixels containing objects at $j$ depths. Let a 2D image of the scene $\vect{x}$ at depth $j$ be represented by a real one-dimensional vector $\vect{x}_{Ij} \in \mathbb{R}^{n}$. 
The $l^{\mathrm{th}}$ pixel in $\vect{x}_{Ij}$ is represented as $\vect{x}_{Ij}^{l}$.
The scene at depth $j$ (i.e. $\vect{x}_{Ij}$) yields a time-dependent heterodyne signal of $\sum_{l=1}^n \vect{x}_{Ij}^l = \epsilon_0cA_{\mathrm{LO}}A_{j}\sin\left(2\pi\nu_jt+\phi_j\right)$, where $\nu_j=\Delta\nu\tau_j/T$. Let us also assume that we are time-sampling at the Nyquist rate needed to see an object at a maximum depth $L$ having a beat-note frequency of $2\Delta\nu L / (Tc)$, meaning we acquire $2N$ data points when sampling at a frequency of $4\Delta\nu L / (Tc)$ over a period $T$. This is the minimum sampling requirement. Within an experiment, sampling faster than the Nyquist rate will enable the accurate recovery of both the frequency and amplitude at depth $L$.
Let a single pattern to be projected onto the imaged scene be chosen from the sensing matrix $\vect{A}\in\mathbb{R}^{m\times n}$ and be represented by a real one-dimensional vector $\vect{A}_k \in \mathbb{R}^n$ for $k = 1,2,...m$, where $m < n$. For efficent compression, we require $m\ll n$. When including the $2N$ time samples, the heterodyne signal is now represented as a matrix of $n$ pixels by $2N$ time points, $\vect{P}_{\mathrm{Scope}}(t)\in\mathbb{R}^{m\times 2N}$, such that
\begin{equation}
\vect{P}_{\mathrm{Scope}}(t) = \epsilon_0 c\sum\limits_{j} A_{\text{LO}}\vect{A}\vect{x}_{Ij}\sin\left(2\pi\frac{\Delta\nu\tau_j}{T}t + \phi_j\right).
\end{equation}

To obtain TOF information, it is helpful to work in the Fourier domain. The absolute value of the first $N$ elements in the Fourier transform of $\vect{P}_{\mathrm{Scope}}(t)$ returns the compressive measurement matrix of positive frequency amplitudes, $\vect{y}_{I}(\nu_+)=\left\lvert\mathscr{F}\left[\vect{P}_{\mathrm{Scope}}(t)\right]_{+}\right\rvert \in\mathbb{R}^{m\times N}$, where $\mathscr{F}\left[\ast\right]_{+}$ returns the positive-frequency components of $\ast$. Thus, $\vect{y}_I(\nu_+)$ consists of $N$ different measurement vectors of length $m$. Each of the $N$ measurement vectors contain compressed transverse information for a particular depth.

While individual images at a particular depth can be reconstructed from $\vect{y}_I(\nu_+)$, it is inefficient based on the size of the measurement matrix and the number of needed reconstructions. A more practical method with a significantly smaller memory footprint and far fewer reconstructions requires that we trace out the frequency dependence and generate only two measurement vectors of length $m$, designated as $\vect{y}_I$ and $\vect{y}_{I\nu}\in\mathbb{R}^m$ such that
\begin{align}
\vect{y}_I &= \sum\limits_{\nu_+ = 1}^N \left\lvert\mathscr{F}\left[\vect{P}_{\mathrm{Scope}}(t)\right]_{+}\right\rvert \label{eq:yi}\\
\vect{y}_{I\nu} &= \sum\limits_{\nu_+ = 1}^N \left\lvert\mathscr{F}\left[\vect{P}_{\mathrm{Scope}}(t)\right]_{+}\right\rvert \vect{\nu_+}, \label{eq:yvi}
\end{align}
where $\vect{\nu_+}\in\mathbb{R}^{N\times N}$ is a diagonal matrix of frequency indices that weights each frequency amplitude of $\vect{y}_I(\nu_+)$ by its own frequency ($\nu$) before summation.

To summarize, the measurement vector $\vect{y}_I$ is formed by summing the positive-frequency amplitudes, per projection, and is equivalent to a typical compressive-imaging measurement. The measurement vector $\vect{y}_{I\nu}$ requires that each positive-frequency amplitude first be weighted by its frequency before summation. A depth map $\vect{d}$ can be extracted by reconstructing the transverse image profiles ($\hat{\vect{x}}_I, \hat{\vect{x}}_{I\nu}\in\mathbb{R}^n$) and then performing a scaled element-wise division as 
\begin{equation}
\vect{d} = \frac{\hat{\vect{x}}_{I\nu}}{\hat{\vect{x}}_I}\frac{T c}{2 \Delta\nu},
\label{eq:depth}
\end{equation}
where $\hat{\vect{x}}_{I\nu}/\hat{\vect{x}}_I$ is an element-wise division for elements $>0$. A similar technique designed for photon-counting with time-tagging was implemented in \cite{howland2011photon}.

\section{Computer Simulation}

\subsection{Simulation parameters} \label{ssec:param}

\begin{figure}
\centering\includegraphics[width=.7\textwidth]{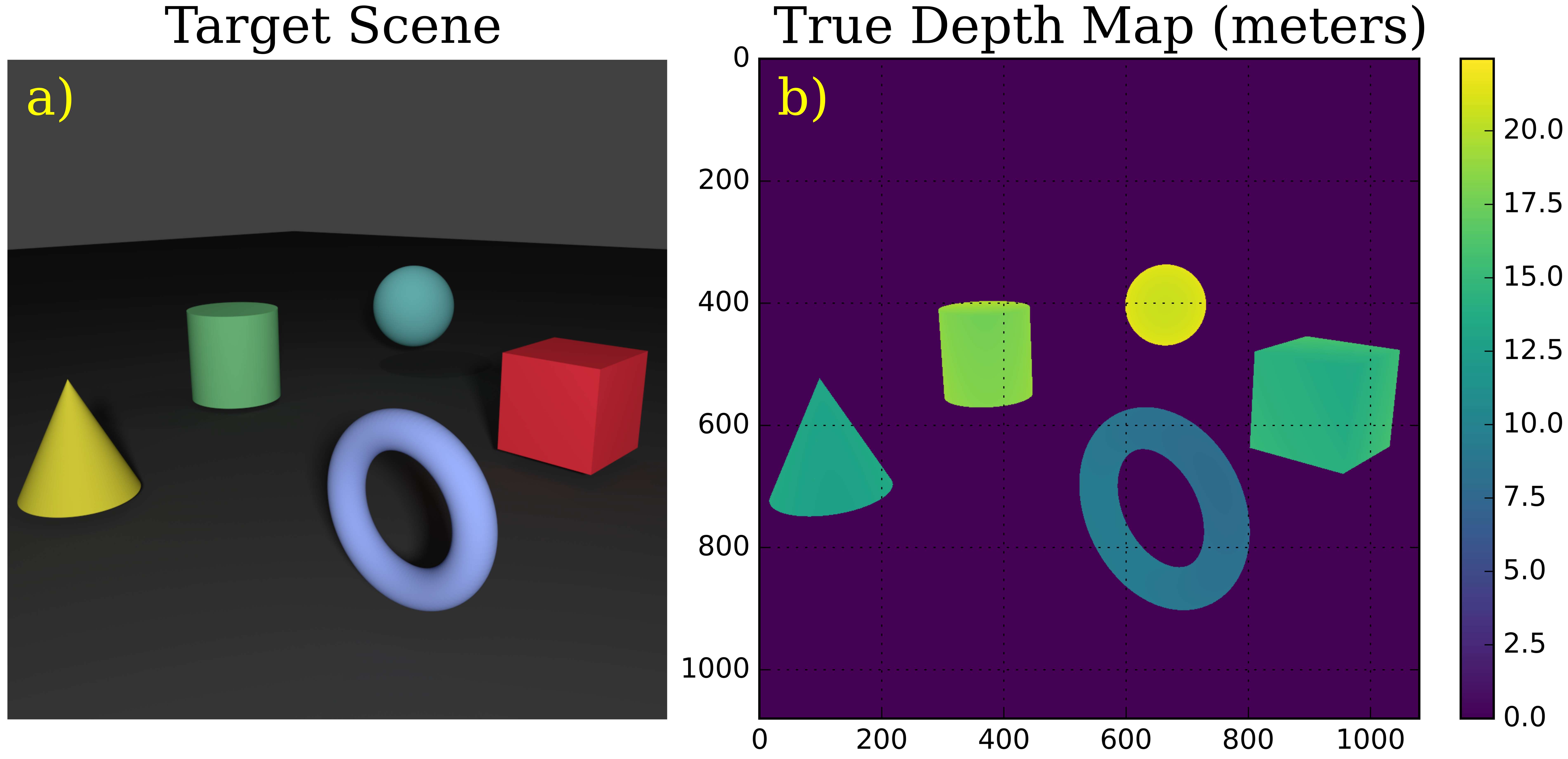}
\caption[]{Image \textbf{(a)} presents a 3-dimensional scene composed of Lambertian-scattering targets. Image \textbf{(b)} presents the depth map we wish to compressively recover.}
\label{fig:original}
\end{figure}

\begin{figure}
\centering\includegraphics[width=.8\textwidth]{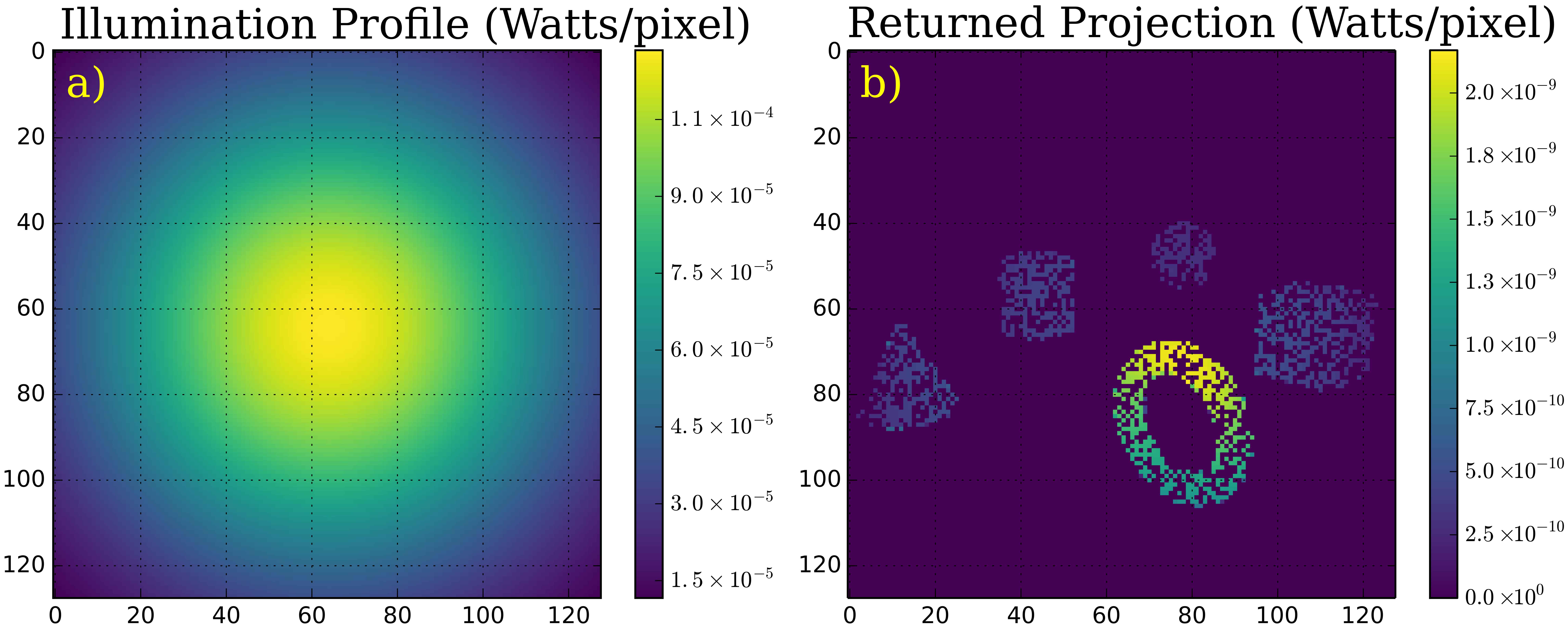}
\caption[Illumination profile and return flux]{Image \textbf{(a)} presents an illumination profile that consists of 1 Watt in total power. The scene is discretized into a $128 \times 128$ pixel resolution scene with a maximum power per pixel of 119 $\mu$W. When using a 2 inch collection optic and modeling each object as a Lambertian scatter, image \textbf{(b)} presents a single projection taken by the DMD of the reflected radiation as seen by one detector. The maximum power per pixel is of order nanowatts.}
\label{fig:illumination}
\end{figure}

A noisy simulation of compressive depth-mapping is presented within this section. The simulation presents detailed steps needed to acquire depth-maps from signals disrupted by both laser-linewidth uncertainty and Gaussian white noise when compressively measuring the scene presented in Fig. \ref{fig:original}. We assume the scene contains Lambertian-scattering objects and is illuminated with 1 W of 780 nm light using an illumination profile shown in Fig. \ref{fig:illumination}. With a large enough angular spread of optical power, the system can considered eye-safe at an appropriate distance from the source. We discretize the scene with a $128\times 128$ pixel resolution DMD ($n = 16384$) operating at 1 kHz and use a 2 inch collection optic according to the diagram in Fig. \ref{fig:setup}. Figure \ref{fig:illumination} presents an example DMD projection of the returned radiation intensities before the balanced heterodyne detection as seen by only one of the detectors (corresponding to a (1,0) projection). A local oscillator of 100 $\mu$W is mixed with the returned image for the balanced heterodyne detection. 

To model the frequency sweep, we use experimental parameters taken from \cite{cheok20103d}  and \cite{Satyan:09} which state that a 100 GHz linear sweep over 1 ms is realizable for various laser sources. To bias the simulation towards worse performance, we model the laser linewidth with a 1 MHz Lorentzian full-width half-maximum (FWHM), meaning the beat note on our detectors will have a 2 MHz FWHM Lorentzian linewidth \cite{nazarathy1989spectral}. A narrower linewidth can be obtained with many lasers on the market today, but we bias the simulation towards worse operating parameters to show the architecture's robustness. The simulated laser has a coherence length of $c/(\pi \Delta\nu_\mathrm{FWHM}) \approx 95$ m, where $\Delta\nu_{\mathrm{FWHM}} = 1$ MHz.

FMCW-LiDAR can use relatively inexpensive measurement electronics while still providing decent resolution at short to medium range. Using a 100 GHz frequency sweep over 1 ms leads to a beat note of 16.67 MHz from an object 25 m away. Keeping with the idea of slower electronics, we model the detection scheme using balanced-detectors and an oscilloscope that can see 16.65 MHz. The simulated oscilloscope sampled at 33.3 MHz over 1 ms, corresponding to a record-length of $33.3\times 10^3$ samples per frequency sweep. The frequency resolution is 1 kHz with a maximum depth resolution of 1.5 mm. The furthest object in our test scene is located at approximately 22 m. 

We also assume a 1 ms integration time per DMD projection, corresponding to one period of the frequency sweep per projection. We model the linear frequency chirp using a sawtooth function for simplicity, as opposed to more commonly used triangle function in experiments. 

The sensing matrix used for the compressive sampling is based on a randomized Sylvester-Hadamard matrix displays values of $\pm 1$. We use two detectors to acquire the information with one projection -- resulting in only $m$ projections for an $m$-row sampling matrix. Alternatively, a cheaper single-detector version can also be implemented, but it must display two patterns for every row of the sensing matrix -- resulting in $2m$ projections.  While the alternate detection scheme requires twice as many measurements, we still operate in the regime where $2m \ll n$.

\subsection{Compressive measurements} \label{ssec:measure}

\begin{figure}
\centering\includegraphics[width=.9\textwidth]{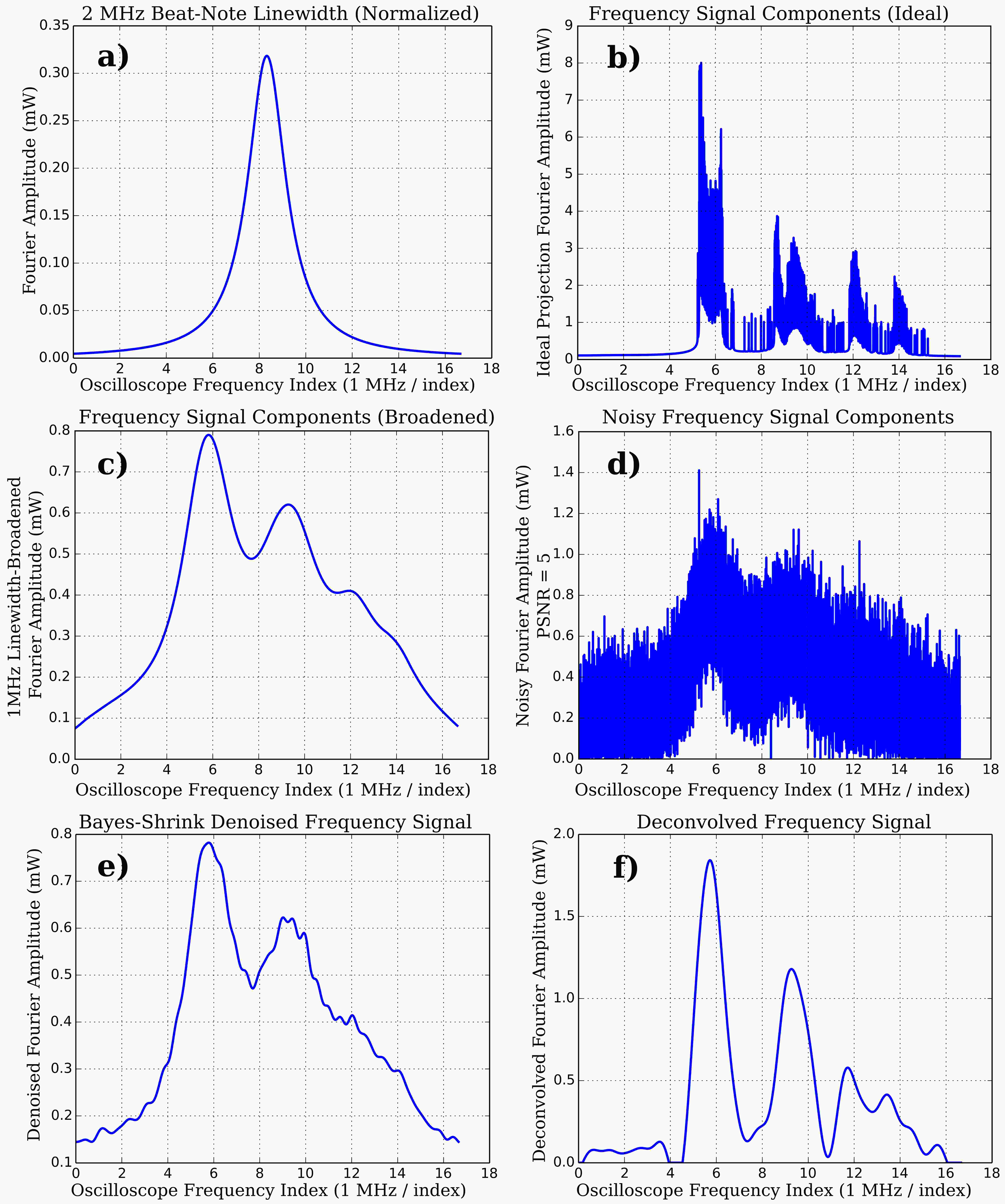}
\caption[Expected FMCW-LiDAR return signal (before and after processing)]{Image \textbf{(a)} shows the 2 MHz Lorentzian linewidth on the beat notes expected from a laser's 1 MHz Lorentzian linewidth as seen by an oscilloscope sampling at 33.3 MHz. Image \textbf{(b)} shows the noiseless, positive-frequency components from a single noiseless projection from Fig. \ref{fig:illumination} when using a zero-linewidth laser. When accounting for a 2 MHz beat-note linewidth, image \textbf{(c)} shows the broadened expected frequency components. When including a peak SNR = 5  (based on the SNR of the brightest object or pixel), image \textbf{(d)} presents a realistic noisy signal one would measure in an experiment. Image \textbf{(e)} presents the result of denoising the signal in \textbf{(d)} with a Bayes-shrink denoising filter using a symlet-20 wavelet decomposition. Image \textbf{(f)} presents the result of deconvolving the 2 MHz Lorentzian linewidth with the denoised signal in \textbf{(e)} using a Weiner filter. Image \textbf{(f)} is the cleaned signal from the (1,0) projection. A similarly cleaned result from the (0,1) signal will then be subtracted from \textbf{(f)} and then used to form $\vect{y}_I$ and $\vect{y}_{I\nu}$. }
\label{fig:signal}
\end{figure}

To simulate data acquisition in a noisy environment, we add Gaussian white noise in addition to laser-linewidth uncertainty. This section explains how the Fourier-transformed oscilloscope amplitudes are processed before applying the operations found in Eqns. (\ref{eq:yi}) and (\ref{eq:yvi}). 

Figure \ref{fig:signal} explains the simulated noise addition and removal process. Image (a) depicts a 2 MHz normalized beat note Lorentzian linewidth uncertainty in terms frequency components. Image (b) presents a single ideal DMD projection, using a 1 ms integration time, that has not been distorted by any form of noise. Image (b) clearly depicts 5 main peaks, one per object. However, beat-note linewidth uncertainty will blur these features together such that image (c) contains frequency uncertainty. In addition to laser-linewidth, varying levels of Gaussian white noise are injected into image (c). Image (d) presents a typical noisy signal one would expect to see for a peak signal-to-noise ratio (PSNR) of 5, meaning the SNR of the brightest Lorentzian-broadened Fourier component has an SNR equal to 5. SNR is defined as the signal's boradened amplitude divided by the standard deviation of Gaussian white noise. Note that the farthest object has an SNR at or below unity.

To clean a typical signal seen by only one detector, as in image (d), we use a BayesShrink filter \cite{BayesShrink} with a symlet-20 wavelet decomposition (see the PyWavelets library for more details). BayesShrink is a filter designed to suppress Gaussian noise using a level-dependent soft-threshold at each level of a wavelet decomposition. Because a symlet-20 wavelet is reasonably close in shape to the original signal components found in image (b), it has excellent denoising properties for these particular data sets. Alternate wavelets or denoising algorithms can be implement and may demonstrate better performance. Image (e) is the result of applying a symlet-20 based BayesShrink filter. A Weiner filter \cite{wiener1949extrapolation} is then used to deconvolve the known Lorentzian linewidth of image (a) with the denoised signal in (e) to produce signal in (f). Image (f) constitutes only one of the needed projections, i.e. (1,0). The process must also be applied to the signal from the second detector consisting of (0,1) projection. The two cleaned signals are then subtracted to produce a (1,-1) projection. The differenced signal is then used to form a single element of each measurement vector $\vect{y}_{I}$ and $\vect{y}_{I\nu}$ using Eqns. \ref{eq:yi} and \ref{eq:yvi}. 


\subsection{Depth-map reconstruction} \label{ssec:reconstruct}

\begin{figure}
\centering\includegraphics[width=\textwidth]{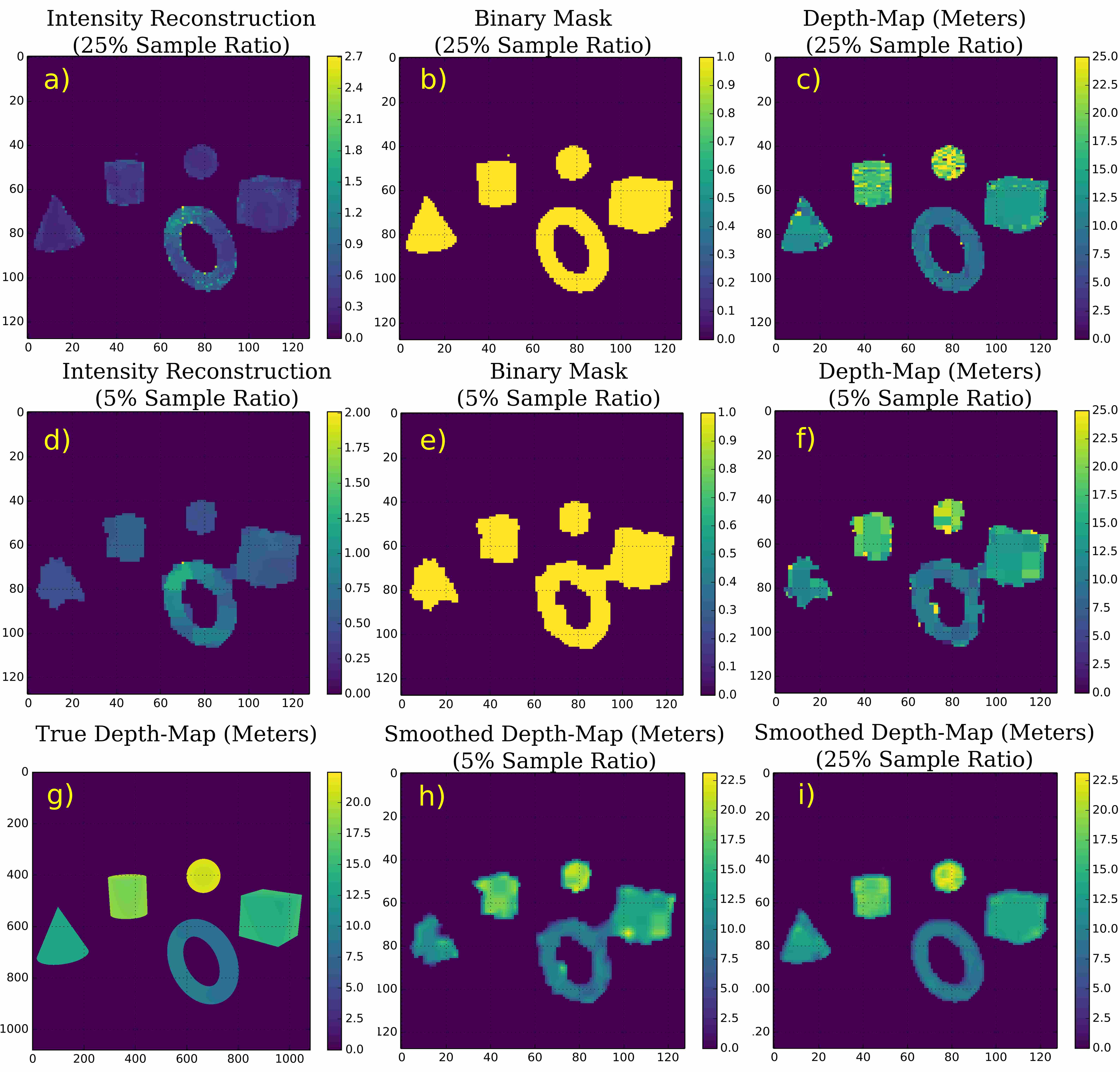}
\caption[Depth-map reconstructions]{When using a PSNR = 5, image \textbf{(a)} shows a typical total-variation minimization reconstruction from a $25\%$ sample-ratio intensity-measurement vector $\vect{y}_I$. Shapes are identified, but the pixel values are incorrect. Image \textbf{(b)} is a binary mask generated by hard-thresholding image \textbf{(a)}. After representing image \textbf{(b)} in a sparse basis, such as a Haar-wavelet decomposition, least-squares can be performed on the $m/3$ largest signal components to construct $\vect{x}_I$ and $\vect{x}_{I\nu}$. After applying Eq. (\ref{eq:depth}), a depth map is presented in image \textbf{(c)}. Images \textbf{(d)}, \textbf{(e)}, and \textbf{(f)} demonstrate the same process for a $5\%$ sample ratio, again with PSNR = 5. Image \textbf{(g)} is the true depth-map presented for easy comparison. Images \textbf{(h)} and \textbf{(i)} are smoothed depth-maps after applying a $4\!\times\! 4$ pixel averaging kernel to depth-maps \textbf{(f)} and \textbf{(c)}, respectively. }
\label{fig:results}
\end{figure}

Once the process outlined in section \ref{ssec:measure} has been repeated $m$ times, depth maps are reconstructed from $\vect{y}_{I}$ and $\vect{y}_{I\nu}$ using the procedure outlined in section \ref{ssec:ReconTheory}. Details to the recovery of $\hat{\vect{x}}_I$ and $\hat{\vect{x}}_{I\nu}$ are presented in this section.

Ideally, one would only solve Eqn. (\ref{eq:TV}) twice to obtain $\hat{\vect{x}}_I$ and $\hat{\vect{x}}_{I\nu}$ and then apply Eq. (\ref{eq:depth}) to extract high-resolution depth-maps. Unfortunately, TV-minimization will not always return accurate pixel values, particularly at low sample percentages, and will result in depth-maps with well-defined objects but at incorrect depths. While TV-minimization excels at finding objects when making $\alpha>1$, the least-squares term that ensure accurate pixel values suffers. To obtain accurate depth maps, we must locate objects accurately \emph{and} obtain accurate pixel values.

Instead, we only apply one TV-minimization to locate the objects within $\hat{\vect{x}}_{I}$. After hard thresholding the image returned from TV-minimization, if needed to eliminate background noise, we are left with an image of objects with incorrect pixel values. An example of this step can be found in images (a) and (d) of Fig. \ref{fig:results}, where image (a) was obtained from $m=.25n$ measurements and image (d) was obtained from $m=.05n$ measurements. From $\hat{\vect{x}}_{I}$, we generate a binary mask $\vect{M}$, as demonstrated in images (b) and (e) that tells us the locations of interest.

Once we have a binary mask $\vect{M}$, a unitary transform $\Psi$ converts $\vect{M}$ into a sparse representation $\vect{s}$ such that $\vect{s} = \Psi\vect{M}$. In our simulations, $\Psi$ is a wavelet transformation using a Haar wavelet. We tested other wavelets with similar results. We then keep only $m/3$ of the largest signal components of $\vect{s}$ to form a vector $\vect{s}_{m/3} = \vect{P}\vect{s}$, where the operator $\vect{P}\in\mathbb{R}^{m/3\times n}$ is a sub-sampled identity matrix that extracts the $m/3$ largest signal components of $\vect{s}$. We chose $m/3$ rather than a larger percentage of $m$ based on empirical results. In essence, $\vect{s}_{m/3}$ is the support of $\vect{s}$, i.e. the significant nonzero components of $\vect{s}$ containing the transverse spatial information of our depth-map. Because we have a vector $\vect{s}_{m/3}\in\mathbb{R}^{m/3}$ and two measurement vectors $\vect{y}_{I}\in\mathbb{R}^m$ and $\vect{y}_{I\nu}\in\mathbb{R}^m$, we can perform least-squares on the now-overdetermined system. Specifically, we apply the operations
\begin{align}
\vect{s}_{m/3} &= \vect{P}\Psi\hat{\mathrm{M}}\left(\text{arg}\,\min\limits_{\vect{x}_{I}\in\mathbb{R}^n}\parallel \vect{A}\vect{x}_I - \vect{y}_I \parallel_2^2 + \alpha \mathrm{TV}\left(\vect{x}_I\right)\right) \label{eq:TVxI}\\
\hat{\vect{x}}_I &= \vect{M}\Psi^{-1}\vect{P}^T\left(\mathrm{arg}\,\min\limits_{\vect{s}_{m/3}}\parallel \vect{A}\Psi^{-1}\vect{P}^T\vect{s}_{m/3}-\vect{y}_I\parallel_2^2\right) \label{eq:LSxI} \\
\hat{\vect{x}}_{I\nu} &= \vect{M}\Psi^{-1}\vect{P}^T\left(\mathrm{arg}\,\min\limits_{\vect{s}_{m/3}}\parallel \vect{A}\Psi^{-1}\vect{P}^T\vect{s}_{m/3}-\vect{y}_{I\nu}\parallel_2^2\right) \label{eq:LSxInu},
\end{align}
where the operation $\hat{\mathrm{M}}$ returns a mask $\vect{M}$. Note that $\vect{P}^T$ is the transpose of $\vect{P}$ and $\Psi^{-1}$ is the inverse wavelet transform. Also notice that $\vect{s}_{m/3}$ and $\vect{P}$ contain the transverse spatial information of interest within our depth-map and are used in both Eqn. (\ref{eq:LSxI}) and Eqn. (\ref{eq:LSxInu}). They allow us to obtain an accurate heterodyne amplitude image $\hat{\vect{x}}_I$ (from $\vect{y}_I$) and an accurate frequency weighted heterodyne amplitude image $\hat{\vect{x}}_{I\nu}$ (from $\vect{y}_{I\nu}$). 

The total variation minimization was performed with a custom augmented Lagrangian algorithm written in Python. It is based on the work in \cite{yin2010practical} but is modified to use fast-Hadamard transforms. Additionally, the gradient operators within the TV operator use periodic boundary conditions that enables the entire algorithm to function with only fast-Hadamard transforms, fast-Fourier transforms, and soft-thresholding. The least-squares minimizations in Eqns. (\ref{eq:LSxI}) and (\ref{eq:LSxInu}) are performed by the Broyden-Fletcher-Goldfarb-Shanno (BFGS) algorithm \cite{fletcher2013practical} -- a fast quasi-Newton method. Depth maps are obtained by applying Eq. (\ref{eq:depth}). Example depth maps are found in images (c) and (f) of Fig. \ref{fig:results}. If desired, light smoothing can be applied to obtain images (h) and (i).

The reconstruction algorithms recovered the low-spatial-frequency components of the depth-maps in Fig. \ref{fig:results}, and arguably recovered a low-resolution depth-map. However, missing details can be retrieved when using a higher resolution DMD, sampling with higher sampling ratios, or applying adaptive techniques \cite{AdaptiveCS}.   

\subsection{Results}

\begin{figure}
\centering\includegraphics[width=.7\textwidth]{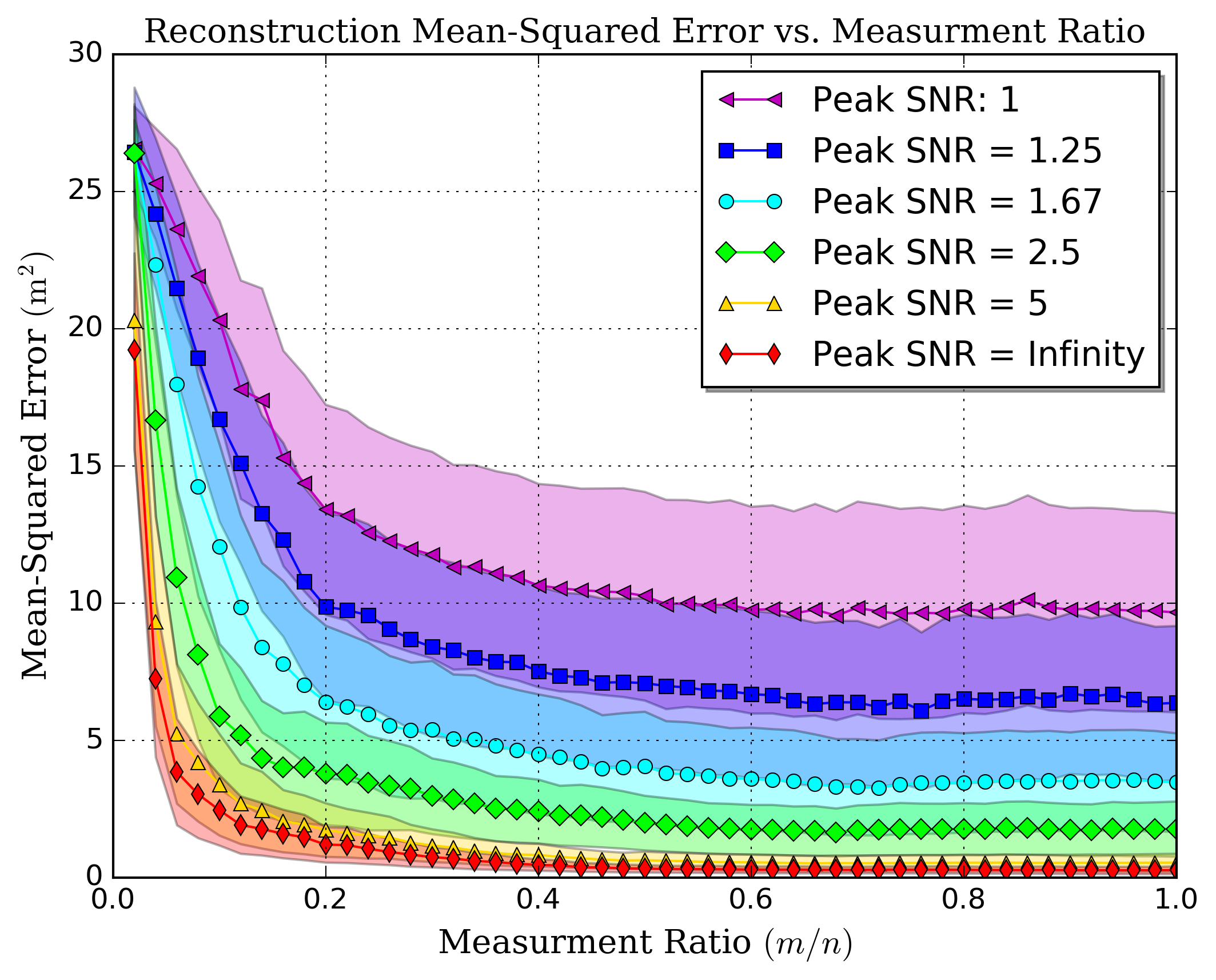}
\caption[Mean-squared error in reconstructed depth-maps as a function of PSNR and measurments $M$]{\textbf{Reconstruction Mean-Squared Error:} Six different peak-SNR scenarios were considered. Within each peak-SNR scenario, 10 compressive simulations using a 100\% sample ratio were generated. Reconstructions were then performed by varying the measurment ratio in increments of $2\%$, starting from $m = .02n$ and ending at $m = n$. Each data-point represents an average mean-squared error from 10 reconstructions. The error in the averaged mean-squared error is given by the shaded regions. We can see that the error jumps after a peak-SNR of 5, likely due to the farthest object having an SNR of approximately 1 at peak-SNR = 5. The plot demonstrates that a measurment ratio of $m = .2n$ is sufficient to generate a reasonably accurate depth-map for the scene in Fig. \ref{fig:original}.}
\label{fig:statistics}
\end{figure}

To test the robustness and accuracy of the reconstructions with respect to noise and sample ratios, varying levels of Gaussian white noise were introduced into the simulation. Six different noise levels are presented in Fig. \ref{fig:statistics}. For each PSNR level, 10 critically sampled ($m=n$) measurement vectors were generated using different sensing matrices but with the same parameters listed in section \ref{ssec:param}. Moving in increments of $2\%$, $m$ was adjusted from $.02n$ to $n$ for each measurement vector. Reconstructions were performed at each sample percentage and the non-smoothed reconstructed depth-map were compared against the true depth-map to generate a mean-squared error. Thus, every data point in Fig. \ref{fig:statistics} is the average mean-squared error of 10 reconstructions while the solid colored regions above and below each marker designate the error in the average. As seen in Fig. \ref{fig:results}, the mean-squared error increases rapidly for PSNR$<5$, probably because the SNR of distant objects is well below unity. Additionally, we see that a sample ratio of roughly $m/n = .2$ is sufficient for depth-map generation at almost all noise levels. For PSNR$>5$, sample ratios of approximately $m/n = .1$ are sufficient for accurate depth maps.

\section{Conclusion}

FMCW-LiDAR is a valuable technique for ranging because of its accuracy and relatively low cost. It is a natural extension to apply FMCW-LiDAR systems to depth-mapping, whether through raster scanning or other means. The compressive architecture presented in this article is an efficient means of depth-mapping that is inexpensive and can be easily implemented with off-the-shelf components. It can be easily rendered as eye-safe compared to pulsed or raster-based ranging systems. Finally, we demonstrate how to minimize the memory overhead from the sampling process. Instead of storing $m$ high-dimensional spectrograms (consisting of $m\times N$ numbers), we demonstrate how only $2m$ recorded numbers can be effectively used to recover high-dimensional depth-maps. 

This architecture's main limitation is in capturing quickly moving objects. The problem can be alleviated by moving to faster frame rates arising from fewer measurements. Our simulations assumed a 1 ms integration time per projection. When using a $20\%$ sampling ratio for a $128\times 128$ pixel resolution depth-map, only 3.3 sec of light-gathering time are needed when using the setup presented in Fig. \ref{fig:setup}. Perhaps the easiest way to reduce the acquisition time is to reduce the resolution. Reducing the resolution from $128\times 128$ pixels to $64\times 64$ pixels while maintaining the same active area on the DMD offers a factor of at least 4 timing improvement because the SNR will increase with larger super-pixel sizes. Given the short range of this LiDAR system and that some off-the-shelf DMD's can operate at 20 kHz, the 1 ms integration time can also be readily reduced by a factor of 10. Finally, the Python-based reconstruction algorithms take approximately 10 sec using a single thread on a 4.1 GHz Intel i7-4790K CPU, preventing real-time video. Moving the reconstruction algorithms from our serial implementation to a faster programming language, executed on a parallel architecture that takes advantage of multi-core CPUs or graphics cards, will significantly alleviate timing overhead from the reconstructions.

This chapter presents an architecture to build a compressive FMCW-LiDAR depth-mapping system. Simulations support our conclusion that this system will be robust to noise when operating in the field. While there is room for improvement within the experimental design and reconstruction algorithms, we present a viable architecture that simplifies the data-taking process and significantly reduces memory and computation overhead. These advances serve to reduce the overall expense of such a system while simultaneously increasing the resolution for real-time depth-mapping.   

\chapter{Experimental Demonstration of Quantum Data Locking}\label{ch5}

In this chapter, classical information is quantum-locked in quantum states for quantum-secure direct communication. 
Quantum data locking thrives on high-dimensional systems -- gaining better encryption rates with increasing dimensionality. An introduction to classical encryption and quantum data locking is given before diving into an experimental demonstration published in \cite{PhysRevA.94.022315}. Up to this point in time, technological limitations have prevented an experimental demonstration. The work presented here \cite{PhysRevA.94.022315} along with \cite{PhysRevA.94.020301} constitute the first experimental demonstrations of quantum data locking.



\section{Introduction}
\label{sec:intro}  

Quantum data locking (QDL) is a quantum effect whereby a message is first encoded into quantum states and then encrypted with a $k$-bit key. QDL states that the correlations between a message and the quantum-encoded ciphertext can decrease by more than $k$-bits without access to the $k$-bit key. This phenomenon is purely quantum with no classical analogue. 

To understand the implications of the locking effect, Sec. \ref{sec:ClassicalEncryption} introduces the classical mutual information and its role in encryption, particularly in the development of one-time pad encryption \cite{shannon1949communication}. Next, Sec. \ref{QDL} presents a generalized QDL protocol to securely communicate directly across a quantum channel. The generalized protocol stems from the results published in several works by Lloyd, Lupo, and Wilde \cite{lupo2015quantumArXiv,lupo2014robust,lupo2015pull}. Those works detail how to calculate the key consumption rate to secure a message against an eavesdropper. To describe the generalized locking protocol in a more intuitive manner,
an introductory example, first presented by DiVincenzo et al. \cite{divincenzo2004locking}, is given in Sec. \ref{sec:example}.
Finally, Sec. \ref{QDLexperiment} presents the experimental demonstration published in \cite{PhysRevA.94.022315}.

Two experimental demonstrations of QDL have been performed to date. The experiments were performed independently, yet at the same time. In one demonstration, a fiber-based experiment was performed by Liu et al. \cite{PhysRevA.94.020301} using the protocols proposed by DiVincenzo et al. \cite{divincenzo2004locking} and Fawzi et al. \cite{Fawzi2013}. This chapter presents details of the other demonstration -- an experiment that used free-space optics and a high-dimensional detector array \cite{PhysRevA.94.022315}. 
Specifically, we present the fundamentals behind how our system works with phase modultion and demonstrate how to account for additional key consumption to cover the redundancy found with forward error correction codes -- an essential criteria for direct secure communication.

\section{Background: Classical Encryption}
\label{sec:ClassicalEncryption}
Before diving into quantum data locking, it is beneficial to first present a brief overview of encryption from a classical-information-theory point of view. To determine how much information is passed between parties, information must first be quantized. This quantization is based on the probability of transmitting a particular alphabet. Given a discrete variable $x_i \in \{X\}$ where $X$ is any text and $x_i$ are the string of variables (possibly letters and numbers) that comprise the text, the probability that $x_i$ is present is represented as $p(x_i)$. One way of quantizing the amount of information present is through the Shannon entropy and is defined as
\begin{equation}
H(X) = -\sum\limits_{i=1}^n p(x_i) \log_2\big(p(x_i)\big)
\label{eq:entropy}
\end{equation}
for $i\in\{n\}$ and where the base 2 of the logarithm designates we are specifying the information in bits. $H(X)$ is a measure of the uncertainty about the possible text $X$. By design, Eq. \ref{eq:entropy} states that information is positive, is additive, and there exists the potential for more information when there is greater uncertainty in $X$ (i.e. $n\gg 1$ while $p(x_i)\rightarrow 1/n \,\, \forall \,\, i $). After all, information cannot be transferred unless uncertainty is present -- meaning everything would already be known about $X$ without uncertainty. 

For communication purposes, it is also necessary to quantify the correlations that exist between a transmitted text $X$ and a received message $Y$. This metric requires the use of a conditional probability -- knowing $x_i$ was sent while having received a variable $y_i \in Y$ -- and is represented by $p(x_i|y_j)$. The information shared by two parties is then quantized by the mutual information and can be defined as
\begin{equation}
I(X;Y) = H(X) - H(X|Y) = H(Y) - H(Y|X).
\label{eq:mutualinfo}
\end{equation}
where $H(X|Y)$ is a statement of the remaining uncertainty about $X$ once $Y$ is known. The conditional information $H(X|Y)$ is calculated using Eq. (\ref{eq:entropy}) but with $p(x_i|y_i)$ in place of $p(x_i)$. Because $I(X;Y)=I(Y;X)$, the mutual information is a statement about the available information that one variable yields about the another. Furthermore, the mutual information is additive and nonnegative, meaning the addition of $c$-bits about $Y$, where $c>0$, cannot increase the mutual information between $X$ and $Y$ by more than $c$-bits. This is an important point that will be emphasized later. 

In the case where the text is identical to the received message, $X = Y$ means $H(X|Y) = H(Y|X) = 0$. Thus, the mutual information between $X$ and $Y$ is the available information in the original text $I(X;Y) = H(X)$.

Now consider the simple case case where a unitary encryption key operator $\hat{U}\in\mathbb{R}^{N\times N}$ generates an encrypted message, a ciphertext, $\hat{U}X = Y$ whereby $X,Y\in\mathbb{R}^N$ and $\hat{U} \neq \mathds{1}$. The mutual information is a metric about how much information a ciphertext $Y$ reveals about the original text $X$. Having possession of the encryption key allows for the operation $X = \hat{U}^{-1}Y$ which reverts to the case previously discussed. Finding conditions on an encryption key $\hat{U}$ that minimizes $I(X;Y)$ will result in an even stronger encryption. 

Claude Shannon found the conditions for an encryption key $\hat{U}$ that, if kept secret, results in the elimination of the mutual information \cite{shannon1949communication}. If mutual information no longer exists between a text and its ciphertext, there exists no available correlations that can be used to break the encryption. Consequently, every possible decrypted message becomes equally likely. Such systems are referred to as \emph{information-theoretically secure}. Shannon's criteria for information-theoretically secure encryption states that $\hat{U}$ must be random, $\hat{U}$ must be at least as long as the text, and $\hat{U}$ can only be used once. This encryption, particularly with the use of modulo two addition, is often referred to as one-time pad or the Vernam cipher \cite{vernam1926cipher, bennett1992quantum}.

To implement a one-time pad effectively, a secret key must first be securely distributed between legitimate parties. The field of quantum key distribution (QKD) utilizes quantum uncertainty to effectively tackle this problem \cite{RevModPhys.81.1301}. By utilizing a quantum channel to transmit quantum states, correlations between quantum measurements are used to establish a secret key. However, it is important to note that the actual information used to establish a privately shared encryption key according to QKD is transmitted over a \emph{classical} channel. QDL is used for communication by transmitting information directly across the \emph{quantum} channel.

\section{Quantum Data Locking}
\label{QDL}
\subsection{Protocol}
\label{protocol}
Quantum data locking (QDL) is a quantum phenomenon that arises when the classical mutual information between a message and a quantum encoded ciphertext decreases by more than $k$ bits when the $k$-bit encryption key is unavailable. 
The result is a quantum encryption scheme that uses a short encryption key to encrypt a large amount of information and, under appropriate assumptions, maintains information theoretic security.  

Communications systems based on QDL are often referred to as quantum enigma machines \cite{GuhaPhysRevX,lloyd2013quantum,PhysRevA.92.062312}, alluding to the polyalphabetic ciphers used in the early 1900's. The original enigma machines also used a short encryption key and, as discussed in Sec. \ref{sec:ClassicalEncryption}, were not information theoretically secure. QDL stems from quantifying the classical correlations between a message and the outcome of a quantum measurement via the accessible information. Letting $Q$ represent an encrypted quantum state, an eavesdropper not having access to the key must attempt to maximize the classical mutual information between the original text $X$ and a result of a quantum measurement on $Q$, denoted as $\hat{M}(Q)$. More formally, the accessible information is represented as
\begin{equation}
I_{acc}(X;Q) = \max_{\hat{M}}I\left(X;\hat{M}\left(Q\right)\right)
\end{equation}
where $I(*)$ is the classical mutual information and the maximization is focused on obtaining the optimal quantum measurement. In general, the maximization is carried over a generalized class of quantum measurements known as positive operator valued measurements (POVM's) as POVM's are considered optimal measurements for quantum state discrimination \cite{barnett2009quantum,barnett2009quantumbook,wilde2011classical}. In fact, QDL is a direct consequence of the indistinguishability of non-orthogonal quantum states \cite{divincenzo2004locking}. While POVM's are optimal at state discrimination, they are not perfect and the resulting uncertainty is what enables the reduction in the accessible information. 

Using the formalism presented by Lupo and Lloyd \cite{lupo2015quantumArXiv}, a typical QDL protocol requires that two legitimate parties, presented here as Alice and Bob, first establish a secret encryption key. They then attempt to communicate an encrypted message while guaranteeing that an eavesdropper, Eve, cannot decrypt the message without access to the key. QDL requires that the $\log_2(M)$-bit message, which we assume is one of $M$ equiprobable messages, is first encoded onto a quantum system. Given an orthonormal set of quantum states $\left\{\ket{x}\right\}$ that can each store $\log_2(c)$ classical bits, the individual quantum states are designated as $\ket{x_c}$ for $c = 1,2,...,c$. Letting the $\log_2(M)$-bit message be divided into $n$ packets of $\log_2(c)$ bits, such that $\log_2(M) = n\log_2(c)$, $n$ quantum states are required to carry the entire message across a quantum channel. Alice first prepares the following quantum state
\begin{equation}
\ket{\psi} = \bigotimes\limits_{j=1}^n\ket{x_{j,c}}.
\end{equation}
Alice and Bob must already share a secret $\log_2(K)$-bit key that specifies one of $K$ possible unitary operations $\hat{U}^{(s)}$ for $s = 1,2,...,K$ that operate on $\ket{\psi}$. As such, the unitary operation takes the form
\begin{equation}
\hat{U}^{(s)} = \bigotimes\limits_{j=1}^n \hat{U}_j^{(s)}.
\end{equation}
Alice uses the key to select a unitary operator $\hat{U}^{(s)}$ and operates on her quantum system $\ket{\psi}$ to generate the encrypted quantum state 
\begin{equation}
\ket{\psi^{(s)}} = \hat{U}^{(s)}\ket{\psi} = \bigotimes\limits_{j=1}^n \hat{U}_j^{(s)}\ket{x_{j,c}}.
\end{equation} 
Alice then transmits her quantum state to Bob. Since Bob has the key, he can apply the inverse unitary $\hat{U}^{(s)^{-1}}$ and recover the maximum possible accessible information of $\log_2(M)$ bits by measuring in the original basis $\left\{\ket{x}\right\}$. Eve, not having access to the key, must instead make an optimal measurement to maximize her mutual information. The length of the key must be chosen such that
\begin{equation}
I_{acc}\left(X;Q\right) \ll I_{acc}\left(X;KQ\right)
\end{equation}
where the $KQ$ refers to the system that has the key and $Q$ is the system without the key. 

In addition to reducing the accessible information, there are different conditions, or capacities, under which the key length may be chosen \cite{GuhaPhysRevX} -- called the \emph{weak} and the \emph{strong} locking capacity. The strong locking capacity considers the possibility whereby Eve obtains a perfect copy of the encrypted quantum state, i.e. the signal was not altered by the channel environment. The weak-locking capacity presents the scenario where Eve must also consider the possibility that environmental noise altered the quantum state. The amount of information that can be locked in the weak locking capacity is greater than or equal to the strong locking capacity. 

Intuitively, securing a system according to a strong locking capacity also guarantees security according to a weak-locking capacity. 
The case of a noiseless channel was studied extensively by Lupo and Lloyd \cite{lupo2015quantumArXiv} and a key rate necessary to accomplish a strong locking capacity was derived such that
\begin{equation}
I_{acc}\left(X;Q\right) \leqslant \mathcal{O}\left(\epsilon \log_2 \left(d^n\right )\right)
\label{eq:IaccLimit}
\end{equation} 
where a $d$-dimensional quantum channel was used $n$ times and $\epsilon$ is a small constant that grows as a sublinear function of $n$. 
In the same work, key consumption rates were derived such that Eq. \ref{eq:IaccLimit} is satisfied provided the key length $K$ is of length
\begin{equation}
K \geqslant \max
\begin{cases} 2 \left(\frac{2d}{d+1}\right)^n\left(\frac{1}{\epsilon^2} \ln M +\frac{2}{\epsilon^3}\ln\frac{5}{\epsilon}\right) \\[1em]
\frac{d^n}{M}\frac{4 \ln 2 \, \ln d^n}{\epsilon^2}\end{cases}
\label{eq:KeySize}
\end{equation}
where $d$ is the dimension of the quantum channel, $\epsilon = 2^{-n^\alpha}$ for $0<\alpha < 1$, $M$ is the length of the message, and $n$ is the number of channel uses.
Efficient data locking occurs when both $d \approx M \gg 1$ and $K \ll M$.

Again, it should be emphasized that QDL results as a consequence of the indistinguishability of nonorthogonal quantum states. This relationship will, perhaps, be clearer after presenting the following introductory example of QDL -- first presented by DiVincenzo et al. \cite{divincenzo2004locking}.

\subsection{Introductory example}
\label{sec:example}
Consider the problem of discriminating single-photon polarizations in either one of two possible nonorthogonal polarization bases when using projective measurements. The problem presented uses the same protocol devised by Bennett and Brassard in 1984 (BB84) \cite{bennett2014quantum}. However, in this case, classical information is encoded onto the photon's polarization state such that the information is transmitted via the quantum channel. Within the problem, there exist two polarization bases such that one basis is designated as $+$ and consists of horizontal and vertical polarization states $\ket{H}$ and $\ket{V}$ while the other basis is rotated 45 degrees with respect to the $+$ basis, designated as $\times$, and consists of diagonal and antidiagonal polarization states $\ket{D}$ and $\ket{A}$ such that 
\begin{align}
\ket{D} &= \frac{1}{\sqrt{2}}\left(\ket{H}+\ket{V}\right)\\
\ket{A} &= \frac{1}{\sqrt{2}}\left(\ket{H}-\ket{V}\right) \\
\langle H | V \rangle &= \langle D | A \rangle = 0.
\end{align} 
The measurement apparatus used by all parties requires a polarizing beamsplitter oriented in either the $+$ or $\times$ basis to perform either the measurement $\hat{O}_+$ or $\hat{O}_\times$. These are Hermitian operators defined as 
\begin{align}
\hat{O}_{+} &= H\ket{H}\bra{H}+V\ket{V}\bra{V} \\
\hat{O}_{\times } &= D\ket{D}\bra{D}+A\ket{A}\bra{A}
\label{eq:operators}
\end{align} 
and returns an eigenvalue $H$, $V$, $D$, or $A$. Operator $\hat{O}_+$ ($\hat{O}_\times$) performs perfect state discrimination when discerning between states in the $+$ ($\times$) basis, but return either $D$ or $A$ ($H$ or $V$) with $50\%$ probability when operating on states in the $\times$ ($+$) bases, respectively. The protocol requires that a legitimate party, say Alice and Bob, communicate an encrypted $n$-bit string by transmitting $n$ photons (where 1 photon encodes 1 bit) while sharing a private 1-bit encryption key such that the full $n$-bit string of classical information remains unaccessible to an illegitimate eavesdropper (Eve) on average. In this instance, the 1-bit key specifies the basis to encode the string of $n$ quantum states (in either $+$ or $\times$).

To gain an intuitive understanding of the locking effect, first consider the scenario whereby Alice transmits two copies of the same $n$-bit message -- sending $n$ photons to Bob and $n$ photons to Eve. Without access to the key, Bob and Eve must guess an optimal basis and make their measurements. Due to the nature of the unknown quantum state, they cannot clone their state (according to the no-cloning theorem \cite{wootters1982single}) and are left with a string of 1's and 0's after their measurements with no recourse but to alter their measurement apparatus and try again due to the destructive nature of their initial measurement on the quantum state. Given the limit of an infinite number of trials, they will each guess the optimal basis correctly $50\%$ of the time. Hence, the accessible information that Alice shares with Bob and Eve is $n/2$ bits on average. 

Now consider the scenario whereby Alice gives Bob the 1-bit secret encryption key \emph{before} Bob measures, and consequently destroys, his quantum state. Now having 1-bit of additional information about the optimal basis for his quantum measurement, he can unlock an additional $n/2$ bits of accessible information on average. Hence, Alice and Bob share $n$ bits of accessible information while Alice and Eve share only $n/2$ bits of accessible information. One might say that 1 bit of side information (the key) unlocked a disproportionately large amount of accessible information. Alternatively, 1 bit of side information effectively locked a disproportionately large amount of accessible information. Both statements appear to contradict the additive nature of mutual information discussed in Sec. \ref{sec:ClassicalEncryption}. This is the locking effect and presents one of the most significant fundamental differences between classical and quantum information.          

\subsection{Discussion}

The example presented in Sec. \ref{sec:example} presents a means by which a BB84 QKD system may be altered to accomplish QDL. Many QKD systems can be altered to perform QDL. In fact, QDL has been studied as an alternative to QKD due to the theoretically higher key rates \cite{lupo2014PhysRevLett}. Of course, environmental losses and errors induced by realistic noisy channels are, perhaps, one of the greatest limitations. Because QDL has a fundamental reliance on the quantum channel, this problem cannot be circumvented easily. Additionally, many key rate derivations have assumed uniform probability distributions for the list of possible messages. Any deviation from these assumptions will require a higher key rate. Still, there exists the potential for high-throughput secure communication and many of the current technologies within QKD can be allocated to QDL systems. 

Given the similarities between QDL and QKD, a logical question might be What is the fundamental difference between QDL and QKD? The answer is in the security definitions. As previously stated, QDL relies on limiting the accessible information. QKD limits the mutual information between a ciphertext and the original message by limiting the Holevo information $\chi$, defined as 
\begin{equation}
\chi = S(\rho) - \sum\limits_i p_i S(\rho_i)
\end{equation} 
where $\{\rho_i\}$ is a set of quantum states occurring with probability $p_i$ (such that $\rho = \sum_i p_i \rho_i$) and $S(\rho)=-\text{Tr}[\rho\ln\rho]$ is the von Neumann entropy.
This limit is accomplished by ensuring that the quantum state Eve recovers is considered $\epsilon$-close, for small $\epsilon$, to a constant state when defining the distance according to the operator trace norm (for an operator $\hat{O}$, the trace norm is $\text{Tr}(\sqrt{\hat{O}^\dagger\hat{O}})$), regardless of what message was sent by Alice \cite{wilde2011classical,lupo2014PhysRevLett}. This is equivalent to stating that the message sent by Alice is to within a probability $\epsilon$ of being perfectly secure or $\epsilon$ close to being indistinguishable from a random state. Additionally, the Holevo information obeys the property of total proportionality \cite{divincenzo2004locking} which states that $c$ bits of side information cannot increase the Holevo information by more than $c$ bits. Consequently, a system secured by limiting the Holevo information prevents that system from utilizing QDL. 

QDL suffers from the fact that a small amount of side information can jeopardize the security of a QDL protocol because of the potential to unlock a disproportionately large amount of mutual information \cite{PhysRevLett.98.140502}. Thus, QDL cannot provide composable security without additional assumptions. Composable security asserts that if different security protocols are combined, the final system is also secure \cite{canetti2001foundations}. Because Eve has the potential to store her state for an indefinite amount of time with a quantum memory, she can wait until she gains a small amount of side information via alternate means before making her quantum measurements. 

The security of the accessible information depends on the existence of the quantum state. Any side information obtained by Eve can reveal how she could make a better quantum measurement and extract more information. If we assume that Eve must measure and destroy her quantum state shortly after receiving it, then QDL can provide composable security. This statement is equivalent to assuming there exists no perfect quantum memory -- a logical assumption given today's technology.   

\section{Quantum Enigma Machine}
\label{QDLexperiment}

Due to the challenges with generating quantum states on demand, transmitting them over imperfect quantum channels, and then performing measurements with high detection efficiencies, there have been only two published experimental demonstrations of QDL to date. An experimental demonstration by Liu et al. \cite{PhysRevA.94.020301} used single photons over fiber-based system to implement the introductory example given in Sec. \ref{sec:example} along with a loss tolerant scheme proposed by Fawzi et al. \cite{Fawzi2013}. The following sections focus on our work \cite{PhysRevA.94.022315}, which relies on free-space propagation of single photons, phase modulation with spatial light modulators (SLM's), and photon detection with an $8\!\times\! 8$ cryogenically cooled single-photon-counting detector array. With our system, we were able to encode 6-bits per photon and, after incorporating Reed-Solomon forward error correction codes, transmitted messages with success probabilities approaching 100\%. 

\subsection{Free space propagation and phase modulation}
\label{subsec:propagation}
Our experimental demonstration of QDL used SLMs to both encode and decode classical information via phase modulation of single photons. The approach relies on a fundamental principle of optical propagation through a lens according to Fourier optics \cite{goodman2005introduction}. Consider an input electric field $E(x,y)$ with transverse coordinates $x,y$, a wavelength $\lambda$, and wavenumber $k = 2\pi/\lambda$ placed a distance $d$ from a spherical thin lens with focal length $f$. Letting the electric field propagate through the lens to the focus, the output field $E(x_f,y_f)$, where $x_f$ and $y_f$ designate the transverse profile one focal length beyond the lens, can be expressed in the paraxial approximation as  
\begin{equation}
\label{eq:FourierTrans}
E(x_f,y_f) = \frac{i}{\lambda f}e^{-i\frac{k}{2f}\left(1-\frac{d}{f}\right)\left(x_f^2+y_f^2\right)}\iint\limits_{-\infty}^{\infty}E(x,y)e^{i\frac{2\pi}{\lambda f}\left(xx_f+yy_f\right)}dx \, dy
\end{equation}
Equation \ref{eq:FourierTrans} states that the electric field at the focal plane is proportional to a Fourier transform with a quadratic phase factor. Fortunately, we are only interested in the intensity of the output ($~|E(x_f,y_f)|^2$). 
Thus, the intensity profile at the focus of the lens is approximately the modulus squared of a scaled Fourier transform. Presenting the shift theorem associated with Fourier transforms and using shorthand notation for the Fourier transform as $\mathscr{F}(G(f_x,f_y))=g(x,y)$, we see
\begin{equation}
\mathscr{F}_{_{f_x\rightarrow x}^{f_y\rightarrow y}}\left(G\left(f_x,f_y\right)e^{-2\pi i\left(f_x a + f_y b\right)}\right) = g\left(x-a,y-b\right)
\end{equation}
which states that a phase shift on an optical beam will result in the shift in position at the focus of a lens.

Spatial light modulators were used to encode these phase shifts. While there exist a significant number of possible linear phases we could use to encode information, the detector array limits the quantum channel to a dimension of $d = 64$. The detector is an $8\!\times\! 8$ cryogenically cooled single-photon-detecting nanowire array developed at NIST \cite{allman2015near}. It is one of the largest detector arrays to date where each detector has a high detection efficiency with a low dark-count rate -- an essential characteristic for any detector used for quantum communication. We chose our encoding alphabet such that photons could be phase modulated, transmitted over the quantum channel, and then focused onto the detector array. With 64 different phase encodings, we could focus photons onto each of the 64 detectors. Given that each letter in the alphabet was equally likely ($1/64$), we were able to encode 6 bits per photon.

\subsection{Experiment}

\subsubsection{Setup}

\begin{figure}
\begin{center}
\begin{tabular}{c}
\includegraphics[width=\textwidth]{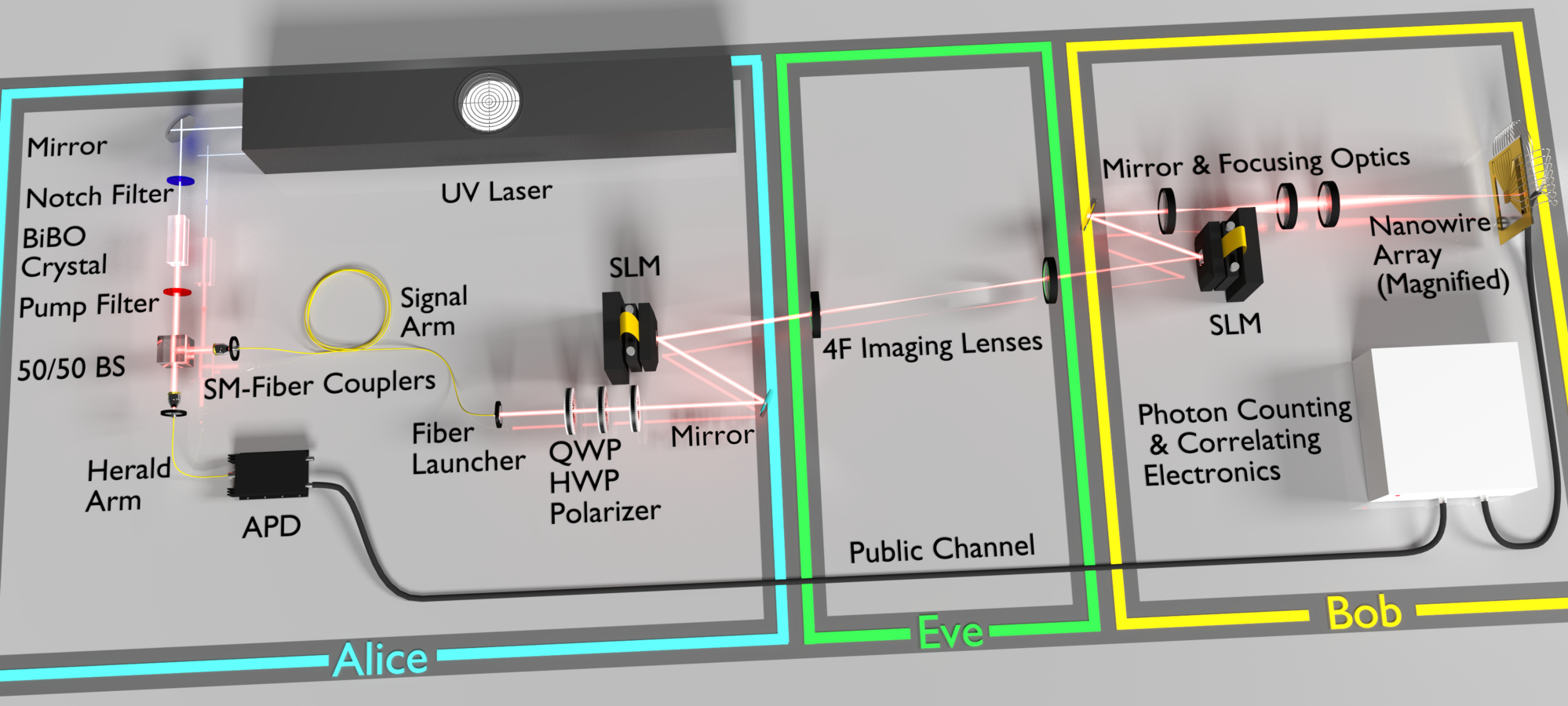}
\end{tabular}
\end{center}
\caption[Experimental diagram]{
\textbf{Experimental Diagram: } A frequency filtered ultraviolet laser at 325 nm pumps a bismuth triborate (BiBO) crystal and generates degenerate single photons, called signal and herald photons, at 650 nm by type-I spontaneous parametric down-conversion. The herald photon is detected by an avalanche photodiode (APD) and heralds the presence of a signal photon used to encode messages. Alice prepares the photon in the necessary polarization state using a quarter and half wave plate (QWP, HWP) along with a polarizer. She then phase modulates her photon with a spatial light modulator (SLM) and using imaging optics to transmit her photon to Bob's SLM. Bob uses the secret key to generate the necessary pattern to phase modulate the photon before focusing the photon onto an $8\!\times\! 8$ cryogenically-cooled single-photon-detecting nanowire array. 
\label{fig:diagram}
} 
\end{figure}

A simplified experimental diagram is presented in Fig. \ref{fig:diagram}. We used single photon states generated from type-I spontaneous parametric down-conversion using a nonlinear bismuth triborate (BiBO) crystal that is pumped by a 325 nm laser. A 50/50 beamsplitter was used to separate the resulting 650 nm photons, called signal and herald. The herald photon was detected by a Perkin Elmer avalanche photodiode (APD) and the resulting electrical signal was used to herald the arrival of the signal photon within the experiment. 

After preparing the photon in the necessary polarization state for use with the spatial light modulators, an SLM was used to encode both the linear phase and the scrambling operation at the same time. 64 different $512\!\times\!512$ pixel linear phase patterns were used to encode 6 bits of classical information to generate the state $\ket{x_c}$ for $c = 1,2,...,64$. A different phase scrambling unitary operator $U_j^{(s)}$ was applied for each of $j$ photons that comprise the entire message. The random encoding pattern was a $128\!\times\! 128$ superpixel binary phase mask composed of 0 and $\pi$ phase shifts that were sized to fill the SLM. Since the SLM has a resolution of $512\!\times \! 512$ pixels, each superpixel was composed of $4\!\times \! 4$ pixels. 

\begin{figure}
\begin{center}
\begin{tabular}{c}
\includegraphics[width=.95\textwidth]{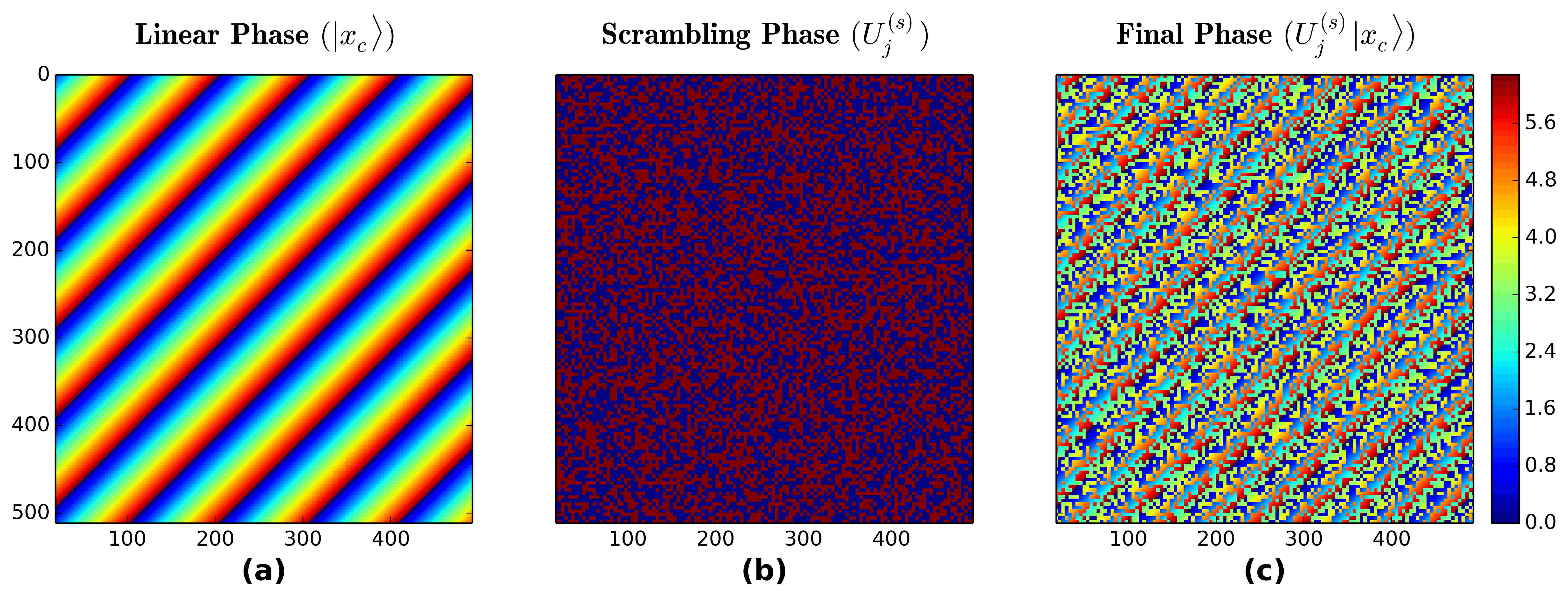}
\end{tabular}
\end{center}
\caption[Example phase encoding]{
\textbf{Phase Patterns: }\textbf{(a)} A $512\!\times\!512$ pixel linear phase pattern is used to encode 6 bits of classical information. \textbf{(b)} A scrambling phase operator, composed of random 0 or $\pi$ phase shifts, randomly generated with a seed specified by the secret key, is used to distribute the single photon over many possible optical modes by disrupting the phase front. It is composed of $128\! \times\! 128$ superpixels that together form a $512\!\times\!512$ pixel pattern. This is also the pattern that Bob will display on his SLM to undo the scrambling operation and flatten the single-photon wavefront leaving the linear phase specified by Alice. \textbf{(c)} Alice combines (a) and (b) to yield the pattern she displays on her SLM.   
\label{fig:phases}
} 
\end{figure}

The $n$ patterns were generated according to a seeded random number generator. The $n$ seeds were provided by specifying a line within a public code book. This code book contained $K$ lines and each line contained $n$ random numbers. As such, a $\log_2(K)$-bit secret key specified a line within the code book. The scrambling phase pattern was added to the linear phase pattern to generate the final phase operator that Alice used on her SLM to encrypt a single photon. An example pattern can be found in Fig. \ref{fig:phases}.

\subsubsection{QDL results}

\begin{figure}[h]
\centering
\includegraphics[width=.9\textwidth]{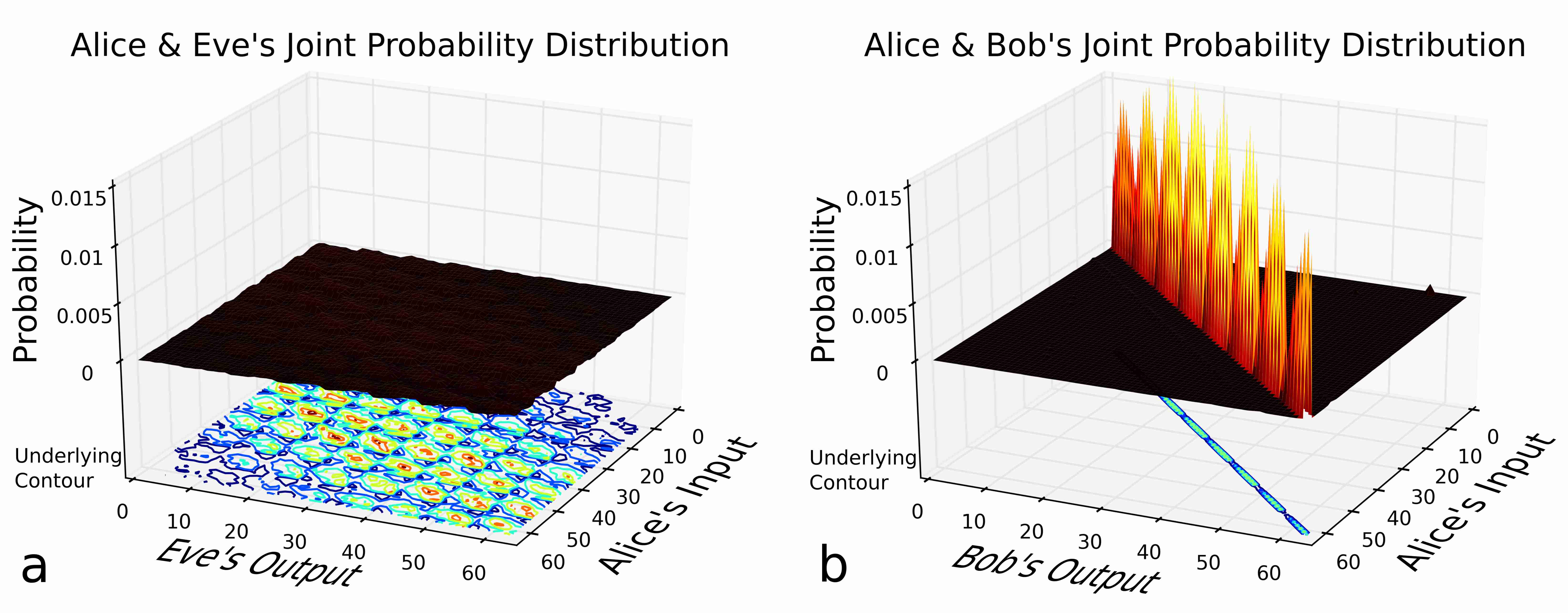}
\caption[Measured joint-probability distributions]{\textbf{Joint Probability Distributions: }The joint probability distributions presented are derived 
from experimental data where Alice scans through 64 messages while locking the information with $128\!\times\! 128$ pixel binary phase masks. \textbf{a}, Eve randomly guesses at binary phase masks in hopes of unlocking the information while being allowed, in a worst-case scenario, to use Bob's properly aligned detector. \textbf{b}, Bob unlocks the messages with binary phase masks prescribed according to a secret key. The distribution with the highest mutual information is a normalized identity matrix with diagonal elements equal to $1/64 \approx .016$. The comb structure seen in (b) is due to a gradient of error rates across the detector array, possibly due to a slight misalignment of a lenslet array placed above the detector array. In summary, Alice and Eve's distribution is highly uncorrelated while Alice and Bob's distribution is highly correlated attesting the effectiveness of this locking method.}
\label{fig:jpd} 
\end{figure}

The effectiveness of our phase patterns were verified by allowing Alice to encrypt each of the 64 possible settings over 600 times while allowing Eve, who did not have access to the secret key, the use of Bob's SLM and detector. We then repeated the experiment with Bob, who had access to the secret key. From the data, probability distributions for each test are presented in Fig. \ref{fig:jpd} where one axis represents Alice's input and the other axis represents the output. Note that the distributions are normalized such that all elements sum to one. Figure \ref{fig:jpd} shows that Alice's messages and the results of Bob's measurements are highly correlated, being a normalized identity matrix with a maximum value of $1/64 \approx .016$. Alternatively, Alice's inputs and Eve's measurement outcomes are highly uncorrelated, corresponding to very little mutual information. 

\begin{figure}[h]
\centering
\includegraphics[width=.4\textwidth]{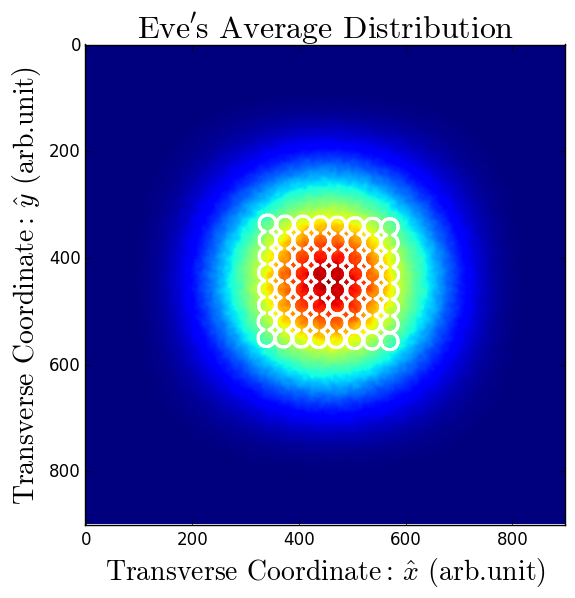}
\caption[Eve's simulated average distribution]{\textbf{Eve's Simulated Average Distribution: }Using experimental parameters, such as the beam width, the SLM pixel size, and the actual phase encodings, we were able to calculate the detector array centers (outline in white) and the result of Eve's average distribution within the plane of the detector after Eve's unsuccessful attempts to guess the key. Notice that there exist a significant region in which the photon is not likely to hit the detector.}
\label{fig:evedistrib} 
\end{figure}

In addition to estimating the mutual information, a computer simulation using experimental parameters was done to test the spread of Eve's average distribution in a plane of her $8\!\times\! 8$ detector. The linear phases used to encode messages in the experiment, the measured beam radius, and the SLM pixel sizes were used to calculate the detector's nanowire centers and Eve's distribution relative the nanowire centers. Eve's average distribution can be seen in Fig. \ref{fig:evedistrib}. We see that a significant portion of Eve's distribution does not allow for a photon to strike the detector.     

\subsubsection{Key rates}

To determine the necessary key requirements, the system is easily mapped to the theory presented in Sec. \ref{protocol}. As already mentioned, the individual quantum states transmitted over the channel take the form of phase-encoded photons $\ket{x_c}$ for $c = 1,2,...,64$ meaning $M$ = 64 and $\log_2(M) = 6$. Within \cite{PhysRevA.94.022315}, we discuss two arguments for the dimension $d$ of our quantum channel. The main results presented here limit the channel dimension to the same dimension as our detector $d = M = 64$. 
To arrive at a sufficient key rate, we need to decide how many channel uses $n$ will be required. Because of both the message encoding and the limitation from forward error correction, it is easier to first decide on the key rate and then pick the number of channel uses. 

We first consider what we would like to encode. QDL has the benefit that a short encryption key can encode a long message. Thus, there exist room to encrypt both new secret keys and messages. In this manner, a short secret key needs to only be used once before QDL is used for secret key generation. However, the success of this protocol necessitates that forward error correction codes be used to fix potential errors. Otherwise, future messages cannot be decoded with error-riddled keys. We used various Reed-Solomon error-correction codes because of their robustness against errors occurring in bursts \cite{reedsolomon1960polynomial}. 

\begin{figure}
\begin{center}
\begin{tabular}{c}
\includegraphics[width=.5\textwidth]{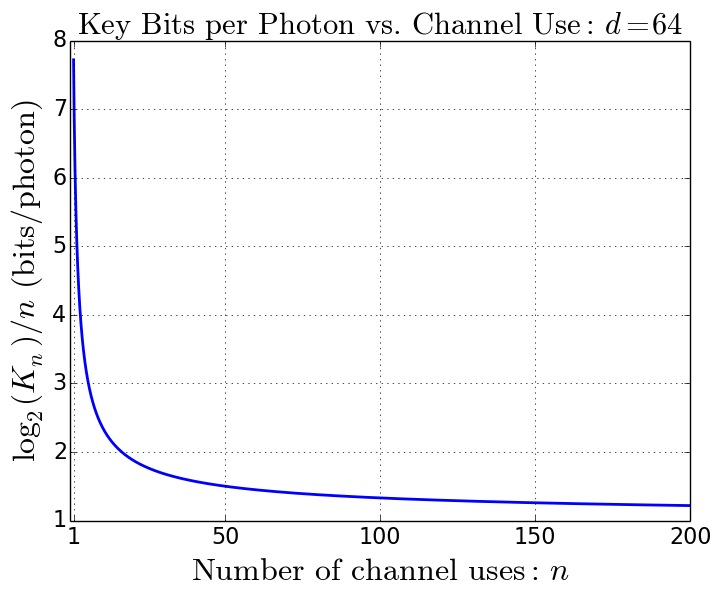}
\end{tabular}
\end{center}
\caption[Key consumption rate per photon]{
\textbf{Key Consumption Rate per Photon: } For a $d = 64$ dimensional channel and no error correction, Eq. \ref{eq:KeySize} states that we need approximately 1.5 bits per photon for $n = 63$ channel uses (to transmit 378 bits) in order to bound the accessible information to be no more than $\mathcal{O}\left(1.5\right)$ bits. 
\label{fig:keyrate}
} 
\end{figure}

Reed-Solomon codes perform error correction on $s$-bit symbols. Thus, multiple bit errors within a single symbol are treated as a single error. Reed-Solomon codes also require that data be transmitted in packets of $2^s-1$ symbols. Since we encode 6-bits per photon, we naturally consider a 6-bit alphabet. Each symbol is now represented by a single photon and the most efficient Reed-Solomon protocol requires that we transmit in packets of 63 symbols (photons). For this reason, we chose the number of channel uses to be a multiple of 63 photons. In the results presented here, we consider $n = 63$ uses of a $d = 64$ dimension channel. Using these experimental parameters, we plot Eq. \ref{eq:KeySize} in Fig. \ref{fig:keyrate} to ascertain our key rate. Figure \ref{fig:keyrate} suggests that we need approximately 1.5 bits per photon to limit the accessible information of the transmitted 378 bits to be no greater than $\mathcal{O}\left(6n 2^{-\sqrt{n}}\right) = \mathcal{O}(1.5)$ bits.

\subsubsection{Key rates with forward error correction}

\begin{figure}[h]
\centering
\includegraphics[width=.7\textwidth]{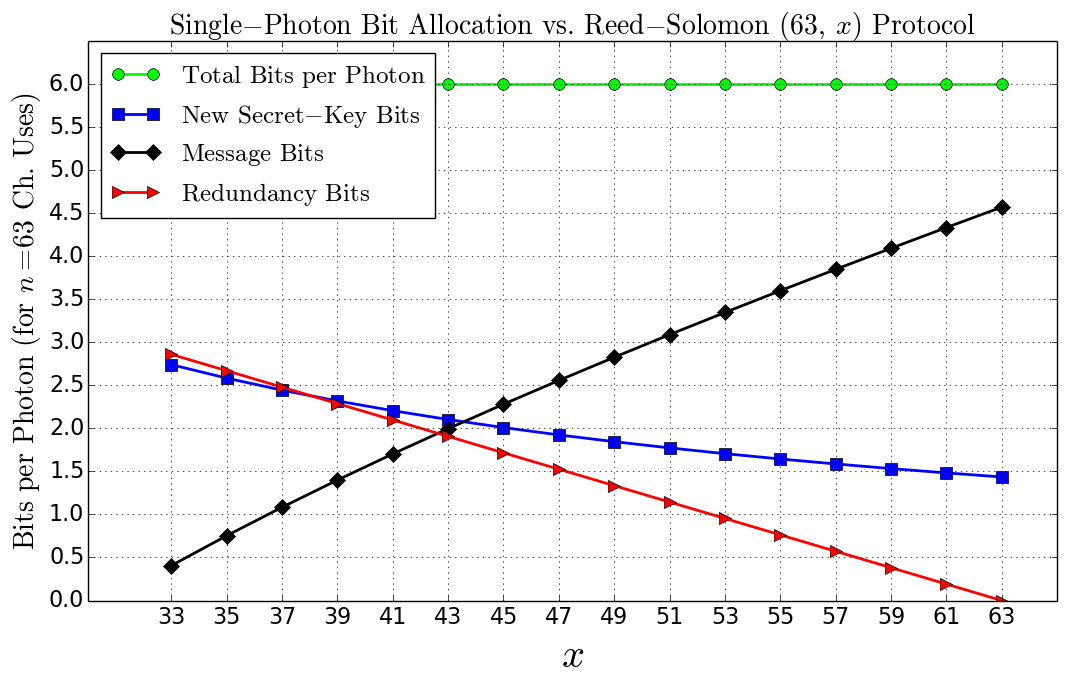}
\caption[Single-photon bit allocation -- message, new secret-key, and forward error correction]{\textbf{Single-Photon Bit Allocation (for $n = 63$ channel uses): } The plot above lists how many secret-key bits (blue squares), FEC redundancy bits (red triangles), and message bits (black diamonds) are encoded within a single 6-bit photon as a function of Reed-Solomon (63,\,$x$) codes. The new secret-key plot shows the necessary allocation of bits to replenish the consumed key. 
Any FEC redundancy requires the consumption of more secret key. Messages bits may be allocated to either additional key or secret messages for direct communication.} 
\label{fig:bit} 
\end{figure}

Fig. \ref{fig:keyrate} is only useful in the scenario where no forward error correction is used. Forward error correction adds redundancy into the message -- redundancy that may help Eve increase her accessible information. Fortunately, this redundancy can be covered by the use of more secret key. We used varying levels of Reed-Solomon error correction codes, designated as a Reed-Solomon (63,\,$x$) code, where $x<63$ is the number of symbols in the original message. The remaining $63-x$ symbols are allocated to redundancy. A Reed-Solomon (63,\,$x$) code can correct up to $(63-x)/2$ symbol errors. Thus, the fractional redundancy is $(63-x)/x$. This redundancy is covered by making the secret key $1+(63-x)/x$ times larger. The bit allocation per photon, for $n = 63$ uses of a $d=64$ dimensional channel is presented in Fig. \ref{fig:bit}.

Figure \ref{fig:bit} takes into account the allocation of messages, redundancy, and new secret key needed to refresh the secret keys. One can immediately see within Fig. \ref{fig:bit} that the available capacity for message bits decreases with decreasing $x$, corresponding to increasing redundancy, as both the number of redundancy bits and key bits increase.

\subsubsection{QDL with forward error correction results}

We carried out a proof-of-principle experimental demonstration of QDL utilizing forward error correction where we transmitted 420 packets of 63 symbols ($\approx$ 19 kB) containing messages and new secret keys. Ideally, a single photon source should emit a single photon on demand. This protocol requires that a single SLM pattern be applied to a single photon. However, we were forced to rely on the random generation of photon pairs from SPDC. In addition, there were several sources of loss. Because of these losses and uncertainties, we were unable to guarantee that a single photon was transmitted through the channel for a given SLM patten. The result is that Eve may have access to more than one photon per channel use. 

The experiment performed was meant as a proof-of-principle demonstration and we modified our detection scheme to account for a more accurate simulation of what a truly secure demonstration would entail. This included only utilizing first detection events within our detector for a given pattern on Alice and Bob's respective SLM. Even under these circumstances, Alice and Bob were able to communicate with a success rate approaching $100\%$ for different Reed-Solomon codes. 

\begin{figure}[h] 
\centering
\includegraphics[width=.7\textwidth]{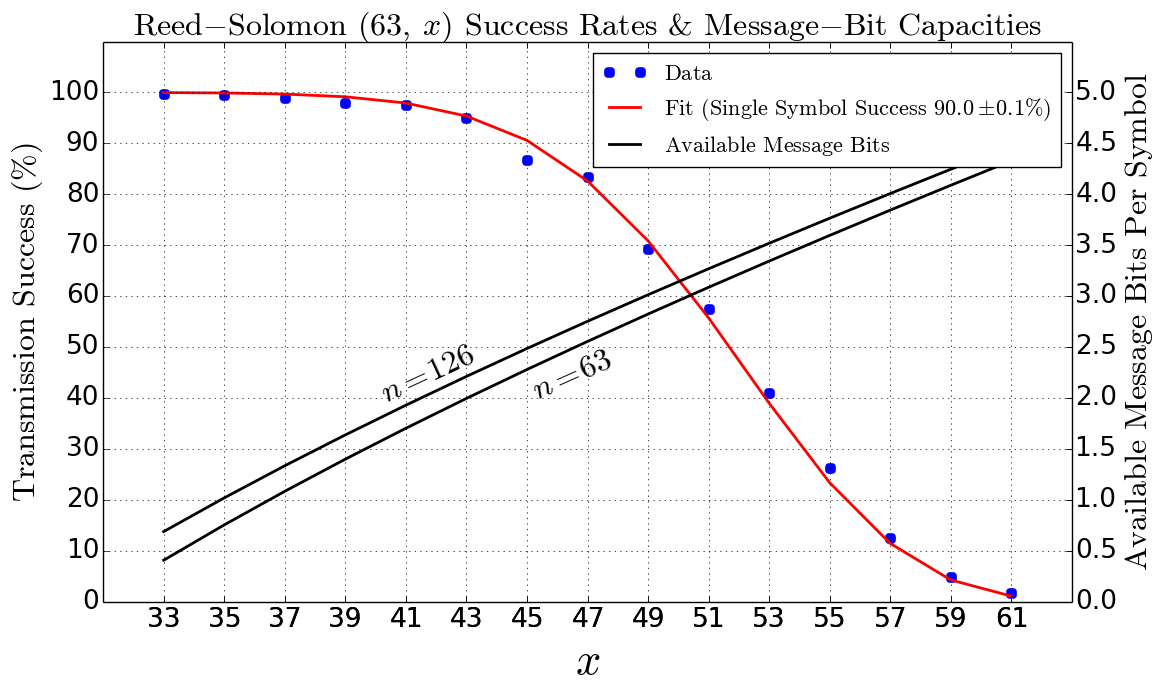}
\caption[Reed-Solomon forward error correction success rates]{\textbf{Reed-Solomon Error Correction Success Rates: } The success rates for a Reed-Solomon (63,\,$x$) code are presented, where $x$ is the number of 6-bit symbols containing the new key and message and 63 is the total number of packet symbols after the addition of $63-x$ redundancy bits. This data was obtained after transmitting packets of 63 photons 420 times. The plot also depicts the available capacity for message bits per photon as a function of $x$ and the number of channel uses $n$.} 
\label{fig:FEC} 
\end{figure}

The results for varying Reed-Solomon (63,\,$x$) codes are presented in Fig. $\ref{fig:FEC}$ where we also plot the available message capacity per photon when considering $n = 63$ and $n=126$ uses of the quantum channel. For example, we present a success rate of $99.5\%$ for a Reed-Solomon (63,35) code where each photon was divided into 1 bit of message, 2.3 bits of secret key and 2.7 bits of redundancy. Hence 2.3 bits of secret key per photon locked the entire 6-bit photon. While the bit-error ratio from our data is larger than the standard tolerance within most telecommunication systems ($<10^{-6}$ bit errors per bits transmitted), this is a proof-of-principle demonstration. The development of future technologies may lead to lower-loss quantum channels and higher efficiency QDL protocols to become a reliable means of direct secure quantum communication.

\section{Conclusion}
Alongside a fiber-based demonstration performed by Liu et al. \cite{PhysRevA.94.020301}, we present one of the first experimental demonstrations of QDL. In addition, we demonstrate how to incorporate forward error correction and calculate the necessary key rate needed to limit an eavesdropper's accessible information to be no more than $\mathcal{O}\left(2^{-\sqrt{n}}\log\left(d^n\right)\right)$ bits using the formalism found in Lupo and Lloyd's work \cite{lupo2015quantumArXiv}. For $n=63$ uses of a $d = 64$ dimension quantum channel, we transmitted, per photon, 1 bit of message, 2.3 bits of secret key, and 2.7 bits of redundancy with a success rate of $99.5\%$ when using a key of only 2.3 bits to secure the 6-bit photon. Having transmitted 420 packets of 63 photons, this corresponds to having transmitting 25.8 kB of message, 59.4 kB of new secret key, and 69.8 kB of forward error correction while limiting the accessible information to $\mathcal{O}(0.6)$ kB. While our experimental demonstration suffers from the use of randomly generated single photon states and losses from free-space optical components, the first-event detection success rates found in Fig. \ref{fig:FEC} demonstrate the potential applicability for QDL systems if the technological limitations associated with communicating over a quantum channel can be overcome.

\chapter{Summary and Concluding Remarks}

In this dissertation, compressive sensing and quantum data locking are deployed to solve fundamental problems in high-dimensional computational-imaging, entanglement characterization, and efficient large-alphabet quantum-secure communication.

In terms of introductory material, chapter 1 showed that the results of compressive sensing can be trusted if enough random projections are obtained. Chapter 2 shows how position-momentum entangled photons may be generated in a laboratory setting. Additionally, a discussion of entanglement characterizations is presented to emphasize the need for the density matrix.

Due to the difficulty of obtaining a high-dimensional density matrix, chapter 3 presented an alternative means of witnessing high-dimensional entanglement based solely on observing high-dimensional correlations. The poor scaling of the joint-space sensing matrix made sampling at adequate resolutions a difficult task. Raster-based and even unstructured random sensing matrices used to sample each subspace quickly results in the inability to either measure or reconstruct the joint-space correlations. This problem is laid to rest by using randomly permuted Sylvester-Hadamard matrices -- with the reconstruction algorithms relying on efficient fast-Hadamard transforms, as shown in the appendix.

Chapter 4 addresses the difficult task of efficiently obtaining high-resolution depth-images from one of the most accurate ranging systems in existence -- a frequency-modulated continuous-wave (FMCW) LiDAR. Previous compressive imaging depth-mapping systems relied on pulsed LiDAR. FMCW-LiDAR is less expensive while also having the capability of obtaining higher precision. Compressive sensing is presented as an efficient means of quickly obtaining precise high-resolution depth-maps with affordable equipment.

Finally, chapter 5 presents a unique solution for large-alphabet quantum-secure direct communication. A single photon's transverse phase is modulated by a high-dimensional spatial light modulator while an $8\times 8$ single-photon-detector array allows for the transmission of 6-bits per photon across a quantum channel. Under proper assumptions quantum data locking stands as an efficient means of information-theoretic-secure communication. While the work presented in chapter 5 is a proof of principle demonstration, it states that the technology is becoming readily available to work in conjunction with quantum key distribution systems. It also shows that the key to more efficient locking protocols is in even higher-dimensional systems.

By taking advantage of efficient measurement schemes from the signal processing community, ideas from the quantum cryptography community, and new measurement devices and techniques from the experimental physics community, interesting solutions to difficult problems are being readily obtained. Thus, it is reasonable to predict that a growing class of noteworthy problems will be solved through multi-disciplinary means. To conclude on this note, the reader is encouraged to pursue research interests into unfamiliar fields with the hope of applying new skills to short-lived problems.

\chapter{Appendix: Constrained optimization via the alternating direction method of multipliers}\label{appendix}

\section{Introduction}

This section will focus on particularly robust class of $L^1$ minimization algorithms, namely those solved by augmented Lagrangian methods. Note that this is not the same as the method of Lagrange multipliers often encountered in constrained optimization, yet there are similarities. First, the motivation behind augmented Lagrangians is introduced and naturally leads to a discusson of variable splitting. ADMM is then applied to the LASSO problem for different conditions placed on the sensing matrices and sparse-basis transform operators.

\subsection{Augmented Lagrangian Methods}

To motivate the use of an augmented Lagrangian, we will show how they evolved from penalty methods used for constrained optimization \cite{fiacco1990nonlinear}. These minimization problems take constrained optimization problems, such as 
\begin{align*}
&\min\limits_{\vect{x}\in\mathbb{R}^N}f(\vect{x}) \\
&\mathrm{subject\,\, to}\,\,\,c(\vect{x})=0,
\end{align*}
where $f:\mathbb{R}^n\rightarrow\mathbb{R}$ is convex and $c(\vect{x})=0$ is a nonlinear constraint. Penalty methods convert these problems into unconstrained optimization problems such as
\begin{align*}
P(\vect{x},\rho) &= f(\vect{x})+\frac{\rho}{2}\norm{c(\vect{x})}_2^2 \\
\vect{x}(\rho) &= \arg\min\limits_{\vect{x}\in\mathbb{R}^n}P(\vect{x},\rho),
\end{align*}
where $\rho$ is a positive scalar variable and the optimal solution $\hat{\vect{x}}$ occurs in the limit
\begin{equation*}
\lim\limits_{\rho\rightarrow\infty}\vect{x}(\rho) = \hat{\vect{x}}.
\end{equation*}
The optimization is often accomplished by first minimizing $P(\vect{x},\rho)$ for a small $\rho$, and then using the solution as a seed for the next iteration, but with a larger $\rho$. The process is repeated while letting $\rho\rightarrow\infty$, which can lead to instability. The idea behind the augmented Lagrangian is to first relax the constraint using a Lagrange multiplier $\vect{\lambda}$ in the following problem
\begin{align*}
&\min\limits_{\vect{x}\in\mathbb{R}^N}f(\vect{x}) \\
&\mathrm{subject\,\, to}\,\,\,c(\vect{x})-\frac{1}{\rho}\vect{\lambda}=0.
\end{align*}
The Lagrange multiplier allows the problem to converge without taking $\rho$ to infinity. Repeating the process results in the following equations:
\begin{align*}
\vect{x}(\rho) &= \arg\min\limits_{\vect{x}\in\mathbb{R}^n} f(\vect{x})+\frac{\rho}{2}\norm{c(\vect{x})-\frac{1}{\rho}\vect{\lambda}}_2^2 \\
&= \arg\min\limits_{\vect{x}\in\mathbb{R}^n} f(\vect{x})+\frac{\rho}{2}\norm{c(\vect{x})}_2^2 - \mean{c(\vect{x})|\vect{\lambda}} \\
&= \arg\min\limits_{\vect{x}\in\mathbb{R}^n} \mathcal{L}_\rho(\vect{x},\vect{\lambda}),
\end{align*}
where $\mathcal{L}_\rho(\vect{x},\vect{\lambda})$ is the augmented Lagrangian to be minimized.

\subsection{Alternating Direction Method of Multipliers}
The difficulty with augmented Lagrangian methods is that they require the simultaneous minimization of multiple variables. This section introduces the concept of \emph{variable splitting}. Applying variable splitting to an augmented Lagrangian splits the minimization problem into several variables, each of which can be minimized separately, while also forcing the split variables to converge to the same value with each iteration. The method is known as the
alternating direction method of multipliers (ADMM) \cite{boyd2011distributed} and was designed as a robust means of minimizing an augmented Lagrangian associated with the problem
\begin{align}
\mathrm{minimize} \,\,\, &f(\vect{x}) \nonumber \\
\mathrm{subject\, to} \,\,\, &\vect{A\,x} = \vect{b},
\label{prob:ConvOpt}
\end{align}
with the variables $\vect{x}\in\mathbb{R}^N$, 
$\vect{A}\in\mathbb{R}^{M \times N}$, and $f:\mathbb{R}^N\rightarrow \mathbb{R}$ is convex.
The Lagrangian for problem (\ref{prob:ConvOpt}) is
\begin{equation}
\mathcal{L}(\vect{x},\vect{\lambda}) = f(\vect{x}) + \langle \vect{\lambda} | \vect{A\,x} - \vect{b}\rangle,
\label{eq:Lagrangian}
\end{equation}
where $\langle \vect{\lambda} | \vect{\xi} \rangle$ represents the inner product between the Lagrange multiplier $\vect{\lambda} \in \mathbb{R}^M$ and $\vect{\xi} \in \mathbb{R}^M$.

The augmented Lagrangian from problem (\ref{prob:ConvOpt}) is
\begin{equation}
\mathcal{L}_\rho(\vect{x},\vect{\lambda}) = f(\vect{x}) + \langle \vect{\lambda} | \vect{A\,x} - \vect{b} \rangle + \frac{\rho}{2}\parallel\! \vect{A\,x}-\vect{b} \!\parallel_2^2,
\label{eq:AugLagrangian} 
\end{equation}
where $\rho > 0$ is a penalty parameter.
Within Eq. (\ref{eq:AugLagrangian}), the $L^2$-norm is a least-squares condition. Equation (\ref{eq:AugLagrangian}) is the Lagrangian (unaugmented) to the equivalent problem of
\begin{align}
\mathrm{minimize} \,\,\, &f(\vect{x}) + \frac{\rho}{2}\parallel\! \vect{A\,x} - \vect{b}\!\parallel_2^2 \nonumber \\
\mathrm{subject\, to} \,\,\, &\vect{A\,x} = \vect{b},
\label{prob:AugConvOpt}
\end{align}
because the optimal $\vect{x}$ results in adding zero to the objective. It has been shown that problem presented in Eq. (\ref{prob:AugConvOpt}) can be minimized under less stringent conditions (such as nonsmooth convex functions) compared to the original problem (\ref{prob:ConvOpt}) \cite{eckstein2012augmented}. The steps necessary to minimize $\mathcal{L}_\rho(\vect{x},\vect{\lambda})$ at iteration $k$ take the form
\begin{align}
\vect{x}^{k+1} &\in \text{arg}\,\min\limits_{\vect{x}}\, \mathcal{L}_p(\vect{x},\vect{\lambda}^k) \label{eq:xmin}\\
\vect{\lambda}^{k+1} &= \vect{\lambda}^k + \rho\left(\vect{A\,x}^{k+1}-\vect{b}\right) .
\end{align}
Depending on $f(\vect{x})$, the ease of optimizing Eq. (\ref{eq:AugLagrangian}) can vary drastically. Optimization of Eq. (\ref{eq:AugLagrangian}) is normally accomplished via calculating the gradient. The augmented Lagrangian form is more robust since we can allow $f$ to be nonsmooth and relax the strict convexity (having a second derivative that is greater than zero) requirement. Yet, this robustness results in an expression that cannot be minimized easily. ADMM was introduced to make the minimization of Eq. (\ref{eq:AugLagrangian}) easier by factoring the problem and then splitting it into multiple minimization problems, each of which are easier to minimize individually. The ADMM solves problems of the form
\begin{align}
\mathrm{minimize} \,\,\, &f(\vect{x}) + g(\vect{z}) \nonumber \\
\mathrm{subject\, to} \,\,\, &\vect{A\,x} + \vect{B\,z} = \vect{c},
\label{prob:AugConvOpt}
\end{align}
and has an augmented Lagrangian
\begin{equation}
\mathcal{L}_\rho(\vect{x},\vect{z},\vect{\lambda}) = f(\vect{x}) + g(\vect{z}) + \langle \vect{\lambda} | \vect{A\,x} + \vect{B\,z} - \vect{c} \rangle + \frac{\rho}{2}\parallel\! \vect{Ax + Bz - c}\! \parallel_2^2.
\label{eq:ADMM_Lagrangian}
\end{equation}
The algorithm consists of the iterations
\begin{align}
\vect{x}^{k+1} &\in \text{arg}\,\min\limits_{\vect{x}}\, \mathcal{L}_p(\vect{x},\vect{z}^k,\vect{\lambda}^k) \label{eq:xmin}\\
\vect{z}^{k+1} &\in \text{arg}\,\min\limits_{\vect{z}}\, \mathcal{L}_p(\vect{x}^{k+1},\vect{z},\vect{\lambda}^k) \label{eq:xmin}\\
\vect{\lambda}^{k+1} &= \vect{\lambda}^k + \rho\left(\vect{A\,x}^{k+1}+\vect{B\,z}^{k+1}- \vect{c}\right).
\end{align}
In contrast to the alternating direction method of multipliers for Eq. (\ref{eq:ADMM_Lagrangian}), the normal method of multipliers for Eq. (\ref{eq:ADMM_Lagrangian}) requires the iterations
\begin{align}
(\vect{x}^{k+1},\vect{z}^{k+1}) &\in \text{arg}\,\min\limits_{\vect{x,z}}\, \mathcal{L}_p(\vect{x},\vect{z}^k,\vect{\lambda}^k) \label{eq:xmin}\\
\vect{\lambda}^{k+1} &= \vect{\lambda}^k + \rho\left(\vect{A\,x}^{k+1}+\vect{B\,z}^{k+1}- \vect{c}\right),
\end{align}
where the more difficult minimization is taken jointly over variables $\vect{x}$ and $\vect{z}$ simultaneously.

ADMM is a convex optimization algorithm with a rich history. Similar ideas began in the mid-1950's but the algorithm was formally introduced in the 1970's \cite{boyd2011distributed}. The method's robustness and simplicity have led to its application and popularity in multiple fields including signal processing, engineering, and physics. It should be noted that extremely high accuracy is not normally obtained via the ADMM. It has been shown that the algorithm exhibits a slow tail convergence beyond the first few iterations \cite{eckstein2012augmented}. Yet, this behavior is suitable to large scale problems in that it approaches an approximate solution quickly.

\subsection{ADMM-LASSO}

Drawing from \cite{eckstein2012augmented,boyd2011distributed,yang2011alternating}, ADMM is applied to the least absolute shrinkage and selection operator (LASSO) mimiziation problem. This problem is, perhaps, more commonly known as basis pursuit denoising when $m < n$ for a sensing matrix $\vect{A} \in \mathbb{R}^{m\times n}$ \cite{donoho2006compressed}. We will adopt a more general framework and let the sensing matrix, signal, and sparse basis transform $\Psi$ be complex valued; i.e. $\vect{A}\in\mathbb{C}^{m\times n}$, $\vect{x}\in\mathbb{C}^{n}$, and $\Psi\in\mathbb{C}^{n'\times n}$. Note that the sparse basis transform $\vect{\Psi}$ is not necessarily unitary. Additionally, $n'$ is not necessaritly equal to $n$. From the measurement vector $\vect{b} = \vect{Ax}$, The LASSO problem takes the form
\begin{equation}
\min_{\hat{\vect{x}}\in\mathbb{C}^n} \frac{1}{2}\parallel\!\vect{A}\hat{\vect{x}} - \vect{b}\! \parallel^2_2 + \, \beta \parallel\! \vect{\Psi}\hat{\vect{x}} \! \parallel_1.
\label{prob:LASSO}
\end{equation}
In ADMM form, LASSO takes the form
\begin{align}
\mathrm{minimize} \,\,\, &f(\hat{\vect{x}}) + g(\hat{\vect{z}}) \nonumber \\
\mathrm{subject\, to} \,\,\, & \Psi\hat{\vect{x}} - \hat{\vect{z}} = 0,
\label{prob:ADMM_LASSO}
\end{align}
where $f(\hat{\vect{x}}) = (1/2)\parallel \!\vect{A}\hat{\vect{x}} - \vect{b}\!\parallel_2^2$ and $g(\hat{\vect{z}}) = \beta\parallel\! \hat{\vect{z}}\! \parallel_1$.

Problem $\ref{prob:ADMM_LASSO}$ has an augmented Lagrangian that takes the form
\begin{equation}
\mathcal{L}_\rho(\hat{\vect{x}},\hat{\vect{z}},\vect{\lambda}) = f(\hat{\vect{x}}) + g(\hat{\vect{z}}) + \langle \vect{\lambda} | \Psi\hat{\vect{x}} - \hat{\vect{z}} \rangle + \frac{\rho}{2}\parallel \Psi\hat{\vect{x}} - \hat{\vect{z}} \!\parallel_2^2.
\label{eq:ADMM_Lasso_Lagrangian}
\end{equation}
By solving for $\hat{\vect{x}}$ within $\partial\mathcal{L}_\rho(\hat{\vect{x}},\hat{\vect{z}},\vect{\lambda})/\partial \hat{\vect{x}}= 0$ and solving for $\partial\mathcal{L}_\rho(\hat{\vect{x}},\hat{\vect{z}},\vect{\lambda})/\partial \hat{\vect{z}}= 0$, we find the necessary iterations to minimize Eq. (\ref{eq:ADMM_Lasso_Lagrangian}) as  
\begin{align}
\hat{\vect{x}}^{k+1} &= \left(\vect{A}^\dagger \vect{A}+\rho \Psi^\dagger\Psi\right)^{-1}\left(\vect{A}^\dagger \vect{b}+\rho\Psi^{\dagger}\left(\hat{\vect{z}}^k-\vect{\lambda}^k/\rho\right)\right) \label{eq:MinX} \\
\hat{\vect{z}}^{k+1} &= S_{\beta/\rho}\left(\hat{\Psi\vect{x}}^{k+1} + \vect{\lambda}^k/\rho\right) \label{eq:SoftThresh}\\
\vect{\lambda}^{k+1} &= \vect{\lambda}^k + \rho\left(\Psi\hat{\vect{x}}^{k+1} - \hat{\vect{z}}^{k+1}\right) \label{eq:LambUpdata},
\end{align}
where $\dagger$ is the conjugate transpose operation. Within Eq. (\ref{eq:SoftThresh}), $S_{\beta/\rho}$ is an element-wise soft-thresholding function defined as
\begin{equation}
S_{\alpha}(\vect{x}) =
\begin{cases} 
\left(|x_j| - \alpha\right)
\exp\left(i
\arctan\left[\frac{\mathrm{Im}\left(x_j\right)}
{\mathrm{Re}\left(x_j\right)}\right]
\right) & |x_j| > \alpha \\
0 & |x_j|\leq \alpha \\ 
\left(|x_j| + \alpha\right)
\exp\left(i
\arctan\left[\frac{\mathrm{Im}\left(x_j\right)}
{\mathrm{Re}\left(x_j\right)}\right]
\right) & |x_j| < -\alpha \end{cases},
\end{equation}
where $\mathrm{Re}(x)$ and $\mathrm{Im}(x)$ are the real and imaginary parts of $x$, respectively. 
Looking closer at Eq. (\ref{eq:MinX}) through Eq. (\ref{eq:LambUpdata}), we see that Eq. (\ref{eq:MinX}) will require the most computational operations due to the matrix inversion. As explained in \cite{boyd2011distributed}, we can make the inversion easier using the matrix-inversion lemma, which states
\begin{equation}
(P+\rho A^T A)^{-1} = P^{-1}-\rho P^{-1}A^T\left(\iden+\rho A P^{-1}A^T\right)^{-1}AP^{-1}.
\label{eq:InvLemma}
\end{equation}
Within Eq (\ref{eq:InvLemma}), $A^TA = \vect{A}^\dagger\vect{A}$ and $P = \Psi^\dagger\Psi$.
The expression simplifies considerably if $\Psi$ is unitary, i.e. $\Psi^\dagger\Psi = \iden$. Knowing that $(\vect{A}^{\dagger}\vect{A}+\rho \iden) \in \mathbb{C}^{n\times n}$, the matrix-inversion lemma states that
\begin{equation}
(\vect{A}^{\dagger}\vect{A} + \rho \iden)^{-1} = \frac{1}{\rho}\left[\iden - \frac{1}{\rho} \vect{A}^{\dagger}\left(\iden + \frac{1}{\rho}\vect{A}\vect{A}^{\dagger}\right)^{-1}\vect{A}\right]
\label{eq:submatrix}
\end{equation}
and we find that the inversion must be performed on the submatrix $\left(\iden + \frac{1}{\rho}\vect{A}\vect{A}^{\dagger}\right) \in \mathbb{C}^{m\times m}$ -- a significantly smaller matrix when $m \ll n $.

While finding the inverse of the submatrix in Eq. (\ref{eq:submatrix}) is more feasible, a more accurate solution exists by iteratively solving for $\vect{\mu}$ within the equation
\begin{align}
\left(\iden + \frac{1}{\rho} \vect{A}\vect{A}^{\dagger}\right)\vect{\mu} &= \vect{A}\left(\vect{A}^\dagger \vect{b}+\rho\Psi^{\dagger}\left(\hat{\vect{z}}^k-\vect{\lambda}^k/\rho\right)\right)\label{eq:subinverse} \\
\hat{\vect{x}}^{k+1} &= \frac{1}{\rho}\left[\iden-\vect{A}^{\dagger}\vect{\mu}\right], \\
\end{align}
where $\vect{\mu}\in\mathbb{C}^{m}$. The solution to (\ref{eq:subinverse})
can be obtained through iterative means such as conjugate-gradient or a lower-upper (LU) decomposition of $(\iden + 1/\rho\,\,\vect{AA}^{\dagger})$.
An LU decomposition is initially expensive but needs to be done only once since the decomposition can be saved and quickly applied for each iteration.
 
\subsubsection{Parameter tuning for ADMM-LASSO}

In order for the algorithm to converge quickly, the parameters $\beta$ and $\rho$ should be tuned to the problem at hand. However, tuning parameters for optimal performance may require a significant amount of time or require prior expertise. In addition, every time parameter $\rho$ is altered, the LU decomposition must be recalculated. We present a method introduced in \cite{boyd2011distributed} that automatically tunes the parameter $\rho$ for a given $\beta$. When the algorithm is designed to be self-tuning, only the parameter $\beta$ is required. An algorithm of this form is easier to use and addresses the heart of the problem within Eq. (\ref{prob:LASSO}), i.e. how much weight should be assigned to the penalty term when compared to the least-squares solution. Using an initial guess $\hat{\vect{x}}^0 = \vect{A}^{\dagger}\vect{b}$ and a user supplied $\beta$, the first $\rho$ parameter is defined as
\begin{equation}
\rho = \frac{\beta}{\mathrm{mean}\left(|\Psi\hat{\vect{x}}^0|\right)}.
\end{equation}
Within the first $k$ iterations, the algorithm will threshold away the mean value of the first guess $\hat{\vect{x}}^0$ at Eq. (\ref{eq:SoftThresh}) in favor of a sparse solution.
After $k$ iterations, we analyze the rates of change associated with $\vect{z}^{k+1} - \vect{z}^{k}$ and $\vect{x}^{k+1} - \vect{z}^{k+1}$. We require $\hat{\vect{z}} \rightarrow \hat{\vect{x}}$, but we want it to occur quickly. Specifically, we want the convergence of $\parallel \!\rho(\hat{\vect{z}}^{k+1} - \hat{\vect{z}}^{k})\! \parallel_2$ and $\parallel\! (\hat{\vect{x}}^{k+1} - \hat{\vect{z}}^{k+1})\! \parallel_2$ to be within a factor of 10 of one another. This is accomplished by adjusting $\rho$ after $k$ iterations according to the conditions
\begin{equation}
\rho^{k+1} =
\begin{cases}
2\rho^{k} & \parallel\! (\hat{\vect{x}}^{k+1} - \hat{\vect{z}}^{k+1})\! \parallel_2 > 10\parallel \!\rho(\hat{\vect{z}}^{k+1} - \hat{\vect{z}}^{k})\! \parallel_2 \\
\frac{1}{2}\rho^{k} & \parallel\! \rho(\hat{\vect{z}}^{k+1} - \hat{\vect{z}}^{k})\! \parallel_2 > 10\parallel \!(\hat{\vect{x}}^{k+1} - \hat{\vect{z}}^{k+1})\! \parallel_2 \\
\rho^{k}   & \mathrm{otherwise}
\end{cases}.
\label{eq:rho}
\end{equation}
The number of iterations $k$ to be executed before calling an evaluation of Eq. (\ref{eq:rho}) should be a small fraction of the total number of allowed iterations $k_{\mathrm{max}}$. In this manner, the number of LU decompositions remains small while adjusting parameters results in a faster convergence and an overall fewer number of necessary iterations to achieve a desired accuracy for any given stopping criterion. 

\subsubsection{ADMM-LASSO pseudocode}

The previous sections are summarized in the following pseudocode below.

\begin{algorithm}[H]
  \caption{ADMM-LASSO}
  \label{pseudocode}
   \begin{algorithmic}[1]
   \State $\hat{\vect{x}}^0 = \vect{A}^{\dagger}\vect{b}$
   \State $\rho^0 = \frac{\beta}{\mathrm{mean}\left(|\Psi\hat{\vect{x}}^0|\right)}$
   \State $\hat{\vect{z}}^{0} = S_{\beta/\rho^0}\left(\Psi\hat{\vect{x}}^0\right)$
   \State $\vect{\lambda}^{0} = \rho\left(\Psi\hat{\vect{x}}^0 - \hat{\vect{z}}^0\right)$
   \While{Not Converged}   
        \State $\hat{\vect{x}}^{k+1} \leftarrow \left(\vect{A}^\dagger \vect{A}+\rho^k \Psi^\dagger\Psi\right)^{-1}\left(\vect{A}^\dagger \vect{b}+\rho^k\Psi^{\dagger}\left(\hat{\vect{z}}^k-\vect{\lambda}^k/\rho^k\right)\right)$
        \State $\hat{\vect{z}}^{k+1} \leftarrow S_{\beta/\rho^k}\left(\Psi\hat{\vect{x}}^{k+1} + \vect{\lambda}^k/\rho^k\right)$
        \State $\vect{\lambda}^{k+1} \leftarrow \vect{\lambda}^k + \rho^k\left(\Psi\hat{\vect{x}}^{k+1} - \hat{\vect{z}}^{k+1}\right)$
        \If{$k =$ Large Enough}
            \State $\rho^{k+1} =
\begin{cases}
2\rho^k & \parallel\! (\hat{\vect{x}}^{k+1} - \hat{\vect{z}}^{k})\! \parallel_2 > 10\parallel \!\rho(\hat{\vect{z}}^{k+1} - \hat{\vect{z}}^{k})\! \parallel_2 \\
\frac{1}{2}\rho^k & \parallel\! \rho(\hat{\vect{z}}^{k+1} - \hat{\vect{z}}^{k})\! \parallel_2 > 10\parallel \!(\hat{\vect{x}}^{k+1} - \hat{\vect{z}}^{k})\! \parallel_2 \\
\rho^k   & \mathrm{otherwise}
\end{cases}$
        \EndIf 
   \EndWhile
   \end{algorithmic}
\end{algorithm}

\subsection{ADMM-LASSO with fast transforms and a unitary $\Psi$ }

In the previous section, we considered the case of a generalized sensing matrix $\vect{A}$. A significant speed enhancement can result from sampling with sensing matrices based on randomly permuted unitary matrices that have known fast-transform operations. Common examples include a fast-Fourier transform and a fast-Hadamard transform. This section will demonstrate how to design the sensing matrix and reconstruction algorithm such that we can take advantage of the fast transforms.

Let the matrix having a fast transform be represented as $\vect{F}\in\mathbb{C}^{n\times n}$. To randomize the matrix (making it less coherent) we introduce a row permutation matrix $\vect{P_r}\in\mathbb{R}^{m\times n}$ and a column permutation matrix $\vect{P_c}\in\mathbb{R}^{n\times n}$. The row permutation matrix $\vect{P_r}$ is formed by randomly choosing $m$ rows from $\iden\in\mathbb{R}^{n\times n}$ while the column permutation matrix is formed by randomly permuting the $n$ columns of $\iden\in\mathbb{R}^{n\times n}$. Thus, the resulting sensing matrix is $\vect{A} = \vect{P_r F P_c}$.

We can use the previous algorithm efficiently as long as we can use the matrix inversion lemma. Within the $\hat{\vect{x}}^k$ update, $\vect{A}^\dagger\vect{A} = \vect{P_c}^T\vect{F}^\dagger\vect{P_r}^T\vect{P_r F P_c}$. Typical fast transform operations are unitary or satisfy the condition $\vect{F}^\dagger\vect{F} = c\iden$, where $c$ is a constant. Additionally, $\vect{P_c}^T\vect{P_c} = \vect{P_c}\vect{P_c}^T = \iden$ while $\vect{P_r}^T\vect{P_r}\neq \iden$. The matrix inversion lemma will reverse the ordering and result in the expression $\vect{P_r}\vect{P_r}^T= \iden$. Thus, efficient matrix inversion in the previous algorithm is only possible if $\Psi$ is unitary and results in the expression
\begin{align}
\hat{\vect{x}}^{k+1} &= \left[\vect{A}^\dagger \vect{A}+\rho \Psi^\dagger\Psi\right]^{-1}\left[\vect{A}^\dagger \vect{b}+\rho\Psi^{\dagger}\left(\hat{\vect{z}}^k-\vect{\lambda}^k/\rho\right)\right] \\
&= \frac{1}{\rho}\left[\iden - \frac{1}{\rho}\vect{A}^{\dagger}\left(\iden + \frac{1}{\rho}\vect{A}\vect{A}^{\dagger}\right)^{-1}\vect{A}\right]\left[\vect{A}^\dagger \vect{b}+\rho\Psi^{\dagger}\left(\hat{\vect{z}}^k-\vect{\lambda}^k/\rho\right)\right] \\
&= \frac{1}{\rho}\left[\iden - \frac{1}{\rho}\vect{A}^{\dagger}\left(\iden + \frac{c}{\rho}\iden\right)^{-1}\vect{A}\right]\left[\vect{A}^\dagger \vect{b}+\rho\Psi^{\dagger}\left(\hat{\vect{z}}^k-\vect{\lambda}^k/\rho\right)\right] \\
& = \frac{1}{\rho}\left[\iden - \left(\rho+c\right)^{-1}\vect{A}^{\dagger}\vect{A}\right]\left[\vect{A}^\dagger \vect{b}+\rho\Psi^{\dagger}\left(\hat{\vect{z}}^k-\vect{\lambda}^k/\rho\right)\right],
\end{align}
where we have assumed $\vect{F}^\dagger\vect{F} = c\iden$. The update step is now trivial and based almost entirely on fast transform operations, aside from the $\vect{P_r}$, $\vect{P_c}$, and $\Psi$ operations.

\subsection{ADMM-LASSO with fast transforms and a non-unitary $\Psi$ }

In the case of small problems with feasible LU decompositions, the previous algorithm will still suffice when $\Psi$ is non-unitary. However, if the problem becomes too large, as is the case with megapixel image reconstructions, we must be clever to avoid the case of inverting matrices. In these instances, even inverting a submatrix with dimensions $m\times m$ can be a prohibitively expensive operation. Instead, we will leverage an additional degree of variable splitting and use fast transforms almost exclusively, as demonstrated in \cite{yin2010practical}. Although, the final expression will require a matrix inversion of the expression $(\alpha \iden + \beta\Psi^\dagger\Psi)^{-1}$, where $\alpha$ and $\beta$ are constants, within the $\hat{\vect{x}}$ update. Ideally, $\Psi$ should have a structure that enables the easy inversion of this expression. An example where structure is present is given in the next section when discussing total-variation minimization.

For the problem at hand, we wish to minimize
\begin{align}
\mathrm{minimize} \,\,\, &f(\hat{\vect{z}}) + g(\hat{\vect{u}}) \\
\mathrm{subject \, to} \,\,\,& \vect{F}\vect{P_c}\hat{\vect{x}} - \hat{\vect{u}} = 0 \\
& \Psi\hat{\vect{x}} - \hat{\vect{z}} = 0
\end{align}
where 
$f(\hat{\vect{z}}) = \alpha\parallel \hat{\vect{z}} \parallel_1$ and 
$g(\hat{\vect{u}}) = (1/2)\parallel \vect{P_r}\hat{\vect{u}}-\vect{b} \parallel_2^2$.
 
Knowing that $\vect{A} = \vect{P_rFP_c}$, we have effectively split the variables $\vect{A}\hat{\vect{x}}
 = \vect{P_r}\hat{\vect{u}} = (\vect{P_r})(\vect{FP_c}\hat{\vect{x}})$. The corresponding augmented Lagrangian is
\begin{align}
\mathcal{L}(\hat{\vect{x}},\hat{\vect{z}},\hat{\vect{u}},\vect{\lambda}) &=
\frac{1}{2}\norm{\vect{P_r}\hat{\vect{u}}-\vect{b}}_2^2 + \alpha\norm{\hat{\vect{z}}}_1 + \mean{\vect{\lambda_1}|\vect{FP_c}\hat{\vect{x}}-\hat{\vect{u}}} \nonumber \\ 
&+ \frac{\beta}{2}\norm{\vect{FP_c}\hat{\vect{x}}-\hat{\vect{u}}}_2^2+
\mean{\vect{\lambda_2}|\Psi\hat{\vect{x}}-\hat{\vect{z}}} + \frac{\gamma}{2}\norm{\Psi\hat{\vect{x}}-\hat{\vect{z}}}_2^2. \label{eq:Lagrange2}
\end{align}
Equation (\ref{eq:Lagrange2}) is minimized with the following operations:
\begin{align}
\hat{\vect{x}}^{k+1} &= \left(\beta\vect{P_c}^T\vect{F}^\dagger\vect{F}\vect{P_c}+\gamma\Psi^\dagger\Psi\right)^{-1}\left[\vect{P_c}^T\vect{F}^\dagger\left(\beta\hat{\vect{u}}^k-\vect{\lambda}_1^k\right) + \Psi^\dagger\left(\gamma\hat{\vect{z}}^k-\vect{\lambda}_2^k\right)\right] \\
\hat{\vect{u}}^{k+1} &= \left(\vect{P_r}^T\vect{P_r} + \beta\iden\right)^{-1}\left[\vect{P_r}^T\vect{b} + \beta\vect{FP_c}\hat{\vect{x}}^{k+1} + \vect{\lambda}^k_1 \right] \\
\hat{\vect{z}}^{k+1} &= S_{\alpha/\gamma}\left(\Psi\hat{\vect{x}}^{k+1} + \vect{\lambda}_2^k/\gamma\right) \\
\vect{\lambda}_1^{k+1} &= \vect{\lambda}_1^k + \beta\left(\vect{FP_c}\hat{\vect{x}}^{k+1} - \hat{\vect{u}}^{k+1}\right) \\
\vect{\lambda}_2^{k+1} &= \vect{\lambda}_2^{k} + \gamma\left(\Psi\hat{\vect{x}}^{k+1} - \hat{\vect{z}}^{k+1}\right).
\end{align}
The $\hat{\vect{u}}$ update is trivial because $\left(\vect{P_r}^T\vect{P_r} + \beta\iden\right)$ is diagonal. The only complicated update is the $\hat{\vect{x}}$ update. Yet, it can be simplified slightly if $\vect{F}^\dagger\vect{F} = c\iden$ such that
\begin{equation}
\beta\vect{P_c}^T\vect{F}^\dagger\vect{F}\vect{P_c}+\gamma\Psi^\dagger\Psi = c\beta\iden + \gamma\Psi^\dagger\Psi.
\label{eq:DifficultInverse}
\end{equation}
In the next section, we use the fact that if $\Psi$ obeys periodic boundary conditions, the combination $\Psi^\dagger\Psi$ is block circulant and the inverse of Eq. (\ref{eq:DifficultInverse}) can be easily diagonalized with a fast-Fourier transform \cite{ng1999fast}. 

The update parameters to improve performance are identical as in the previous algorithm and are not shown here. Instead, they may be found in the following algorithm for total-variation minimization for video sequences.

\subsection{ADMM-TV-Video with Fast-Hadamard Transforms}

This section introduce how to implement a total-variation minimization algorithm on a set of images found in a video. In addition to minimizing the total-variation in spatial coordinates, the variation in the time is also considered to reduce reconstruction artifacts that may appear frame to frame. To convert this algorithm such that it only works on a single image, merely omit the temporal component of the finite gradient operator and only use 2D reshaping operations with a 2D fast-Fourier transform for the $\hat{\vect{x}}$ update.   

We tackle a more general problem of sampling and reconstructing a vector-representation of a video $\vect{x}\in\mathbb{R}^{n_F\cdot n}$ composed of $n_F$ frames / images, with $n$ pixels per frame. Drawing from \cite{eckstein2012augmented,boyd2011distributed,yang2011alternating} and total-variation methods introduced in \cite{yin2010practical,chan2011augmented}, ADMM is applied to the total-variation (TV) minimization problem represented as
\begin{equation}
\min_{\vect{x}\in\mathbb{R}^{n_Fn}} \frac{1}{2}\parallel\! \vect{A}\,\vect{x} - \vect{b}\! \parallel^2_2 + \, \alpha \, TV(\vect{x}).
\label{eq:TV}
\end{equation}
where $\vect{A}$ is the sensing matrix, $\vect{b}$ is the data, and $TV(\vect{x})$ is the total-variation of $\vect{x}$ discussed later.
Instead of circulant matrices comprising the sensing matrix as presented in \cite{yin2010practical}, $\vect{A}\in \mathbb{R}^{n_F\cdot m\times n_F\cdot n}$ uses randomized Sylvester-ordered Hadamard matrices. Letting $\vect{F}\rightarrow \vect{H}$ (for Hadamard), the sensing matrix \vect{A} is based on Sylvester-ordered Hadamard matrices $\vect{H}\in\mathbb{R}^{m\times n}$ because of their speed of implementation arising from a fast-Hadamard transform. The $n_F$-frame sensing matrix takes the form
\begin{equation}
\vect{A} =
\left[
\begin{array}{cccc}
\vect{P_r}_1 \vect{H} \vect{P_c}_1 & 0 & \cdots & 0 \\
0 & \vect{P_r}_2 \vect{H} \vect{P_c}_2 & \cdots & 0 \\
\vdots & \vdots & \ddots & \vdots  \\
0 & 0 & \cdots & \vect{P_r}_F \vect{H} \vect{P_c}_F
\end{array}
\right],
\end{equation}
where $\vect{Pr}_i\in\mathbb{R}^{m\times n}$ is a random-subsampling row-permutation matrix and $\vect{P_c}_i\in\mathbb{R}^{n\times n}$ is a random column-permutation matrix, as before. 
Thus, each frame is sampled by a randomized and subsampled Hadamard sensing matrix $\vect{P_r}_i\,\vect{H}\,\vect{P_c}_i\in\mathbb{R}^{m\times n}$ for $i\in \{n_F\}$ frames. Within Eq. (\ref{eq:TV})$, \vect{b} \in \mathbb{R}^{n_FM}$ is the resulting measurement vector, $\alpha$ is a sparsity promoting constant, and $TV(\vect{x})$ is an anisotropic total-variation function defined as
\begin{equation}
TV(\vect{x}) = \parallel \nabla \vect{x} \parallel_1 =
\left\lVert \left[
\begin{array}{c}
 \omega_x\nabla_x  \\
 \omega_y\nabla_y  \\
 \omega_t\nabla_t 
\end{array}
\right] \vect{x}
\right\rVert_1,
\end{equation}
where $\omega_x$, $\omega_y$, and $\omega_t$ are are weighting coefficients assigned to stress the importance of one gradient over another.
We wish to exploit spatial sparsity that exists within the gradient of each image, as done in standard TV-minimization. Additionally, we also exploit sparsity that exists in the time-gradient between frames. Thus, $\nabla \in \mathbb{R}^{3n_F\cdot n \times n_F\cdot n}$ is composed of $\nabla_x$, $\nabla_y$, and $\nabla_t \in \mathbb{R}^{n_F\cdot n \times n_F \cdot n}$.  Each $\nabla_j$ (for $j \in \{x,y,t\}$) are first-order finite gradient operators that obey periodic bounary conditions. To define each $\nabla_j$, first let each $N$-pixel frame have a spatial resolution of $n = n_y \times n_x$, and then define the submatrices $\vect{D_x}\in\mathbb{R}^{n_x\times n_x}$, $\vect{D_y}\in\mathbb{R}^{n_y\times n_y}$, and $\vect{D_t}\in\mathbb{R}^{n_F\times n_F}$ such that
\begin{equation}
\vect{D_j} = 
\left[
\begin{array}{ccccc}
1       & -1  &        &         & 0 \\
        &  1  & -1     &         &   \\
        &     & \ddots & \ddots  &  \\
 0      &     &        &    1     & -1 \\
-1       &  0   &        &         & 1
\end{array}
\right] \text{ such that } \vect{j} \in \{ x,y,t \}.
\label{eq:D}
\end{equation}
Using the Kronecker product ($\otimes$), the difference operators are defined as
\begin{align}
\nabla_x &= \iden_{n_F} \otimes \left( \iden_{n_y} \otimes \vect{D_x} \right) \\
\nabla_y &= \iden_{n_F} \otimes \left( \vect{D_y} \otimes \iden_{n_x} \right) \\
\nabla_t &= \vect{D_t} \otimes \iden_{n_x \times n_y},
\end{align}
where $\iden_n\in\mathbb{R}^{n\times n}$ is an identity matrix. 

In ADMM form, TV-minimization takes the form
\begin{align}
\mathrm{minimize} \,\,\, &f(\vect{u}) + g(\vect{z}) \nonumber \\
\mathrm{subject\, to} \,\,\, & \vect{H\,P_c}\,\vect{x} - \vect{u} = \vect{0}, \nonumber \\
& \nabla\vect{x} - \vect{z} = \vect{0}.
\label{prob:ADMM_TV}
\end{align}
where $f(\vect{u}) = (1/2)\parallel \!\vect{P_r}\vect{u} - \vect{b}\!\parallel_2^2$ and $g(\vect{z}) = \alpha\parallel\! \vect{z}\! \parallel_1$.

Equation ($\ref{prob:ADMM_TV}$) has the following augmented Lagrangian:
\begin{align}
L_{\beta,\gamma}(\vect{x},\vect{z},\vect{u},\vect{\lambda}_1,\vect{\lambda}_2) &= \frac{1}{2}\parallel\vect{P_r}\,\vect{u}-\vect{b}\parallel_2^2 + \alpha\parallel\vect{z}\parallel_1 \nonumber \\
&+ \langle \vect{\lambda}_1 | \vect{H\,P_c\,x} - \vect{u} \rangle + \frac{\beta}{2}\parallel \! \vect{H\,P_c\,x} - \vect{u} \!\parallel_2^2 \label{eq:ADMM_Lasso_Lagrangian} \\ 
&+ \langle \vect{\lambda}_2 | \nabla\vect{x}-\vect{z} \rangle + \frac{\gamma}{2}\parallel \nabla\vect{x}-\vect{z} \parallel_2^2. \nonumber
\end{align}

By minimizing $L_{\beta,\gamma}(\vect{x},\vect{z},\vect{u},\vect{\lambda}_1,\vect{\lambda}_2)$ with respect to each variable, we find the necessary iterations to minimize Eq. (\ref{eq:ADMM_Lasso_Lagrangian}) at each iteraion $k$ as  
\begin{align}
\vect{z}^{k+1} &= S_{\alpha/\gamma}\left(\nabla\vect{x}^{k}+\frac{\vect{\lambda}_2^k}{\gamma}\right) \label{eq:zmin}\\
\vect{u}^{k+1} &= \left(\beta\iden + \vect{P_r}^T\vect{Pr}\right)^{-1}\left(\vect{P_r}^{T}\vect{b}+\beta\vect{H\,P_c\,x}^k+\vect{\lambda}_1^k\right) \label{eq:umin} \\
\vect{M}\vect{x}^{k+1} &= \left[\vect{P_c}^T\,\vect{H}^T\,\left(\beta\vect{u}^{k+1}-\vect{\lambda}_1^k\right)+\nabla^T\left(\gamma \vect{z}^{k+1}-\vect{\lambda}_2^k\right)\right] \label{eq:xmin} \\
\vect{\lambda}_{1}^{k+1} &= \vect{\lambda}_1^k + \beta\left(\vect{H\,P_c\,x}^{k+1} - \vect{u}^{k+1}\right) \\
\vect{\lambda}_{2}^{k+1} &= \vect{\lambda}_2^k + \gamma\left(\nabla\vect{x}^{k+1} - \vect{z}^{k+1}\right)
\end{align}
where $\vect{M} = \beta\vect{P_c}^T\,\vect{H}^T\,\vect{H}\,\vect{P_c}+\gamma\nabla^{T}\,\nabla$ and $^T$ is the transpose operation. Within Eq. (\ref{eq:zmin}), $S_{\alpha/\gamma}$ is an element-wise soft-thresholding function defined as
\begin{equation}
S_{\alpha/\gamma}(\vect{x}) =
\begin{cases} 
x_i - \frac{\alpha}{\gamma}\,\mathrm{sgn}(x_i) & |x_i| > \alpha/\gamma \\
0 & |x_i| \leq \alpha/\gamma \\ 
\end{cases},
\end{equation}
where $\mathrm{sgn}(x_i)$ returns the sign of the $i^{\mathrm{th}}$ element in $\vect{x}$.

Looking closer at Eq. (\ref{eq:zmin}) through Eq. (\ref{eq:xmin}), we see that the $\vect{z}$ update in Eq. (\ref{eq:zmin}) is trivially minimized by soft thresholding. The $\vect{u}$ update in Eq. (\ref{eq:umin}) is easily minimized becasue the matrix to be inverted is diagonal, i.e. $\vect{P_r}^T\vect{P_r} = \iden_M$. The $\vect{x}$ update within Eq. (\ref{eq:xmin}) is more difficult because $\vect{M}$ is not diagonal. However, $\beta \vect{P_c}^T\vect{H}^T\vect{H}\vect{P_c} = \beta N \iden_{FN}$. As outlined in \cite{rivenson2009practical}, $\nabla$ was constructed with periodic boundary conditions making $\nabla^T\nabla$ a block-circulant matrix. As a circulant matrix can be diagonalized by a Fourier transform, a block-circulant matrix can be diagonalized by a multi-dimensional Fourier transform. Thus, Eq. (\ref{eq:xmin}) can be efficiently solved using fast-Fourier transforms. If not solving iteratively, $\vect{x}$ has the solution
\begin{equation}
\vect{x}^{k+1} = \mathscr{F}^{-1}_{3\mathrm{D}}\left[\frac{\mathscr{F}_{3\mathrm{D}}\left[\vect{P_c}^T\,\vect{H}^T\,\left(\beta\vect{u}^{k+1}-\vect{\lambda}_1^{k}\right)+\nabla^T\left(\gamma \vect{z}^{k+1}-\vect{\lambda}_2^{k}\right)\right]}{\mathscr{F}_{3\mathrm{D}}\left[\vect{M_{\mathrm{col}}}\right]}\right],
\end{equation}  
where $\mathscr{F}_{3\mathrm{D}}$ is a three-dimensional Fourier transform of the first column element of $\vect{M}$, i.e. $\vect{M_{\mathrm{col}}}$. If only considering the total variation of a single image, using only $\nabla_x$ and $\nabla_y$, only a two-dimensional Fourier transform would be required. 

Additionally, we can leverage the block-circulant structure of $\vect{M}$ to make the $\vect{x}$ update more efficient. After expanding and simplifying,
\begin{equation}
\vect{M} = \beta N \iden_{FN} + \gamma\left(\nabla_x^T\nabla_x + \nabla_y^T\nabla_y + \nabla_t^T\nabla_t\right)
\end{equation}
Because a block-circulant matrix is determined by its first column, only the first column of $\nabla_T \nabla$ needs to be constructed. We construct the first column of each $\nabla_j^T\nabla_j$ and add the results together. Consider the column vector of $\nabla_x^T\nabla_x$ first. We extract the first row and column of $\vect{D_x}$, $\vect{D_y}$, and $\vect{D_t}$ within Eq. (\ref{eq:D}), labeled $\vect{D_j}_\mathrm{row}$ and $\vect{D_j}_\mathrm{col}$ for $\vect{j}\in\{\vect{x,y,t}\}$. Next construct the unit vectors $\vect{v}_{n_y} \in \mathbb{R}^{n_y}$, $\vect{v}_{n_x} \in \mathbb{R}^{n_x}$, $\vect{v}_{n_y n_x} \in \mathbb{R}^{n_y n_x}$, and $\vect{v}_{n_F} \in \mathbb{R}^{n_F}$ with only the first entry of each vector being 1. Finally, the first column of $\nabla^T\nabla$ is constructed in the following equations:
\begin{align}
\nabla_x^T\nabla_{x\,\mathrm{col}} &= \vect{v}_{n_F}\otimes\left[\vect{v}_{n_y}\otimes\left(\vect{D_x}_\mathrm{row}+\vect{D_x}_\mathrm{col}\right)\right] \\
\nabla_y^T\nabla_{y\,\mathrm{col}} &= \vect{v}_{n_F}\otimes\left[\left(\vect{D_{y}}_\mathrm{row}+\vect{D_y}_\mathrm{col}\right)\otimes\vect{v}_{n_x}\right] \\
\nabla_t^T\nabla_{t\,\mathrm{col}} &= \left(\vect{D_t}_\mathrm{row}+\vect{D_t}_\mathrm{col}\right)\otimes\vect{v}_{n_y n_x}\\
\nabla^T\nabla_{\mathrm{col}} &= \omega_x^2\, \nabla_x^T\nabla_{x\,\mathrm{col}} + \omega_y^2\, \nabla_y^T\nabla_{y\,\mathrm{col}} + \omega_t^2 \,\nabla_t^T\nabla_{t\,\mathrm{col}} \label{eq:Mcol}
\end{align}
Thus, $\mathcal{F}_{3\mathrm{D}}\left[\vect{M}\right]$ merely involves reshaping the first column of $\vect{M}$, using Eqn. (\ref{eq:Mcol}), into a three-dimensional object ($n_F\times n_x\times n_y$) and then applying a three-dimensional fast-Fourier transform.
 
\subsubsection{ADMM-TV-Video parameter tuning}

Because the $\vect{x}$ update can be quickly calculated, it is of little computational burden to occasionally alter the parameters $\beta$ and $\gamma$ to improve the algorithm's convergence rate.
In order for the algorithm to converge quickly, $\beta$ and $\gamma$ should be tuned to the problem at hand. Again, we use the method presented in \cite{boyd2011distributed} that automatically tunes $\beta$ and $\gamma$ for a given $\alpha$. Because we designed the algorithm to be self-tuning, only the parameter $\alpha$ is required. An algorithm of this form is easier to use and addresses the heart of the problem within Eq. (\ref{eq:TV}), i.e. how much weight should be assigned to the TV term when compared to the least-squares solution. Using an initial guess $\vect{x}^0 = \vect{A}^{T}\vect{b}$ and a user supplied $\alpha$, the first $\beta$ and $\gamma$ parameters can be set to small values, but emperically we found that $\beta = .1$ and $\gamma = \alpha/\mathrm{mean}(|\nabla\vect{x}^0|)$ appear to work well. Because these parameters will be tuned by the algorithm after a predefined set of iterations $k_{adj}$, it is not problematic if the values are suboptimal. 

Within the first $k$ iterations, the algorithm will threshold $\vect{x}^0$ at Eq. (\ref{eq:SoftThresh}) in favor of a sparse solution.
After $k_{adj}$ iterations, we analyze the rates of change associated with $\vect{u}^{k+1} - \vect{u}^{k}$ and $\vect{z}^{k+1} - \vect{z}^{k+1}$. 
We require $\vect{u} \rightarrow \vect{H\,P_c\,x}$ and $\vect{z}\rightarrow\nabla\vect{x}$ but we want it to occur quickly. Specifically, we want the convergence of $\parallel \!\beta ( \vect{u}^{k+1} - \vect{u}^{k})\! \parallel_2$ and 
$\parallel\! (\vect{H\,P_c\,x}^{k+1} - \vect{u}^{k+1})\! \parallel_2$ to be within a factor of 10 of one another. 
Additionally, we want the convergence of $\parallel \!\gamma(\vect{z}^{k+1} - \vect{z}^{k})\! \parallel_2$ and $\parallel\! (\nabla\vect{x}^{k+1} - \vect{z}^{k+1})\! \parallel_2$ to be within a factor of 10 of one another. This is accomplished by adjusting $\beta$ and $\gamma$ after $k$ iterations according to the conditions
\begin{equation}
\beta^{k+1} =
\begin{cases}
2\beta^{k} & \parallel\! (\vect{H\,P_c\,x}^{k+1} - \vect{u}^{k+1})\! \parallel_2 > 10\parallel \!\beta(\vect{u}^{k+1} - \vect{u}^{k})\! \parallel_2 \\
\frac{1}{2}\beta^{k} & \parallel\! \beta(\vect{u}^{k+1} - \vect{u}^{k})\! \parallel_2 > 10\parallel \!(\vect{H\,P_c\,x}^{k+1} - \vect{u}^{k+1})\! \parallel_2 \\
\beta^{k}   & \mathrm{otherwise}
\end{cases}.
\label{eq:beta}
\end{equation}

\begin{equation}
\gamma^{k+1} =
\begin{cases}
2\gamma^{k} & \parallel\! (\nabla\vect{x}^{k+1} - \vect{z}^{k+1})\! \parallel_2 > 10\parallel \!\gamma(\vect{z}^{k+1} - \vect{z}^{k})\! \parallel_2 \\
\frac{1}{2}\gamma^{k} & \parallel\! \gamma(\vect{z}^{k+1} - \vect{z}^{k})\! \parallel_2 > 10\parallel \!(\nabla\vect{x}^{k+1} - \vect{z}^{k+1})\! \parallel_2 \\
\gamma^{k}   & \mathrm{otherwise}
\end{cases}.
\label{eq:gamma}
\end{equation}

The number of iterations $k$ to be executed before calling an evaluation of Eq. (\ref{eq:beta}) and Eq. (\ref{eq:gamma}) should be a small fraction of the total number of allowed iterations $k_{\mathrm{max}}$. In this manner, the number of convergence evaluations and changes to $\vect{M}$ remains relatively small. Additionally, adjusting parameters infrequently results in a faster convergence and an overall fewer number of necessary iterations to achieve a desired accuracy for any given stopping criterion. 

\subsubsection{ADMM-TV-Video pseudocode}

The previous sections are summarized in the following pseudocode below.

\begin{algorithm}[H]
  \caption{ADMM-TV-Video}
  \label{pseudocode}
   \begin{algorithmic}[1]
   \State $\vect{x}^0 = \vect{A}^{T}\vect{b}$
   \State $\beta^0 = .1$
   \State $\gamma^0 = \frac{\alpha}{\mathrm{mean}(|\nabla\vect{x}^0|)}$
   \State $\vect{\lambda}_1^0 = \vect{0}$
   \State $\vect{\lambda}_2^0 = \vect{0}$
   \State Calculate $\vect{M_{\mathrm{col}}}$ from $\vect{M} = \beta^0 N \iden_{FN} + \gamma^k\nabla^T\nabla$
   \While{Not Converged}   
       \State $\vect{z}^{k+1} \leftarrow S_{\alpha/\gamma^k}\left(\nabla\vect{x}^{k}+\frac{\vect{\lambda}_2^k}{\gamma^k}\right)$
       
       \State $\vect{u}^{k+1} \leftarrow \left(\beta^k\iden + \vect{P_r}^T\vect{Pr}\right)^{-1}\left(\vect{P_r}^{T}\vect{b}+\beta^k\vect{H\,P_c\,x}^k+\vect{\lambda}_1^k\right)$
       
       \State $\vect{x}^{k+1} \leftarrow \mathscr{F}^{-1}_{3\mathrm{D}}\left[\frac{\mathscr{F}_{3\mathrm{D}}\left[\vect{P_c}^T\,\vect{H}^T\,\left(\beta^k\vect{u}^{k+1}-\vect{\lambda}_1^{k}\right)+\nabla^T\left(\gamma^k \vect{z}^{k+1}-\vect{\lambda}_2^{k}\right)\right]}{\mathscr{F}_{3\mathrm{D}}\left[\vect{M_{\mathrm{col}}}\right]}\right]$
       
       \State $\vect{\lambda}_{1}^{k+1} \leftarrow \vect{\lambda}_1^k + \beta^k\left(\vect{H\,P_c\,x}^{k+1} - \vect{u}^{k+1}\right)$
       
       \State $\vect{\lambda}_{2}^{k+1} \leftarrow \vect{\lambda}_2^k + \gamma^k\left(\nabla\vect{x}^{k+1} - \vect{z}^{k+1}\right)$

       \If{$k =$ Large Enough}           
            \State $\beta^{k+1} =
\begin{cases}
2\beta^{k} & \parallel\! (\vect{H\,P_c\,x}^{k+1} - \vect{u}^{k+1})\! \parallel_2 > 10\parallel \!\beta^k(\vect{u}^{k+1} - \vect{u}^{k})\! \parallel_2 \\
\frac{1}{2}\beta^{k} & \parallel\! \beta^k(\vect{u}^{k+1} - \vect{u}^{k})\! \parallel_2 > 10\parallel \!(\vect{H\,P_c\,x}^{k+1} - \vect{u}^{k+1})\! \parallel_2 \\
\beta^{k}   & \mathrm{otherwise}
\end{cases}$

            \State $\gamma^{k+1} =
\begin{cases}
2\gamma^{k} & \parallel\! (\nabla\vect{x}^{k+1} - \vect{z}^{k+1})\! \parallel_2 > 10\parallel \!\gamma^k(\vect{z}^{k+1} - \vect{z}^{k})\! \parallel_2 \\
\frac{1}{2}\gamma^{k} & \parallel\! \gamma^k(\vect{z}^{k+1} - \vect{z}^{k})\! \parallel_2 > 10\parallel \!(\nabla\vect{x}^{k+1} - \vect{z}^{k+1})\! \parallel_2 \\
\gamma^{k}   & \mathrm{otherwise}
\end{cases}$

            \If{$\beta^{k+1} \neq \beta^k$ or $\gamma^{k+1} \neq \gamma^k$} 
                \State Calculate $\vect{M_{\mathrm{col}}}$ from $\vect{M} = \beta^k N \iden_{FN} + \gamma^k\nabla^T\nabla$
            \EndIf
        \EndIf 
   \EndWhile
   \end{algorithmic}
\end{algorithm}

\makebibliography{bibliography}



\end{document}